\documentclass[twocolumn,traditabstract,longauth]{aa}
\usepackage{natbib}
\usepackage{url}
\usepackage{graphicx}
\usepackage{subfigure}
\usepackage{amsmath}
\usepackage{txfonts}
\usepackage{color}
\usepackage{listings} 
\usepackage{times}
\usepackage{multirow}
\usepackage{float}
\usepackage{morefloats}
\usepackage{ifthen}
\usepackage{longtable}
\bibpunct{(}{)}{;}{a}{}{,} 
\usepackage[colorlinks=true,linkcolor=red,citecolor=blue, breaklinks=true]{hyperref}

\newcommand{\StokesI}{{\ifmmode I\else $I$\fi}}     
\newcommand{\StokesQ}{{\ifmmode Q\else $Q$\fi}}     
\newcommand{\StokesU}{{\ifmmode U\else $U$\fi}}     

\newcommand{\polint}{{\ifmmode P\else $P$\fi}}                     

\newcommand{\polfrac}{{\it p\/}}                      
							
\newcommand{\polang}{\psi}                  

\def\setsymbol#1#2{\expandafter\def\csname #1\endcsname{#2}}
\def\getsymbol#1{\csname #1\endcsname}

\def\Planck{\textit{Planck}}

\def\all2013resultspapers{\nocite{planck2013-p01, planck2013-p02, planck2013-p02a, planck2013-p02d, planck2013-p02b, planck2013-p03, planck2013-p03c, planck2013-p03f, planck2013-p03d, planck2013-p03e, planck2013-p01a, planck2013-p06, planck2013-p03a, planck2013-pip88, planck2013-p08, planck2013-p11, planck2013-p12, planck2013-p13, planck2013-p14, planck2013-p15, planck2013-p05b, planck2013-p17, planck2013-p09, planck2013-p09a, planck2013-p20, planck2013-p19, planck2013-pipaberration, planck2013-p05, planck2013-p05a, planck2013-pip56, planck2013-p06b}}

\newbox\tablebox    \newdimen\tablewidth
\def\leaderfil{\leaders\hbox to 5pt{\hss.\hss}\hfil}
\def\endPlancktable{\tablewidth=\columnwidth 
    $$\hss\copy\tablebox\hss$$
    \vskip-\lastskip\vskip -2pt}
\def\endPlancktablewide{\tablewidth=\textwidth 
    $$\hss\copy\tablebox\hss$$
    \vskip-\lastskip\vskip -2pt}
\def\tablenote#1 #2\par{\begingroup \parindent=0.8em
    \abovedisplayshortskip=0pt\belowdisplayshortskip=0pt
    \noindent
    $$\hss\vbox{\hsize\tablewidth \hangindent=\parindent \hangafter=1 \noindent
    \hbox to \parindent{$^#1$\hss}\strut#2\strut\par}\hss$$
    \endgroup}
\def\doubleline{\vskip 3pt\hrule \vskip 1.5pt \hrule \vskip 5pt}

\def\L2{\ifmmode L_2\else $L_2$\fi}

\def\DeltaT{\ifmmode \Delta T\else $\Delta T$\fi}
\def\deltat{\ifmmode \Delta t\else $\Delta t$\fi}
\def\fknee{\ifmmode f_{\rm knee}\else $f_{\rm knee}$\fi}
\def\Fmax{\ifmmode F_{\rm max}\else $F_{\rm max}$\fi}
\def\solar{\ifmmode{\rm M}_{\mathord\odot}\else${\rm M}_{\mathord\odot}$\fi}
\def\Msolar{\ifmmode{\rm M}_{\mathord\odot}\else${\rm M}_{\mathord\odot}$\fi}
\def\Lsolar{\ifmmode{\rm L}_{\mathord\odot}\else${\rm L}_{\mathord\odot}$\fi}

\def\inv{\ifmmode^{-1}\else$^{-1}$\fi}
\def\mo{\ifmmode^{-1}\else$^{-1}$\fi}
\def\sup#1{\ifmmode ^{\rm #1}\else $^{\rm #1}$\fi}
\def\expo#1{\ifmmode \times 10^{#1}\else $\times 10^{#1}$\fi}
\def\,{\thinspace}
\def\lsim{\mathrel{\raise .4ex\hbox{\rlap{$<$}\lower 1.2ex\hbox{$\sim$}}}}
\def\gsim{\mathrel{\raise .4ex\hbox{\rlap{$>$}\lower 1.2ex\hbox{$\sim$}}}}

\def\simprop{\mathrel{\raise .4ex\hbox{\rlap{$\propto$}\lower 1.2ex\hbox{$\sim$}}}}
\def\deg{\ifmmode^\circ\else$^\circ$\fi}
\def\pdeg{\ifmmode $\setbox0=\hbox{$^{\circ}$}\rlap{\hskip.11\wd0 .}$^{\circ}
          \else \setbox0=\hbox{$^{\circ}$}\rlap{\hskip.11\wd0 .}$^{\circ}$\fi}
\def\arcs{\ifmmode {^{\scriptstyle\prime\prime}}
          \else $^{\scriptstyle\prime\prime}$\fi}
\def\arcm{\ifmmode {^{\scriptstyle\prime}}
          \else $^{\scriptstyle\prime}$\fi}
\newdimen\sa  \newdimen\sb
\def\parcs{\sa=.07em \sb=.03em
     \ifmmode \hbox{\rlap{.}}^{\scriptstyle\prime\kern -\sb\prime}\hbox{\kern -\sa}
     \else \rlap{.}$^{\scriptstyle\prime\kern -\sb\prime}$\kern -\sa\fi}
\def\parcm{\sa=.08em \sb=.03em
     \ifmmode \hbox{\rlap{.}\kern\sa}^{\scriptstyle\prime}\hbox{\kern-\sb}
     \else \rlap{.}\kern\sa$^{\scriptstyle\prime}$\kern-\sb\fi}
\def\ra[#1 #2 #3.#4]{#1\sup{h}#2\sup{m}#3\sup{s}\llap.#4}
\def\dec[#1 #2 #3.#4]{#1\deg#2\arcm#3\arcs\llap.#4}
\def\deco[#1 #2 #3]{#1\deg#2\arcm#3\arcs}
\def\rra[#1 #2]{#1\sup{h}#2\sup{m}}

\def\dots{\relax\ifmmode \ldots\else $\ldots$\fi}
\def\WHzsr{\ifmmode $W\,Hz\mo\,sr\mo$\else W\,Hz\mo\,sr\mo\fi}
\def\mHz{\ifmmode $\,mHz$\else \,mHz\fi}
\def\GHz{\ifmmode $\,GHz$\else \,GHz\fi}
\def\mKs{\ifmmode $\,mK\,s$^{1/2}\else \,mK\,s$^{1/2}$\fi}
\def\muKs{\ifmmode \,\mu$K\,s$^{1/2}\else \,$\mu$K\,s$^{1/2}$\fi}
\def\muKRJs{\ifmmode \,\mu$K$_{\rm RJ}$\,s$^{1/2}\else \,$\mu$K$_{\rm RJ}$\,s$^{1/2}$\fi}
\def\muKHz{\ifmmode \,\mu$K\,Hz$^{-1/2}\else \,$\mu$K\,Hz$^{-1/2}$\fi}
\def\MJysr{\ifmmode \,$MJy\,sr\mo$\else \,MJy\,sr\mo\fi}
\def\MJysrmK{\ifmmode \,$MJy\,sr\mo$\,mK$_{\rm CMB}\mo\else \,MJy\,sr\mo\,mK$_{\rm CMB}\mo$\fi}
\def\microns{\ifmmode \,\mu$m$\else \,$\mu$m\fi}

\def\muK{\ifmmode \,\mu$K$\else \,$\mu$\hbox{K}\fi}
\def\microK{\ifmmode \,\mu$K$\else \,$\mu$\hbox{K}\fi}
\def\muW{\ifmmode \,\mu$W$\else \,$\mu$\hbox{W}\fi}
\def\kms{\ifmmode $\,km\,s$^{-1}\else \,km\,s$^{-1}$\fi}
\def\kmsMpc{\ifmmode $\,\kms\,Mpc\mo$\else \,\kms\,Mpc\mo\fi}
\setsymbol{LFI:center:frequency:70GHz:units}{70.3\,GHz}
\setsymbol{LFI:center:frequency:44GHz:units}{44.1\,GHz}
\setsymbol{LFI:center:frequency:30GHz:units}{28.5\,GHz}

\setsymbol{LFI:center:frequency:70GHz}{70.3}
\setsymbol{LFI:center:frequency:44GHz}{44.1}
\setsymbol{LFI:center:frequency:30GHz}{28.5}

\setsymbol{LFI:center:frequency:LFI18:Rad:M:units}{71.7\GHz}
\setsymbol{LFI:center:frequency:LFI19:Rad:M:units}{67.5\GHz}
\setsymbol{LFI:center:frequency:LFI20:Rad:M:units}{69.2\GHz}
\setsymbol{LFI:center:frequency:LFI21:Rad:M:units}{70.4\GHz}
\setsymbol{LFI:center:frequency:LFI22:Rad:M:units}{71.5\GHz}
\setsymbol{LFI:center:frequency:LFI23:Rad:M:units}{70.8\GHz}
\setsymbol{LFI:center:frequency:LFI24:Rad:M:units}{44.4\GHz}
\setsymbol{LFI:center:frequency:LFI25:Rad:M:units}{44.0\GHz}
\setsymbol{LFI:center:frequency:LFI26:Rad:M:units}{43.9\GHz}
\setsymbol{LFI:center:frequency:LFI27:Rad:M:units}{28.3\GHz}
\setsymbol{LFI:center:frequency:LFI28:Rad:M:units}{28.8\GHz}
\setsymbol{LFI:center:frequency:LFI18:Rad:S:units}{70.1\GHz}
\setsymbol{LFI:center:frequency:LFI19:Rad:S:units}{69.6\GHz}
\setsymbol{LFI:center:frequency:LFI20:Rad:S:units}{69.5\GHz}
\setsymbol{LFI:center:frequency:LFI21:Rad:S:units}{69.5\GHz}
\setsymbol{LFI:center:frequency:LFI22:Rad:S:units}{72.8\GHz}
\setsymbol{LFI:center:frequency:LFI23:Rad:S:units}{71.3\GHz}
\setsymbol{LFI:center:frequency:LFI24:Rad:S:units}{44.1\GHz}
\setsymbol{LFI:center:frequency:LFI25:Rad:S:units}{44.1\GHz}
\setsymbol{LFI:center:frequency:LFI26:Rad:S:units}{44.1\GHz}
\setsymbol{LFI:center:frequency:LFI27:Rad:S:units}{28.5\GHz}
\setsymbol{LFI:center:frequency:LFI28:Rad:S:units}{28.2\GHz}

\setsymbol{LFI:center:frequency:LFI18:Rad:M}{71.7}
\setsymbol{LFI:center:frequency:LFI19:Rad:M}{67.5}
\setsymbol{LFI:center:frequency:LFI20:Rad:M}{69.2}
\setsymbol{LFI:center:frequency:LFI21:Rad:M}{70.4}
\setsymbol{LFI:center:frequency:LFI22:Rad:M}{71.5}
\setsymbol{LFI:center:frequency:LFI23:Rad:M}{70.8}
\setsymbol{LFI:center:frequency:LFI24:Rad:M}{44.4}
\setsymbol{LFI:center:frequency:LFI25:Rad:M}{44.0}
\setsymbol{LFI:center:frequency:LFI26:Rad:M}{43.9}
\setsymbol{LFI:center:frequency:LFI27:Rad:M}{28.3}
\setsymbol{LFI:center:frequency:LFI28:Rad:M}{28.8}
\setsymbol{LFI:center:frequency:LFI18:Rad:S}{70.1}
\setsymbol{LFI:center:frequency:LFI19:Rad:S}{69.6}
\setsymbol{LFI:center:frequency:LFI20:Rad:S}{69.5}
\setsymbol{LFI:center:frequency:LFI21:Rad:S}{69.5}
\setsymbol{LFI:center:frequency:LFI22:Rad:S}{72.8}
\setsymbol{LFI:center:frequency:LFI23:Rad:S}{71.3}
\setsymbol{LFI:center:frequency:LFI24:Rad:S}{44.1}
\setsymbol{LFI:center:frequency:LFI25:Rad:S}{44.1}
\setsymbol{LFI:center:frequency:LFI26:Rad:S}{44.1}
\setsymbol{LFI:center:frequency:LFI27:Rad:S}{28.5}
\setsymbol{LFI:center:frequency:LFI28:Rad:S}{28.2}

\setsymbol{LFI:white:noise:sensitivity:70GHz:units}{134.7\muKs}
\setsymbol{LFI:white:noise:sensitivity:44GHz:units}{164.7\muKs}
\setsymbol{LFI:white:noise:sensitivity:30GHz:units}{143.4\muKs}

\setsymbol{LFI:white:noise:sensitivity:70GHz}{134.7}
\setsymbol{LFI:white:noise:sensitivity:44GHz}{164.7}
\setsymbol{LFI:white:noise:sensitivity:30GHz}{143.4}

\setsymbol{LFI:white:noise:sensitivity:LFI18:Rad:M:units}{512.0\muKs}
\setsymbol{LFI:white:noise:sensitivity:LFI19:Rad:M:units}{581.4\muKs}
\setsymbol{LFI:white:noise:sensitivity:LFI20:Rad:M:units}{590.8\muKs}
\setsymbol{LFI:white:noise:sensitivity:LFI21:Rad:M:units}{455.2\muKs}
\setsymbol{LFI:white:noise:sensitivity:LFI22:Rad:M:units}{492.0\muKs}
\setsymbol{LFI:white:noise:sensitivity:LFI23:Rad:M:units}{507.7\muKs}
\setsymbol{LFI:white:noise:sensitivity:LFI24:Rad:M:units}{462.2\muKs}
\setsymbol{LFI:white:noise:sensitivity:LFI25:Rad:M:units}{413.6\muKs}
\setsymbol{LFI:white:noise:sensitivity:LFI26:Rad:M:units}{478.6\muKs}
\setsymbol{LFI:white:noise:sensitivity:LFI27:Rad:M:units}{277.7\muKs}
\setsymbol{LFI:white:noise:sensitivity:LFI28:Rad:M:units}{312.3\muKs}
\setsymbol{LFI:white:noise:sensitivity:LFI18:Rad:S:units}{465.7\muKs}
\setsymbol{LFI:white:noise:sensitivity:LFI19:Rad:S:units}{555.6\muKs}
\setsymbol{LFI:white:noise:sensitivity:LFI20:Rad:S:units}{623.2\muKs}
\setsymbol{LFI:white:noise:sensitivity:LFI21:Rad:S:units}{564.1\muKs}
\setsymbol{LFI:white:noise:sensitivity:LFI22:Rad:S:units}{534.4\muKs}
\setsymbol{LFI:white:noise:sensitivity:LFI23:Rad:S:units}{542.4\muKs}
\setsymbol{LFI:white:noise:sensitivity:LFI24:Rad:S:units}{399.2\muKs}
\setsymbol{LFI:white:noise:sensitivity:LFI25:Rad:S:units}{392.6\muKs}
\setsymbol{LFI:white:noise:sensitivity:LFI26:Rad:S:units}{418.6\muKs}
\setsymbol{LFI:white:noise:sensitivity:LFI27:Rad:S:units}{302.9\muKs}
\setsymbol{LFI:white:noise:sensitivity:LFI28:Rad:S:units}{285.3\muKs}

\setsymbol{LFI:white:noise:sensitivity:LFI18:Rad:M}{512.0}
\setsymbol{LFI:white:noise:sensitivity:LFI19:Rad:M}{581.4}
\setsymbol{LFI:white:noise:sensitivity:LFI20:Rad:M}{590.8}
\setsymbol{LFI:white:noise:sensitivity:LFI21:Rad:M}{455.2}
\setsymbol{LFI:white:noise:sensitivity:LFI22:Rad:M}{492.0}
\setsymbol{LFI:white:noise:sensitivity:LFI23:Rad:M}{507.7}
\setsymbol{LFI:white:noise:sensitivity:LFI24:Rad:M}{462.2}
\setsymbol{LFI:white:noise:sensitivity:LFI25:Rad:M}{413.6}
\setsymbol{LFI:white:noise:sensitivity:LFI26:Rad:M}{478.6}
\setsymbol{LFI:white:noise:sensitivity:LFI27:Rad:M}{277.7}
\setsymbol{LFI:white:noise:sensitivity:LFI28:Rad:M}{312.3}
\setsymbol{LFI:white:noise:sensitivity:LFI18:Rad:S}{465.7}
\setsymbol{LFI:white:noise:sensitivity:LFI19:Rad:S}{555.6}
\setsymbol{LFI:white:noise:sensitivity:LFI20:Rad:S}{623.2}
\setsymbol{LFI:white:noise:sensitivity:LFI21:Rad:S}{564.1}
\setsymbol{LFI:white:noise:sensitivity:LFI22:Rad:S}{534.4}
\setsymbol{LFI:white:noise:sensitivity:LFI23:Rad:S}{542.4}
\setsymbol{LFI:white:noise:sensitivity:LFI24:Rad:S}{399.2}
\setsymbol{LFI:white:noise:sensitivity:LFI25:Rad:S}{392.6}
\setsymbol{LFI:white:noise:sensitivity:LFI26:Rad:S}{418.6}
\setsymbol{LFI:white:noise:sensitivity:LFI27:Rad:S}{302.9}
\setsymbol{LFI:white:noise:sensitivity:LFI28:Rad:S}{285.3}

\setsymbol{LFI:knee:frequency:70GHz:units}{29.5\mHz}
\setsymbol{LFI:knee:frequency:44GHz:units}{56.2\mHz}
\setsymbol{LFI:knee:frequency:30GHz:units}{113.7\mHz}

\setsymbol{LFI:knee:frequency:70GHz}{29.5}
\setsymbol{LFI:knee:frequency:44GHz}{56.2}
\setsymbol{LFI:knee:frequency:30GHz}{113.7}

\setsymbol{LFI:knee:frequency:LFI18:Rad:M:units}{16.3\mHz}
\setsymbol{LFI:knee:frequency:LFI19:Rad:M:units}{15.1\mHz}
\setsymbol{LFI:knee:frequency:LFI20:Rad:M:units}{18.7\mHz}
\setsymbol{LFI:knee:frequency:LFI21:Rad:M:units}{37.2\mHz}
\setsymbol{LFI:knee:frequency:LFI22:Rad:M:units}{12.7\mHz}
\setsymbol{LFI:knee:frequency:LFI23:Rad:M:units}{34.6\mHz}
\setsymbol{LFI:knee:frequency:LFI24:Rad:M:units}{46.2\mHz}
\setsymbol{LFI:knee:frequency:LFI25:Rad:M:units}{24.9\mHz}
\setsymbol{LFI:knee:frequency:LFI26:Rad:M:units}{67.6\mHz}
\setsymbol{LFI:knee:frequency:LFI27:Rad:M:units}{187.4\mHz}
\setsymbol{LFI:knee:frequency:LFI28:Rad:M:units}{122.2\mHz}
\setsymbol{LFI:knee:frequency:LFI18:Rad:S:units}{17.7\mHz}
\setsymbol{LFI:knee:frequency:LFI19:Rad:S:units}{22.0\mHz}
\setsymbol{LFI:knee:frequency:LFI20:Rad:S:units}{8.7\mHz}
\setsymbol{LFI:knee:frequency:LFI21:Rad:S:units}{25.9\mHz}
\setsymbol{LFI:knee:frequency:LFI22:Rad:S:units}{15.8\mHz}
\setsymbol{LFI:knee:frequency:LFI23:Rad:S:units}{129.8\mHz}
\setsymbol{LFI:knee:frequency:LFI24:Rad:S:units}{100.9\mHz}
\setsymbol{LFI:knee:frequency:LFI25:Rad:S:units}{38.9\mHz}
\setsymbol{LFI:knee:frequency:LFI26:Rad:S:units}{58.9\mHz}
\setsymbol{LFI:knee:frequency:LFI27:Rad:S:units}{104.4\mHz}
\setsymbol{LFI:knee:frequency:LFI28:Rad:S:units}{40.7\mHz}

\setsymbol{LFI:knee:frequency:LFI18:Rad:M}{16.3}
\setsymbol{LFI:knee:frequency:LFI19:Rad:M}{15.1}
\setsymbol{LFI:knee:frequency:LFI20:Rad:M}{18.7}
\setsymbol{LFI:knee:frequency:LFI21:Rad:M}{37.2}
\setsymbol{LFI:knee:frequency:LFI22:Rad:M}{12.7}
\setsymbol{LFI:knee:frequency:LFI23:Rad:M}{34.6}
\setsymbol{LFI:knee:frequency:LFI24:Rad:M}{46.2}
\setsymbol{LFI:knee:frequency:LFI25:Rad:M}{24.9}
\setsymbol{LFI:knee:frequency:LFI26:Rad:M}{67.6}
\setsymbol{LFI:knee:frequency:LFI27:Rad:M}{187.4}
\setsymbol{LFI:knee:frequency:LFI28:Rad:M}{122.2}
\setsymbol{LFI:knee:frequency:LFI18:Rad:S}{17.7}
\setsymbol{LFI:knee:frequency:LFI19:Rad:S}{22.0}
\setsymbol{LFI:knee:frequency:LFI20:Rad:S}{8.7}
\setsymbol{LFI:knee:frequency:LFI21:Rad:S}{25.9}
\setsymbol{LFI:knee:frequency:LFI22:Rad:S}{15.8}
\setsymbol{LFI:knee:frequency:LFI23:Rad:S}{129.8}
\setsymbol{LFI:knee:frequency:LFI24:Rad:S}{100.9}
\setsymbol{LFI:knee:frequency:LFI25:Rad:S}{38.9}
\setsymbol{LFI:knee:frequency:LFI26:Rad:S}{58.9}
\setsymbol{LFI:knee:frequency:LFI27:Rad:S}{104.4}
\setsymbol{LFI:knee:frequency:LFI28:Rad:S}{40.7}

\setsymbol{LFI:slope:70GHz:units}{$-1.03$\mHz}
\setsymbol{LFI:slope:44GHz:units}{$-0.89$\mHz}
\setsymbol{LFI:slope:30GHz:units}{$-0.87$\mHz}

\setsymbol{LFI:slope:70GHz}{$-1.03$}
\setsymbol{LFI:slope:44GHz}{$-0.89$}
\setsymbol{LFI:slope:30GHz}{$-0.87$}

\setsymbol{LFI:slope:LFI18:Rad:M:units}{$-1.04$\mHz}
\setsymbol{LFI:slope:LFI19:Rad:M:units}{$-1.09$\mHz}
\setsymbol{LFI:slope:LFI20:Rad:M:units}{$-0.69$\mHz}
\setsymbol{LFI:slope:LFI21:Rad:M:units}{$-1.56$\mHz}
\setsymbol{LFI:slope:LFI22:Rad:M:units}{$-1.01$\mHz}
\setsymbol{LFI:slope:LFI23:Rad:M:units}{$-0.96$\mHz}
\setsymbol{LFI:slope:LFI24:Rad:M:units}{$-0.83$\mHz}
\setsymbol{LFI:slope:LFI25:Rad:M:units}{$-0.91$\mHz}
\setsymbol{LFI:slope:LFI26:Rad:M:units}{$-0.95$\mHz}
\setsymbol{LFI:slope:LFI27:Rad:M:units}{$-0.87$\mHz}
\setsymbol{LFI:slope:LFI28:Rad:M:units}{$-0.88$\mHz}
\setsymbol{LFI:slope:LFI18:Rad:S:units}{$-1.15$\mHz}
\setsymbol{LFI:slope:LFI19:Rad:S:units}{$-1.00$\mHz}
\setsymbol{LFI:slope:LFI20:Rad:S:units}{$-0.95$\mHz}
\setsymbol{LFI:slope:LFI21:Rad:S:units}{$-0.92$\mHz}
\setsymbol{LFI:slope:LFI22:Rad:S:units}{$-1.01$\mHz}
\setsymbol{LFI:slope:LFI23:Rad:S:units}{$-0.95$\mHz}
\setsymbol{LFI:slope:LFI24:Rad:S:units}{$-0.73$\mHz}
\setsymbol{LFI:slope:LFI25:Rad:S:units}{$-1.16$\mHz}
\setsymbol{LFI:slope:LFI26:Rad:S:units}{$-0.79$\mHz}
\setsymbol{LFI:slope:LFI27:Rad:S:units}{$-0.82$\mHz}
\setsymbol{LFI:slope:LFI28:Rad:S:units}{$-0.91$\mHz}

\setsymbol{LFI:slope:LFI18:Rad:M}{$-1.04$}
\setsymbol{LFI:slope:LFI19:Rad:M}{$-1.09$}
\setsymbol{LFI:slope:LFI20:Rad:M}{$-0.69$}
\setsymbol{LFI:slope:LFI21:Rad:M}{$-1.56$}
\setsymbol{LFI:slope:LFI22:Rad:M}{$-1.01$}
\setsymbol{LFI:slope:LFI23:Rad:M}{$-0.96$}
\setsymbol{LFI:slope:LFI24:Rad:M}{$-0.83$}
\setsymbol{LFI:slope:LFI25:Rad:M}{$-0.91$}
\setsymbol{LFI:slope:LFI26:Rad:M}{$-0.95$}
\setsymbol{LFI:slope:LFI27:Rad:M}{$-0.87$}
\setsymbol{LFI:slope:LFI28:Rad:M}{$-0.88$}
\setsymbol{LFI:slope:LFI18:Rad:S}{$-1.15$}
\setsymbol{LFI:slope:LFI19:Rad:S}{$-1.00$}
\setsymbol{LFI:slope:LFI20:Rad:S}{$-0.95$}
\setsymbol{LFI:slope:LFI21:Rad:S}{$-0.92$}
\setsymbol{LFI:slope:LFI22:Rad:S}{$-1.01$}
\setsymbol{LFI:slope:LFI23:Rad:S}{$-0.95$}
\setsymbol{LFI:slope:LFI24:Rad:S}{$-0.73$}
\setsymbol{LFI:slope:LFI25:Rad:S}{$-1.16$}
\setsymbol{LFI:slope:LFI26:Rad:S}{$-0.79$}
\setsymbol{LFI:slope:LFI27:Rad:S}{$-0.82$}
\setsymbol{LFI:slope:LFI28:Rad:S}{$-0.91$}

\setsymbol{LFI:FWHM:70GHz:units}{13\parcm01}
\setsymbol{LFI:FWHM:44GHz:units}{27\parcm92}
\setsymbol{LFI:FWHM:30GHz:units}{32\parcm65}

\setsymbol{LFI:FWHM:70GHz}{13.01}
\setsymbol{LFI:FWHM:44GHz}{27.92}
\setsymbol{LFI:FWHM:30GHz}{32.65}

\setsymbol{LFI:FWHM:LFI18:units}{13\parcm39}
\setsymbol{LFI:FWHM:LFI19:units}{13\parcm01}
\setsymbol{LFI:FWHM:LFI20:units}{12\parcm75}
\setsymbol{LFI:FWHM:LFI21:units}{12\parcm74}
\setsymbol{LFI:FWHM:LFI22:units}{12\parcm87}
\setsymbol{LFI:FWHM:LFI23:units}{13\parcm27}
\setsymbol{LFI:FWHM:LFI24:units}{22\parcm98}
\setsymbol{LFI:FWHM:LFI25:units}{30\parcm46}
\setsymbol{LFI:FWHM:LFI26:units}{30\parcm31}
\setsymbol{LFI:FWHM:LFI27:units}{32\parcm65}
\setsymbol{LFI:FWHM:LFI28:units}{32\parcm66}

\setsymbol{LFI:FWHM:LFI18}{13.39}
\setsymbol{LFI:FWHM:LFI19}{13.01}
\setsymbol{LFI:FWHM:LFI20}{12.75}
\setsymbol{LFI:FWHM:LFI21}{12.74}
\setsymbol{LFI:FWHM:LFI22}{12.87}
\setsymbol{LFI:FWHM:LFI23}{13.27}
\setsymbol{LFI:FWHM:LFI24}{22.98}
\setsymbol{LFI:FWHM:LFI25}{30.46}
\setsymbol{LFI:FWHM:LFI26}{30.31}
\setsymbol{LFI:FWHM:LFI27}{32.65}
\setsymbol{LFI:FWHM:LFI28}{32.66}

\setsymbol{LFI:FWHM:uncertainty:LFI18:units}{0.170\arcm}
\setsymbol{LFI:FWHM:uncertainty:LFI19:units}{0.174\arcm}
\setsymbol{LFI:FWHM:uncertainty:LFI20:units}{0.170\arcm}
\setsymbol{LFI:FWHM:uncertainty:LFI21:units}{0.156\arcm}
\setsymbol{LFI:FWHM:uncertainty:LFI22:units}{0.164\arcm}
\setsymbol{LFI:FWHM:uncertainty:LFI23:units}{0.171\arcm}
\setsymbol{LFI:FWHM:uncertainty:LFI24:units}{0.652\arcm}
\setsymbol{LFI:FWHM:uncertainty:LFI25:units}{1.075\arcm}
\setsymbol{LFI:FWHM:uncertainty:LFI26:units}{1.131\arcm}
\setsymbol{LFI:FWHM:uncertainty:LFI27:units}{1.266\arcm}
\setsymbol{LFI:FWHM:uncertainty:LFI28:units}{1.287\arcm}

\setsymbol{LFI:FWHM:uncertainty:LFI18}{0.170}
\setsymbol{LFI:FWHM:uncertainty:LFI19}{0.174}
\setsymbol{LFI:FWHM:uncertainty:LFI20}{0.170}
\setsymbol{LFI:FWHM:uncertainty:LFI21}{0.156}
\setsymbol{LFI:FWHM:uncertainty:LFI22}{0.164}
\setsymbol{LFI:FWHM:uncertainty:LFI23}{0.171}
\setsymbol{LFI:FWHM:uncertainty:LFI24}{0.652}
\setsymbol{LFI:FWHM:uncertainty:LFI25}{1.075}
\setsymbol{LFI:FWHM:uncertainty:LFI26}{1.131}
\setsymbol{LFI:FWHM:uncertainty:LFI27}{1.266}
\setsymbol{LFI:FWHM:uncertainty:LFI28}{1.287}

\setsymbol{HFI:center:frequency:100GHz:units}{100\,GHz}
\setsymbol{HFI:center:frequency:143GHz:units}{143\,GHz}
\setsymbol{HFI:center:frequency:217GHz:units}{217\,GHz}
\setsymbol{HFI:center:frequency:353GHz:units}{353\,GHz}
\setsymbol{HFI:center:frequency:545GHz:units}{545\,GHz}
\setsymbol{HFI:center:frequency:857GHz:units}{857\,GHz}

\setsymbol{HFI:center:frequency:100GHz}{100}
\setsymbol{HFI:center:frequency:143GHz}{143}
\setsymbol{HFI:center:frequency:217GHz}{217}
\setsymbol{HFI:center:frequency:353GHz}{353}
\setsymbol{HFI:center:frequency:545GHz}{545}
\setsymbol{HFI:center:frequency:857GHz}{857}

\setsymbol{HFI:Ndetectors:100GHz}{8}
\setsymbol{HFI:Ndetectors:143GHz}{11}
\setsymbol{HFI:Ndetectors:217GHz}{12}
\setsymbol{HFI:Ndetectors:353GHz}{12}
\setsymbol{HFI:Ndetectors:545GHz}{3}
\setsymbol{HFI:Ndetectors:857GHz}{4}

\setsymbol{HFI:FWHM:Maps:100GHz:units}{9\parcm88}
\setsymbol{HFI:FWHM:Maps:143GHz:units}{7\parcm18}
\setsymbol{HFI:FWHM:Maps:217GHz:units}{4\parcm87}
\setsymbol{HFI:FWHM:Maps:353GHz:units}{4\parcm65}
\setsymbol{HFI:FWHM:Maps:545GHz:units}{4\parcm72}
\setsymbol{HFI:FWHM:Maps:857GHz:units}{4\parcm39}
\setsymbol{HFI:FWHM:Maps:100GHz}{9.88}
\setsymbol{HFI:FWHM:Maps:143GHz}{7.18}
\setsymbol{HFI:FWHM:Maps:217GHz}{4.87}
\setsymbol{HFI:FWHM:Maps:353GHz}{4.65}
\setsymbol{HFI:FWHM:Maps:545GHz}{4.72}
\setsymbol{HFI:FWHM:Maps:857GHz}{4.39}

\setsymbol{HFI:beam:ellipticity:Maps:100GHz}{1.15}
\setsymbol{HFI:beam:ellipticity:Maps:143GHz}{1.01}
\setsymbol{HFI:beam:ellipticity:Maps:217GHz}{1.06}
\setsymbol{HFI:beam:ellipticity:Maps:353GHz}{1.05}
\setsymbol{HFI:beam:ellipticity:Maps:545GHz}{1.14}
\setsymbol{HFI:beam:ellipticity:Maps:857GHz}{1.19}

\setsymbol{HFI:FWHM:Mars:100GHz:units}{9\parcm37}
\setsymbol{HFI:FWHM:Mars:143GHz:units}{7\parcm04}
\setsymbol{HFI:FWHM:Mars:217GHz:units}{4\parcm68}
\setsymbol{HFI:FWHM:Mars:353GHz:units}{4\parcm43}
\setsymbol{HFI:FWHM:Mars:545GHz:units}{3\parcm80}
\setsymbol{HFI:FWHM:Mars:857GHz:units}{3\parcm67}

\setsymbol{HFI:FWHM:Mars:100GHz}{9.37}
\setsymbol{HFI:FWHM:Mars:143GHz}{7.04}
\setsymbol{HFI:FWHM:Mars:217GHz}{4.68}
\setsymbol{HFI:FWHM:Mars:353GHz}{4.43}
\setsymbol{HFI:FWHM:Mars:545GHz}{3.80}
\setsymbol{HFI:FWHM:Mars:857GHz}{3.67}

\setsymbol{HFI:beam:ellipticity:Mars:100GHz}{1.18}
\setsymbol{HFI:beam:ellipticity:Mars:143GHz}{1.03}
\setsymbol{HFI:beam:ellipticity:Mars:217GHz}{1.14}
\setsymbol{HFI:beam:ellipticity:Mars:353GHz}{1.09}
\setsymbol{HFI:beam:ellipticity:Mars:545GHz}{1.25}
\setsymbol{HFI:beam:ellipticity:Mars:857GHz}{1.03}

\setsymbol{HFI:CMB:relative:calibration:100GHz}{$\lsim 1\%$}
\setsymbol{HFI:CMB:relative:calibration:143GHz}{$\lsim 1\%$}
\setsymbol{HFI:CMB:relative:calibration:217GHz}{$\lsim 1\%$}
\setsymbol{HFI:CMB:relative:calibration:353GHz}{$\lsim 1\%$}
\setsymbol{HFI:CMB:relative:calibration:545GHz}{}
\setsymbol{HFI:CMB:relative:calibration:857GHz}{}

\setsymbol{HFI:CMB:absolute:calibration:100GHz}{$\lsim 2\%$}
\setsymbol{HFI:CMB:absolute:calibration:143GHz}{$\lsim 2\%$}
\setsymbol{HFI:CMB:absolute:calibration:217GHz}{$\lsim 2\%$}
\setsymbol{HFI:CMB:absolute:calibration:353GHz}{$\lsim 2\%$}
\setsymbol{HFI:CMB:absolute:calibration:545GHz}{}
\setsymbol{HFI:CMB:absolute:calibration:857GHz}{}

\setsymbol{HFI:FIRAS:gain:calibration:accuracy:statistical:100GHz}{}
\setsymbol{HFI:FIRAS:gain:calibration:accuracy:statistical:143GHz}{}
\setsymbol{HFI:FIRAS:gain:calibration:accuracy:statistical:217GHz}{}
\setsymbol{HFI:FIRAS:gain:calibration:accuracy:statistical:353GHz}{2.5\%}
\setsymbol{HFI:FIRAS:gain:calibration:accuracy:statistical:545GHz}{1\%}
\setsymbol{HFI:FIRAS:gain:calibration:accuracy:statistical:857GHz}{0.5\%}

\setsymbol{HFI:FIRAS:gain:calibration:accuracy:systematic:100GHz}{}
\setsymbol{HFI:FIRAS:gain:calibration:accuracy:systematic:143GHz}{}
\setsymbol{HFI:FIRAS:gain:calibration:accuracy:systematic:217GHz}{}
\setsymbol{HFI:FIRAS:gain:calibration:accuracy:systematic:353GHz}{}
\setsymbol{HFI:FIRAS:gain:calibration:accuracy:systematic:545GHz}{7\%}
\setsymbol{HFI:FIRAS:gain:calibration:accuracy:systematic:857GHz}{7\%}

\setsymbol{HFI:FIRAS:zero:point:accuracy:100GHz:units}{0.8\MJysr}
\setsymbol{HFI:FIRAS:zero:point:accuracy:143GHz:units}{}
\setsymbol{HFI:FIRAS:zero:point:accuracy:217GHz:units}{}
\setsymbol{HFI:FIRAS:zero:point:accuracy:353GHz:units}{1.4\MJysr}
\setsymbol{HFI:FIRAS:zero:point:accuracy:545GHz:units}{2.2\MJysr}
\setsymbol{HFI:FIRAS:zero:point:accuracy:857GHz:units}{1.7\MJysr}

\setsymbol{HFI:FIRAS:zero:point:accuracy:100GHz}{0.8}
\setsymbol{HFI:FIRAS:zero:point:accuracy:143GHz}{}
\setsymbol{HFI:FIRAS:zero:point:accuracy:217GHz}{}
\setsymbol{HFI:FIRAS:zero:point:accuracy:353GHz}{1.4}
\setsymbol{HFI:FIRAS:zero:point:accuracy:545GHz}{2.2}
\setsymbol{HFI:FIRAS:zero:point:accuracy:857GHz}{1.7}

\setsymbol{HFI:unit:conversion:100GHz:units}{0.2415\MJysrmK}
\setsymbol{HFI:unit:conversion:143GHz:units}{0.3694\MJysrmK}
\setsymbol{HFI:unit:conversion:217GHz:units}{0.4811\MJysrmK}
\setsymbol{HFI:unit:conversion:353GHz:units}{0.2883\MJysrmK}
\setsymbol{HFI:unit:conversion:545GHz:units}{0.05826\MJysrmK}
\setsymbol{HFI:unit:conversion:857GHz:units}{0.002238\MJysrmK}

\setsymbol{HFI:unit:conversion:100GHz}{0.2415}
\setsymbol{HFI:unit:conversion:143GHz}{0.3694}
\setsymbol{HFI:unit:conversion:217GHz}{0.4811}
\setsymbol{HFI:unit:conversion:353GHz}{0.2883}
\setsymbol{HFI:unit:conversion:545GHz}{0.05826}
\setsymbol{HFI:unit:conversion:857GHz}{0.002238}

\setsymbol{HFI:colour:correction:alpha=-2:V1.01:100GHz}{0.9893}
\setsymbol{HFI:colour:correction:alpha=-2:V1.01:143GHz}{0.9759}
\setsymbol{HFI:colour:correction:alpha=-2:V1.01:217GHz}{1.0007}
\setsymbol{HFI:colour:correction:alpha=-2:V1.01:353GHz}{1.0028}
\setsymbol{HFI:colour:correction:alpha=-2:V1.01:545GHz}{1.0019}
\setsymbol{HFI:colour:correction:alpha=-2:V1.01:857GHz}{0.9889}

\setsymbol{HFI:colour:correction:alpha=0:V1.01:100GHz}{1.0008}
\setsymbol{HFI:colour:correction:alpha=0:V1.01:143GHz}{1.0148}
\setsymbol{HFI:colour:correction:alpha=0:V1.01:217GHz}{0.9909}
\setsymbol{HFI:colour:correction:alpha=0:V1.01:353GHz}{0.9888}
\setsymbol{HFI:colour:correction:alpha=0:V1.01:545GHz}{0.9878}
\setsymbol{HFI:colour:correction:alpha=0:V1.01:857GHz}{1.0014}

\providecommand{\sorthelp}[1]{}

\newcommand{\smica}{\ensuremath{\tt SMICA}}
\newcommand{\healpix}{\ensuremath{\tt HEALPix}}

\newcommand{\tI}{\ensuremath{I_{353}}}
\newcommand{\hI}{\ensuremath{I_{0.408}}}
\newcommand{\fI}{\ensuremath{I_{\rm H\alpha}}}

\newcommand{\hatI}{{\hat{I}}}
\newcommand{\hatQ}{{\hat{Q}}}
\newcommand{\hatU}{{\hat{U}}}

\newcommand{\nI}{\ensuremath{I_{\nu}}}

\newcommand{\planck}{{\it Planck\/}}
\newcommand{\wmap}{{WMAP\/}}
\newcommand{\iras}{{IRAS\/}}

\newcommand{\hi}{{\sc Hi}}
\newcommand{\spdust}{\ensuremath{\tt SPDUST}}
\newcommand{\kcmb}{{${\rm K}_{\rm CMB}$}}
\newcommand{\krj}{{${\rm K}_{\rm RJ}$}}

\newcommand{\Nside}{\ensuremath{N_{\rm side}}}
\newcommand{\Td}{\ensuremath{T_{\rm d}}}
\newcommand{\betad}{\ensuremath{\beta_{\rm d}}}

\newcommand{\betadmm}{\ensuremath{\beta_{\rm d,mm}}}
\newcommand{\betadmmI}{\ensuremath{\beta_{\rm d,mm}^{\rm I}}}
\newcommand{\betasubmmI}{\ensuremath{\beta_{\rm d,submm}^{\rm I}}}

\newcommand{\betadmmP}{\ensuremath{\beta_{\rm d,mm}^{\rm P}}}

\newcommand{\AaI}{\ensuremath{A_{\rm a}^{\rm I}}}

\newcommand{\AsP}{\ensuremath{A_{\rm s}^{\rm P}}}

\newcommand{\chiI}{\ensuremath{\chisquare_{\rm I}}}
\newcommand{\chiP}{\ensuremath{\chisquare_{\rm P}}}

\newcommand{\alphaInu}{\ensuremath{\alpha_{\nu}^{\rm I}}}
\newcommand{\alphaI}{\ensuremath{\alpha^{\rm I}}}
\newcommand{\alphaPnu}{\ensuremath{\alpha_{\nu}^{\rm P}}}
\newcommand{\alphaPt}{\ensuremath{\alpha_{353}^{\rm P}}}
\newcommand{\alphaP}{\ensuremath{\alpha^{\rm P}}}
\newcommand{\tildealphaInu}{\ensuremath{\tilde{\alpha}_{\nu}^{\rm I}}}
\newcommand{\tildealphaPnu}{\ensuremath{\tilde{\alpha}_{\nu}^{\rm P}}}
\newcommand{\ffd}{\ensuremath{f_{\rm d}}}
\newcommand{\sd}{\ensuremath{s_{\rm d}}}

\newcommand{\RI}{\ensuremath{R_{100}(353,217)}}

\newcommand{\RIT}{\ensuremath{R^{\rm I}_{100}(353,217)}}
\newcommand{\RIIT}{\ensuremath{R^{\rm I}(3000,857)}}
\newcommand{\RIP}{\ensuremath{R^{\rm P}_{100}(353,217)}}

\newcommand{\sigmaI}{\ensuremath{\sigma_{353}^{\rm I}}} 

\newcommand{\sigmaP}{\ensuremath{\sigma_{353}^{\rm P}}}

\newcommand{\chisquare}{\ensuremath{\chi^2}}
\newcommand{\halpha}{\ensuremath{\rm H\alpha}}

\newcommand{\fsky}{{$f_{\rm sky}$}}
\newcommand{\Cl}{\ensuremath{C_{\ell}}}
\newcommand{\nuref}{\ensuremath{\nu_{\rm ref}}}
\newcommand{\nub}{\ensuremath{\nu_{\rm b}}}
\newcommand{\nuc}{\ensuremath{\nu_{\rm c}}}

\begin{document}

\author{\small
Planck Collaboration:
P.~A.~R.~Ade\inst{75}
\and
M.~I.~R.~Alves\inst{51}
\and
G.~Aniano\inst{51}
\and
C.~Armitage-Caplan\inst{78}
\and
M.~Arnaud\inst{64}
\and
F.~Atrio-Barandela\inst{17}
\and
J.~Aumont\inst{51}
\and
C.~Baccigalupi\inst{74}
\and
A.~J.~Banday\inst{80, 10}
\and
R.~B.~Barreiro\inst{58}
% ...
}
\institute{\small
APC, AstroParticule et Cosmologie, Universit\'{e} Paris Diderot, CNRS/IN2P3, CEA/lrfu, Observatoire de Paris, Sorbonne Paris Cit\'{e}, 10, rue Alice Domon et L\'{e}onie Duquet, 75205 Paris Cedex 13, France\\
\and
African Institute for Mathematical Sciences, 6-8 Melrose Road, Muizenberg, Cape Town, South Africa\\
\and
Agenzia Spaziale Italiana Science Data Center, Via del Politecnico snc, 00133, Roma, Italy\\
\and
Agenzia Spaziale Italiana, Viale Liegi 26, Roma, Italy\\
\and
Astrophysics Group, Cavendish Laboratory, University of Cambridge, J J Thomson Avenue, Cambridge CB3 0HE, U.K.\\
\and
Astrophysics \& Cosmology Research Unit, School of Mathematics, Statistics \& Computer Science, University of KwaZulu-Natal, Westville Campus, Private Bag X54001, Durban 4000, South Africa\\
\and
Atacama Large Millimeter/submillimeter Array, ALMA Santiago Central Offices, Alonso de Cordova 3107, Vitacura, Casilla 763 0355, Santiago, Chile\\
\and
CITA, University of Toronto, 60 St. George St., Toronto, ON M5S 3H8, Canada\\
\and
CNR - ISTI, Area della Ricerca, via G. Moruzzi 1, Pisa, Italy\\
\and
CNRS, IRAP, 9 Av. colonel Roche, BP 44346, F-31028 Toulouse cedex 4, France\\
\and
California Institute of Technology, Pasadena, California, U.S.A.\\
\and
Centro de Estudios de F\'{i}sica del Cosmos de Arag\'{o}n (CEFCA), Plaza San Juan, 1, planta 2, E-44001, Teruel, Spain\\
\and
Computational Cosmology Center, Lawrence Berkeley National Laboratory, Berkeley, California, U.S.A.\\
\and
DSM/Irfu/SPP, CEA-Saclay, F-91191 Gif-sur-Yvette Cedex, France\\
\and
DTU Space, National Space Institute, Technical University of Denmark, Elektrovej 327, DK-2800 Kgs. Lyngby, Denmark\\
\and
D\'{e}partement de Physique Th\'{e}orique, Universit\'{e} de Gen\`{e}ve, 24, Quai E. Ansermet,1211 Gen\`{e}ve 4, Switzerland\\
\and
Departamento de F\'{\i}sica Fundamental, Facultad de Ciencias, Universidad de Salamanca, 37008 Salamanca, Spain\\
\and
Departamento de F\'{\i}sica, Universidad de Oviedo, Avda. Calvo Sotelo s/n, Oviedo, Spain\\
\and
Department of Astronomy and Astrophysics, University of Toronto, 50 Saint George Street, Toronto, Ontario, Canada\\
\and
Department of Astrophysics/IMAPP, Radboud University Nijmegen, P.O. Box 9010, 6500 GL Nijmegen, The Netherlands\\
\and
Department of Physics \& Astronomy, University of British Columbia, 6224 Agricultural Road, Vancouver, British Columbia, Canada\\
\and
Department of Physics and Astronomy, Dana and David Dornsife College of Letter, Arts and Sciences, University of Southern California, Los Angeles, CA 90089, U.S.A.\\
\and
Department of Physics and Astronomy, University College London, London WC1E 6BT, U.K.\\
\and
Department of Physics, Gustaf H\"{a}llstr\"{o}min katu 2a, University of Helsinki, Helsinki, Finland\\
\and
Department of Physics, Princeton University, Princeton, New Jersey, U.S.A.\\
\and
Department of Physics, University of California, Santa Barbara, California, U.S.A.\\
\and
Department of Physics, University of Illinois at Urbana-Champaign, 1110 West Green Street, Urbana, Illinois, U.S.A.\\
\and
Dipartimento di Fisica e Astronomia G. Galilei, Universit\`{a} degli Studi di Padova, via Marzolo 8, 35131 Padova, Italy\\
\and
Dipartimento di Fisica e Scienze della Terra, Universit\`{a} di Ferrara, Via Saragat 1, 44122 Ferrara, Italy\\
\and
Dipartimento di Fisica, Universit\`{a} La Sapienza, P. le A. Moro 2, Roma, Italy\\
\and
Dipartimento di Fisica, Universit\`{a} degli Studi di Milano, Via Celoria, 16, Milano, Italy\\
\and
Dipartimento di Fisica, Universit\`{a} degli Studi di Trieste, via A. Valerio 2, Trieste, Italy\\
\and
Dipartimento di Fisica, Universit\`{a} di Roma Tor Vergata, Via della Ricerca Scientifica, 1, Roma, Italy\\
\and
Discovery Center, Niels Bohr Institute, Blegdamsvej 17, Copenhagen, Denmark\\
\and
Dpto. Astrof\'{i}sica, Universidad de La Laguna (ULL), E-38206 La Laguna, Tenerife, Spain\\
\and
European Southern Observatory, ESO Vitacura, Alonso de Cordova 3107, Vitacura, Casilla 19001, Santiago, Chile\\
\and
European Space Agency, ESAC, Planck Science Office, Camino bajo del Castillo, s/n, Urbanizaci\'{o}n Villafranca del Castillo, Villanueva de la Ca\~{n}ada, Madrid, Spain\\
\and
European Space Agency, ESTEC, Keplerlaan 1, 2201 AZ Noordwijk, The Netherlands\\
\and
Helsinki Institute of Physics, Gustaf H\"{a}llstr\"{o}min katu 2, University of Helsinki, Helsinki, Finland\\
\and
INAF - Osservatorio Astronomico di Padova, Vicolo dell'Osservatorio 5, Padova, Italy\\
\and
INAF - Osservatorio Astronomico di Roma, via di Frascati 33, Monte Porzio Catone, Italy\\
\and
INAF - Osservatorio Astronomico di Trieste, Via G.B. Tiepolo 11, Trieste, Italy\\
\and
INAF/IASF Bologna, Via Gobetti 101, Bologna, Italy\\
\and
INAF/IASF Milano, Via E. Bassini 15, Milano, Italy\\
\and
INFN, Sezione di Bologna, Via Irnerio 46, I-40126, Bologna, Italy\\
\and
INFN, Sezione di Roma 1, Universit\`{a} di Roma Sapienza, Piazzale Aldo Moro 2, 00185, Roma, Italy\\
\and
INFN/National Institute for Nuclear Physics, Via Valerio 2, I-34127 Trieste, Italy\\
\and
IPAG: Institut de Plan\'{e}tologie et d'Astrophysique de Grenoble, Universit\'{e} Joseph Fourier, Grenoble 1 / CNRS-INSU, UMR 5274, Grenoble, F-38041, France\\
\and
Imperial College London, Astrophysics group, Blackett Laboratory, Prince Consort Road, London, SW7 2AZ, U.K.\\
\and
Infrared Processing and Analysis Center, California Institute of Technology, Pasadena, CA 91125, U.S.A.\\
\and
Institut d'Astrophysique Spatiale, CNRS (UMR8617) Universit\'{e} Paris-Sud 11, B\^{a}timent 121, Orsay, France\\
\and
Institut d'Astrophysique de Paris, CNRS (UMR7095), 98 bis Boulevard Arago, F-75014, Paris, France\\
\and
Institute for Space Sciences, Bucharest-Magurale, Romania\\
\and
Institute of Astronomy, University of Cambridge, Madingley Road, Cambridge CB3 0HA, U.K.\\
\and
Institute of Theoretical Astrophysics, University of Oslo, Blindern, Oslo, Norway\\
\and
Instituto de Astrof\'{\i}sica de Canarias, C/V\'{\i}a L\'{a}ctea s/n, La Laguna, Tenerife, Spain\\
\and
Instituto de Astronomia, Geof\'{\i}sica e Ci\^{e}ncias Atmosf\'{e}ricas, Universidade de S\~{a}o Paulo, S\~{a}o Paulo, SP 05508-090, Brazil\\
\and
Instituto de F\'{\i}sica de Cantabria (CSIC-Universidad de Cantabria), Avda. de los Castros s/n, Santander, Spain\\
\and
Jet Propulsion Laboratory, California Institute of Technology, 4800 Oak Grove Drive, Pasadena, California, U.S.A.\\
\and
Jodrell Bank Centre for Astrophysics, Alan Turing Building, School of Physics and Astronomy, The University of Manchester, Oxford Road, Manchester, M13 9PL, U.K.\\
\and
Kavli Institute for Cosmology Cambridge, Madingley Road, Cambridge, CB3 0HA, U.K.\\
\and
LAL, Universit\'{e} Paris-Sud, CNRS/IN2P3, Orsay, France\\
\and
LERMA, CNRS, Observatoire de Paris, 61 Avenue de l'Observatoire, Paris, France\\
\and
Laboratoire AIM, IRFU/Service d'Astrophysique - CEA/DSM - CNRS - Universit\'{e} Paris Diderot, B\^{a}t. 709, CEA-Saclay, F-91191 Gif-sur-Yvette Cedex, France\\
\and
Laboratoire Traitement et Communication de l'Information, CNRS (UMR 5141) and T\'{e}l\'{e}com ParisTech, 46 rue Barrault F-75634 Paris Cedex 13, France\\
\and
Laboratoire de Physique Subatomique et de Cosmologie, Universit\'{e} Joseph Fourier Grenoble I, CNRS/IN2P3, Institut National Polytechnique de Grenoble, 53 rue des Martyrs, 38026 Grenoble cedex, France\\
\and
Laboratoire de Physique Th\'{e}orique, Universit\'{e} Paris-Sud 11 \& CNRS, B\^{a}timent 210, 91405 Orsay, France\\
\and
Lawrence Berkeley National Laboratory, Berkeley, California, U.S.A.\\
\and
Max-Planck-Institut f\"{u}r Astrophysik, Karl-Schwarzschild-Str. 1, 85741 Garching, Germany\\
\and
National University of Ireland, Department of Experimental Physics, Maynooth, Co. Kildare, Ireland\\
\and
Niels Bohr Institute, Blegdamsvej 17, Copenhagen, Denmark\\
\and
Observational Cosmology, Mail Stop 367-17, California Institute of Technology, Pasadena, CA, 91125, U.S.A.\\
\and
Optical Science Laboratory, University College London, Gower Street, London, U.K.\\
\and
SISSA, Astrophysics Sector, via Bonomea 265, 34136, Trieste, Italy\\
\and
School of Physics and Astronomy, Cardiff University, Queens Buildings, The Parade, Cardiff, CF24 3AA, U.K.\\
\and
Space Sciences Laboratory, University of California, Berkeley, California, U.S.A.\\
\and
Special Astrophysical Observatory, Russian Academy of Sciences, Nizhnij Arkhyz, Zelenchukskiy region, Karachai-Cherkessian Republic, 369167, Russia\\
\and
Sub-Department of Astrophysics, University of Oxford, Keble Road, Oxford OX1 3RH, U.K.\\
\and
UPMC Univ Paris 06, UMR7095, 98 bis Boulevard Arago, F-75014, Paris, France\\
\and
Universit\'{e} de Toulouse, UPS-OMP, IRAP, F-31028 Toulouse cedex 4, France\\
% ...
}

\title{{\Planck} intermediate results. XXII. Frequency dependence of thermal emission from Galactic dust in intensity and polarization}

\abstract{\planck\ has mapped the intensity and polarization of the sky at microwave frequencies with unprecedented sensitivity. We use these data to characterize
the frequency dependence of dust emission. We make use of the \planck\ 353\,GHz \StokesI, \StokesQ, and \StokesU\ Stokes maps as dust templates, and cross-correlate 
them with the \planck\ and \wmap\ data at 12 frequencies from $23$ to $353$\,GHz, over circular patches with 10\deg\ radius. The cross-correlation analysis is performed 
for both intensity and polarization data in a consistent manner. The results are corrected for  the chance correlation between the templates and the anisotropies of the cosmic 
microwave background.  We use a mask that focuses our analysis on the diffuse interstellar medium at intermediate Galactic latitudes. We determine the spectral indices of 
dust emission in intensity and polarization between $100$ and $353$\,GHz, for each sky patch. Both indices are found to be remarkably constant over the sky. The mean values,
$1.59\pm0.02$ for polarization and $1.51\pm0.01$ for intensity, for a mean dust temperature of 19.6\,K, are close, but significantly different ($3.6 \, \sigma$).  We determine the mean spectral 
energy distribution (SED) of the microwave emission, correlated with the $353$\,GHz dust templates,  by averaging the results of the correlation over all sky patches.
We find that the mean SED  increases for decreasing frequencies at $\nu < 60$\,GHz for both intensity and polarization. The rise of the polarization SED towards low frequencies
may be accounted for by a synchrotron component correlated with dust, with no need for any polarization of the anomalous microwave emission.  We use a spectral model to
separate the synchrotron and dust polarization and to characterize the spectral dependence of the dust polarization fraction.  The polarization fraction (\polfrac) of the dust 
emission decreases by $(21\pm6)$\,\% from 353 to $70$\,GHz.  We discuss this result within the context of existing dust models.  The decrease in \polfrac\ could indicate 
differences in polarization efficiency among components of interstellar dust (e.g., carbon versus silicate grains). Our observational results provide inputs to quantify and optimize 
the separation between Galactic and cosmological polarization.}


\titlerunning{Frequency dependence of thermal emission from Galactic dust  in intensity and polarization} 

\authorrunning{\Planck\ Collaboration} 

\maketitle


\clearpage

\section{Introduction} \label{sec:intro}

\planck \footnote{\planck~(\url{http://www.esa.int/Planck}) is a project of the European Space Agency (ESA) with instruments provided by two scientific consortia funded by ESA 
member states (in particular the lead countries France and Italy), with contributions from NASA (USA) and telescope reflectors provided by a collaboration between ESA and a 
scientific consortium led and funded by Denmark.}  \citep{Tauber:2010, planck2011-1.1} has mapped the polarization of the sky emission in seven channels at microwave 
frequencies from 30 to $353$\,GHz. The data open new opportunities for investigating the astrophysics of Galactic polarization. In this paper, we use these data to characterize the 
frequency dependence of dust polarization from the diffuse interstellar medium (ISM).

At microwave frequencies, dust emission components include the long-wavelength tail of thermal dust emission \citep{Draine_Li:2007,Meny:2007,Compeigne:2011,Jones:2013}, 
the anomalous microwave emission (AME, \citealt{Kogut:1996, Leitch:1997, Oliveira-Costa:1999, Banday:2003, Lagache:2003a, Davies:2006, Dobler:2008a, MAMD:2008, 
Ysard:2010, planck2011-7.2}), and possibly dipolar magnetic emission of ferromagnetic particles \citep{Draine_Lazarian:1999,Draine_Hensley:2012b}.

Thermal dust emission is known to be polarized, but to a different degree for each dust component, owing to differences in the shape and alignment efficiency of grains 
\citep{Hildebrand:1999, Martin:2007, Draine_Fraisse:2009}. The polarization of the 9.7\microns\ absorption feature from silicates is direct evidence that silicate grains are aligned 
\citep{Smith:2000}. The lack of polarization of the 3.4\microns\ absorption feature from aliphatic hydrocarbons (along lines of sight towards the Galactic centre with strong 
polarization in the 9.7\microns\ silicate absorption) indicates that dust comprises carbon grains that are much less efficient at producing interstellar polarization than silicates 
\citep{Chiar:2006}. Observational signatures of these differences in polarization efficiency among components of interstellar dust are expected to be found in the polarization 
fraction (\polfrac) of the far infrared (FIR) and sub-mm dust emission. Spectral variations of polarization fraction have been reported from observations of star-forming molecular 
clouds \citep{Hildebrand:1999,Vaillancourt:2002,Vaillancourt:2008,Vaillancourt:2012}. However, these data cannot be unambiguously interpreted as differences in the intrinsic 
polarization of dust components \citep{Vaillancourt:2002}; they can also be interpreted as correlated changes in grain temperature and alignment efficiency across the clouds. The 
sensitivity of \planck\ to low-brightness extended-emission allows us to carry out this investigation for the diffuse ISM, where the heating and alignment efficiency of grains are far 
more homogeneous than in star-forming regions.

AME is widely interpreted as dipole radiation from small carbon dust particles. This interpretation, first proposed by \cite{Erickson1957} and modelled by 
\citet{Draine_Lazarian:1998}, has been developed into detailed models \citep{Ali-Haimoud:2009, Silsbee:2011,Hoang:2011} that provide a good spectral fit to the data 
\citep{planck2011-7.2,planck2013-XV}. The intrinsic polarization of this emission must be low, owing to the weakness or absence of polarization of the 220 nm bump in the UV 
extinction curve \citep{Wolff:1997}, which is evidence of the poor alignment of small carbon particles. The polarization fraction of the AME could be up to a few percentage
\citep{Lazarian:2000,Hoang:2013}. The {\it Wilkinson Microwave Anisotropy Probe\/} (\wmap, hereafter) data have been used to search for polarization in a few sources with bright 
AME, for example the $\rho$ Ophichus and Perseus molecular clouds \citep{Dickinson:2011, Lopez-Caraballo:2011}, yielding upper limits in the range of 1.5\,\% to a few percent 
on polarization fraction \citep{Rubino:2012}.

Magnetic dipolar emission (MDE) from magnetic grains was first proposed by \citet{Draine_Lazarian:1999} as a possible interpretation of the AME. \citet{Draine_Hensley:2012b} 
have recently revived this idea with a new model where the MDE could be a significant component of dust emission at frequencies from 50 to a few hundred GHz 
\citep{planck2013-XIV} and \citet{planck2013-XVII}, relevant to cosmic microwave background (CMB) studies. Recently, \citet{Liu:2014} have argued that MDE may be contributing to the microwave emission  
of Galactic radio loops, in particular Loop I. This hypothesis may be tested with the \planck\ polarization 
observations. The polarization fraction of MDE is expected to be high for magnetic grains. If the magnetic particles are inclusions within silicates, the polarization directions of the 
dipolar magnetic and electric emissions are orthogonal. In this case the models predict a significant decrease in the polarization fraction of dust emission at frequencies below 
$350$\,GHz.

\wmap\ provided the first all-sky survey of microwave polarization. Galactic polarization was detected on large angular scales at all frequencies from 23 to 94\,GHz. The data have 
been shown to be consistent with a combination of synchrotron and dust contributions \citep{kogut2007,page2007,MAMD:2008,Macellari:2011}, but they do not constrain the 
spectral dependence of dust polarization.

The spectral dependence of the dust emission at \planck\ frequencies has been determined in the Galactic plane and at high Galactic latitudes by \citet{planck2013-XIV} and 
\citet{planck2013-XVII}.  In this paper, we use the high signal-to-noise $353$\,GHz \planck\ Stokes \StokesI, \StokesQ, \StokesU\ maps as templates to characterize the spectral 
dependence of dust emission in both intensity and polarization.  Our analysis also includes the separation of dust emission from CMB anisotropies. We extract the dust-correlated 
emission in intensity (\StokesI) and polarization (\polint) by cross-correlating the $353$\,GHz maps with both the \planck\ and \wmap\ data.  For the intensity, we also use the 
\halpha\ and $408$\,MHz maps as templates of the free-free and synchrotron emission.  The \polint\ and \StokesI\ spectra are compared and discussed in light of the present 
understanding and questions about microwave dust emission components introduced in \citet{planck2013-XVII}.  We aim to characterize the spectral shape and the relative 
amplitude of Galactic emission components in polarization.  In doing so we test theoretical predictions about the nature of the dust  emission in intensity and polarization.  We also 
provide information that is key to designing and optimizing the separation of the polarized CMB signal from the polarized Galactic dust emission.

The paper is organised as follows. In Sect.~\ref{sec:data}, we introduce the data sets used in this paper. Our methodology for the data analysis is described in the following three sections. We define the part of the sky we analyse in Sect.~\ref{sec:mask}. We describe how we apply the cross-correlation analysis (hereafter CC) to the intensity and polarization data in Sect.~\ref{sec:ccanaly}.  Section~\ref{sec:comp_sep} explains the separation of the dust and CMB emission after data correlation.  The scientific results are presented in Sects.~\ref{sec:betaI} and \ref{sec:em_SED} for  intensity, and  Sects.~\ref{sec:betaP} and~\ref{sec:pol_SED} for  polarization. The dust SEDs,  \StokesI\  and \polint,  are compared and discussed with relation to models of dust emission in Sect.~\ref{sec:discussion}.  Section~\ref{sec:conclusion} summarizes the main results of our work. We detail the derivation of the correlation coefficients in Appendix~\ref{sec:math_cc}. Appendix~\ref{sec:simul} describes the Monte Carlo simulations we have performed to show that our data analysis is unbiased. Appendix~\ref{sec:dust_scattering} describes the dependence of the dust \StokesI\ SED on the correction of the \halpha\ map, used as template of the free-free emission, for dust extinction and scattering.  The power spectra of the maps used as templates of dust, free-free and synchrotron emission are presented in Appendix~\ref{sec:power_spectra} for a set of  Galactic masks.

\section{Data sets used} \label{sec:data}

Here we discuss the \planck, \wmap, and ancillary data used in the paper and listed in Table~\ref{tab:2.1}. 

\begin{table*}[tmb]
\begingroup
\newdimen\tblskip \tblskip=5pt
\caption{\label{tab:2.1} Summary of \planck, \wmap\ and ancillary data used in this paper for both intensity and polarization.}
\nointerlineskip
\vskip -3mm
\setbox\tablebox=\vbox{
   \newdimen\digitwidth 
   \setbox0=\hbox{\rm 0} 
   \digitwidth=\wd0 
   \catcode`*=\active 
   \def*{\kern\digitwidth}
   \newdimen\signwidth 
   \setbox0=\hbox{+} 
   \signwidth=\wd0 
   \catcode`!=\active 
   \def!{\kern\signwidth}
   \newdimen\pointwidth
   \setbox0=\hbox{{.}}
   \pointwidth=\wd0
   \catcode`?=\active
   \def?{\kern\pointwidth}
\halign{
\hbox to 1.0 in{#\leaderfil}\tabskip=1.8em&
\hfil #\hfil&
\hfil #\hfil&
\hfil #\hfil&
\hfil #\hfil\tabskip=0pt\cr
\noalign{\doubleline \vskip 2pt}
\omit Telescope/Survey\hfil& Frequency& Resolution& Reference\cr
\omit& [GHz]& [arcmin]& \omit\cr
\noalign{\vskip 4pt\hrule\vskip 6pt}
          Haslam& ***0.408& 60?**& \citet{Haslam:1982}\cr
 \wmap\ $9$-year& **23?***& 48.42& \citet{Bennett:2012}\cr
         \planck& **28.4**& 32.23& \citet{planck2013-p01}\cr
 \wmap\ $9$-year& **33?***& 37.44& \citet{Bennett:2012}\cr
 \wmap\ $9$-year& **41?***& 28.62& \citet{Bennett:2012}\cr
         \planck& **44.1**& 27.01& \citet{planck2013-p01}\cr
 \wmap\ $9$-year& **61?***& 19.56& \citet{Bennett:2012}\cr
         \planck& **70.4**& 13.25& \citet{planck2013-p01}\cr
 \wmap\ $9$-year& **94?***& 12.30& \citet{Bennett:2012}\cr
         \planck& *100?***& *9.65& \citet{planck2013-p01}\cr
         \planck& *143?***& *7.25& \citet{planck2013-p01}\cr
         \planck& *217?***& *4.99& \citet{planck2013-p01}\cr
         \planck& *353?***& *4.82& \citet{planck2013-p01}\cr
         \planck& *545?***& *4.68& \citet{planck2013-p01}\cr
         \planck& *857?***& *4.32& \citet{planck2013-p01}\cr
           DIRBE& 3000?***& 50?**& \citet{Hauser:1998}\cr
         \halpha&         & 60?**& \citet{Dickinson:2003}\cr 
         LAB \hi&         & 36?**& \citet{Kalberla:2005}\cr
\noalign{\vskip 5pt\hrule\vskip 3pt}}}
\endPlancktablewide
\endgroup
\end{table*}

\subsection{\planck\ data}

\subsubsection{Sky maps}

\planck\  is the third generation space mission to characterize the anisotropies of the CMB.  It observed the sky in seven frequency bands from $30$ to $353$\,GHz for polarization, 
and in two additional bands at $545$ and $857$\,GHz for intensity, with an angular resolution from 31\arcmin\ to 5\arcmin\ \citep{planck2013-p01}. The in-flight performance of the 
two focal plane instruments, the HFI (High Frequency Instrument) and the LFI (Low Frequency Instrument), are given in \cite{planck2011-1.5} and \cite{planck2011-1.4}, 
respectively. The data processing and calibration of the HFI and LFI data used here are described in \cite{planck2013-p03f} and \cite{planck2013-p02}, respectively. The data 
processing specific to polarization is given in \cite{planck2013-p03} and \cite{planck2013-p02a}.

For intensity, we use the full \planck\ mission (five full-sky surveys for HFI and eight full-sky surveys for LFI) data sets  between $30$ and $857$\,GHz. The LFI and HFI frequency maps 
are provided in \healpix\footnote{\url{http://healpix.jpl.nasa.gov}} format \citep{Gorski:2005} with resolution parameters $\Nside=1024$ and $2048$, respectively.  The \planck\ sky 
maps between 30 and 353 GHz are calibrated in CMB temperature units, \kcmb, so that the CMB anisotropies have a constant spectrum across frequencies. The two high 
frequency maps of \planck, 545 and 857 GHz, are expressed in \MJysr, calibrated for a power-law spectrum with a spectral index of $-1$, following the \iras\ convention. 
We use \planck\ maps with the zodiacal light emission (ZLE) subtracted \citep{planck2013-pip88} at frequencies $\nu \ge 353$\,GHz, but maps {\it not\/} corrected for ZLE at lower  
frequencies because the extrapolation of the ZLE model is uncertain at microwave frequencies. Further it has not been estimated at  frequencies smaller than $100$\,GHz. 
We do not correct for the zero offset, nor for the residual dipole identified by \citet{planck2013-p06b} at HFI frequencies because it is not necessary for our analysis based on local 
correlations of data sets.

For polarization, we use the same full \planck\ mission data sets, as used for intensity, between $30$ and $353$\,GHz. The \planck\ polarization that we use in this have 
been generated in exactly the same manner as the data publicly released in March 2013, described in \cite{planck2013-p01} and associated papers. Note, however, that the
publicly available data includes data include only temperature maps based on the first two surveys. \cite{planck2013-p11} shows the very good consistency of cosmological 
models derived from intensity only with polarization data at small scale scales (high CMB multipoles). However, as detailed in \citet{planck2013-p03} (see their Fig.~27),
the 2013 polarization data are known to be affected by systematic effects at low multipoles which were not yet fully corrected, and thus these data were not used for cosmology. 
In this paper, we use the latest \planck\ polarization maps (internal data release ``DX11d''), which are corrected from known systematics. 
The full mission maps for intensity as well as for polarization will be described and made publicly available in early 2015.

\subsubsection{Systematic effects in polarization}\label{sec:systematics}

Current \planck\ polarization data are contaminated by a small amount of leakage from intensity to polarization,  mainly due to bandpass mismatch (BPM) and calibration 
mismatch between detectors \citep{planck2014-XIX, planck2013-p03,planck2013-p02a}. The BPM results from slight differences in the spectral response to Galactic emission of 
the polarization sensitive bolometers (PSB)  \citep{planck2013-p03}. In addition, the signal differences leak into polarization. The calibration uncertainties translate into a small 
mismatch in the response of the detectors, which produces a signal leakage from intensity to polarization. As the microwave sky is dominated by the large scale emission from the 
Galaxy and the CMB dipole, systematics affect the polarization maps mainly on large angular scales. We were only able to correct the maps for leakage of Galactic emission due 
to bandpass mismatch.

The observed Stokes $\StokesQ^{\rm obs}_{\nu}$ and $\StokesU^{\rm obs}_{\nu}$ maps at a given frequency $\nu$  can be written as,
\begin{align}
\StokesQ^{\rm obs}_{\nu}& = \ \StokesQ^{\rm c} + \  \StokesQ^{\rm G}_{\nu} +\ \StokesQ^{\rm n}_{\nu}   + \  L_{\nu} (\StokesI \rightarrow \StokesQ)\ ,  \\
\StokesU^{\rm obs}_{\nu}& = \  \StokesU^{\rm c} + \  \StokesU^{\rm G}_{\nu}  + \ \StokesQ^{\rm n}_{\nu}  + \ L_{\nu} (\StokesI \rightarrow \StokesU)\ ,
\end{align}
where the term $L$ corresponds to the BPM leakage map for Galactic emission, offset, and residual dipole. All of them are computed using the coupling coefficient of each detector to the sky emission 
spectrum together with the actual sky scanning strategy. The superscript ${\rm c}$ represents the CMB polarization, ${\rm n}$ represents the noise and the index ${\rm G}$ 
incorporates all the Galactic emission components in intensity at \planck\ frequencies. We restrict our analysis to intermediate Galactic latitudes where the dominant Galactic 
emission at HFI frequencies  is dust emission. The polarized HFI maps we used are corrected for the dust, CO, offset and residual dipole, to a first approximation, using sky measurements of the spectral 
transmission of each bolometer \citep{planck2013-p03d}. At LFI frequencies, we correct for BPM coming from the low frequency Galactic components, i.e., the AME, synchrotron 
and free-free emission \citep{planck2013-p02a}, using sky measurements of the spectral transmission of each bolometer.

To test the results presented in this paper for systematic effects, we use multiple data sets that include the maps made with two independent groups of four PSBs   
(detector sets ``DS1'' and ``DS2'', see Table~3 in \citealt{planck2013-p03}), the half-ring maps (using the first or second halves of the data from each stable pointing period, ``HR1'' 
and ``HR2'') and maps made with yearly surveys (``YR1'' ,``YR2'', etc.). 
The HR1 and HR2 maps are useful to assess the impact on our data analysis of the noise and 
systematic effects on scales smaller than $20\arcmin$. 
The YR1 map is a combination of first two surveys S1 and S2, and YR2 is a combination of surveys S3 and S4, and so on.
The maps made with individual sky surveys are useful to quantify the impact of systematic effects on larger angular scales, particularly from beam ellipticity and far sidelobes \citep{planck2013-p02a,planck2013-p03}.  
For the intensity and polarization HFI data, we use the two yearly maps YR1 and YR2, whereas for LFI 
data, we use the four yearly maps grouped into odd (YR1+YR3) and even (YR2+YR4) pairs because they share the same scanning strategy.

The different data sets are independent observations of the same sky that capture noise and  systematic effects. They provide means to assess the validity and self-consistency of 
our analysis of the \planck\ data.  The  different map combinations highlight different systematic effects on various timescales and across different dimensions.

\begin{itemize}
\item  Half-ring maps share the same scanning strategy and detectors so they have the same leakage from intensity to polarization. The difference between these two maps shows the noise that is not correlated. The removal of glitches induce some noise correlation between the two half-ring maps that affects the data at all multipoles.
\item The differences between two yearly maps is used to check the consistency of the data over the full duration of the \Planck\ mission.
\item Detector set maps have the same combination of scans. The difference between detector set maps show all systematic effects associated with specific detectors.
\end{itemize}

\subsection{\wmap\ data}

We use the \wmap\ nine year data \citep{Bennett:2012} from the Legacy Archive for Microwave Background Data Analysis (LAMBDA)\footnote{\url{http://lambda.gsfc.nasa.gov}} 
provided in the \healpix\ pixelization scheme with a resolution $\Nside=512$. \wmap\ observed the sky in five frequency bands, denoted K, Ka, Q, V, and W, centred at the 
frequencies $23$, $33$, $41$, $61$, and $94$\,GHz, respectively. \wmap\ has ten differencing assemblies (DAs), one for both K and Ka bands, two for Q band, two for V band, and 
four for W band. \wmap\ has frequency-dependent resolution, ranging from 52\arcmin\ (K band) to 12\arcmin\ (W band). Multiple DAs at each frequency for Q, V and W bands are 
combined using simple average to generate a single map per frequency band.

\subsection{Ancillary data}

We complement the \planck\ and \wmap\ data with several ancillary sky maps. We use the 408\,MHz map from \citet{Haslam:1982}, and \halpha\ map from \citet{Dickinson:2003} (hereafter DDD) as tracers of synchrotron and  free-free emission, respectively.  No dust extinction correction ($f_{\rm d}$ = 0.0) has been applied to the DDD \halpha\ map, which is expressed in units of Rayleigh (R). For our simulations we use the  Leiden/Argentine/Bonn (LAB) survey of Galactic \hi\ column density \citep{Kalberla:2005} as a tracer of dust emission \citep{planck2011-7.12,planck2013-XVII}.  Finally,  we use the DIRBE 100\microns\ sky map to determine the dust temperature, like in \citet{planck2013-p06b}.

The 408\,MHz, LAB \hi, and DIRBE 100\microns\ data are downloaded from LAMBDA. We use the DIRBE data corrected for ZLE. We project the DIRBE 100\microns\ map on a \healpix\ grid at $\Nside = 512$ with a Gaussian interpolation kernel that reduces the angular resolution to 50\arcmin.  Both the 408\,MHz and the DDD \halpha\  maps are provided at 1\deg\ resolution. The LAB \hi\ survey and DIRBE 100\microns\  data have angular resolutions of 36\arcmin\ and 50\arcmin, respectively.

\section{Global mask}\label{sec:mask}

\begin{figure}[h!]
\includegraphics[width=8.8cm]{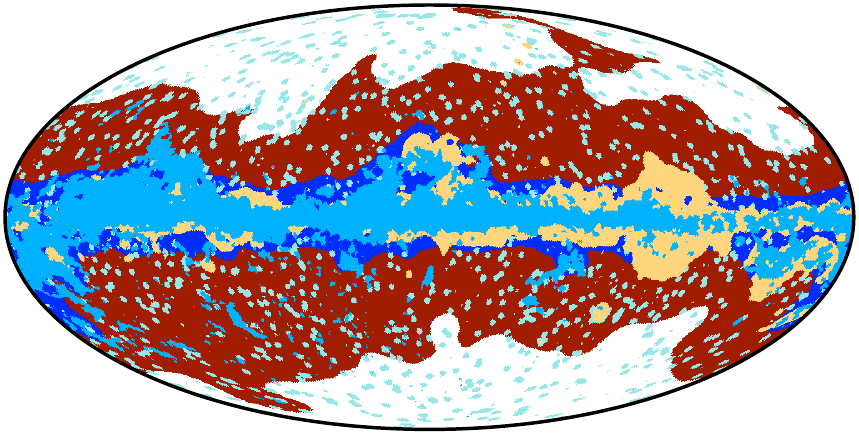}
\caption{Global mask used in the cross-correlation (CC) analysis (Mollweide projection in Galactic coordinates). It comprises the CIB mask (white region), the CO mask (light-blue), the free-free mask (beige), the Galactic mask (deep-blue), and the mask of point sources (turquoise). We use the red regions of the sky. We refer readers to Sect.~\ref{sec:mask} for a detailed description of how the global mask is defined.}
\label{fig:3.1}
\end{figure}

In the data analysis we use a global mask, shown in Fig.~\ref{fig:3.1}, which selects regions of dust emission from the ISM at intermediate Galactic latitudes. We only want to study 
polarization in regions where thermal dust emission dominates.  This means that we need to remove the area around the Galactic plane, where other Galactic contributions are 
significant, and remove the high latitude regions, where the anisotropies of the cosmic infrared background (CIB)  are important with respect to dust emission.
The global mask combines thresholds on several sky emission components (in intensity): carbon monoxide (CO) line emission; free-free; synchrotron; the CIB anisotropies; and 
point sources. We now detail how the global mask is defined.

In the regions of lowest dust column density at high Galactic latitudes, brightness fluctuations from the CIB are significant.  To define the CIB mask, we apply a threshold on the ratio between the root mean square (rms) of the total Galactic emission and of the CIB at $353$\,GHz:
\begin{equation}
\left |\frac{\sigma_{\rm G}}{\sigma_{\rm CIB}}\right | \ < \  9 \ ,
\end{equation}
where $\sigma_{\rm G}$ and $\sigma_{\rm CIB}$ are defined as
\begin{align}
\sigma_{\rm CIB}^2& = \sum_{\ell} \frac{2\ell+1}{4\pi} \ C_{\ell}^{\rm CIB} \ b_\ell^2  \ w_\ell^2, \label{eq:3.1} \\
\sigma_{\rm c}^2& = \sum_{\ell} \frac{2\ell+1}{4\pi} \ C_{\ell}^{\rm c} \ b_\ell^2  \ w_\ell^2, \\
\sigma_{\rm G}^2& = \sigma^2(\tI) - \ \sigma^2_{\rm c} - \ \sigma^2_{\rm CIB} \ .
\end{align}
For this threshold, the CIB contribution to the CC coefficients in Sect.~\ref{sec:cc_em} is smaller than about 1\,\% ($1/9^2$) of that of the total Galactic emission at $353$\,GHz. The 
summation is over the multipole range $15< \ell < 300$ (corresponding to an effective range of angular scales from 1\deg\ to 10\deg). $C_{\ell}^{\rm CIB}$ is the best-fit CIB power 
spectrum at $353$\,GHz \citep{planck2013-pip56}, $C_{\ell}^{\rm c}$ is the best-fit CMB power spectrum \citep{planck2013-p08}, $\tI$ represents the \planck\ $353$\,GHz map, 
$b_{\ell}$ is the beam function and $w_{\ell}$ is the \healpix\ pixel window function. We measure the Galactic to CIB emission ratio over patches with 10\deg\ radius centred on 
\healpix\ pixels at a resolution $\Nside=32$.

The CO, free-free, and synchrotron emission are more important close to the Galactic plane. The first three CO line transitions $J\,{=}\,1\,{\rightarrow}\,0$, $J\,{=}\,2\,{\rightarrow}\,1$, 
and $J\,{=}\,3\,{\rightarrow}\,2$ at 115, 230, and 345\,GHz, respectively, are significant emission components in the \planck\ intensity maps \citep{planck2013-p03a}. The CO mask is 
defined by applying a threshold of 0.5\,K$\kms$ on the ``Type~2" CO $J\,{=}\,1\,{\rightarrow}\,0$, which is extracted using  the \planck\ data between 70 and $353$\,GHz 
\citep{planck2013-p03a}. The free-free emission is weak compared to the CO line emission at $100$\,GHz for most  molecular clouds. In massive star-forming regions and for the 
diffuse Galactic plane emission, free-free emission is significant \citep{planck2013-XIV}.  We take the \wmap\ maximum entropy method free-free map \citep{Bennett:2012} at 
$94$\,GHz and apply a threshold of $10\, \mu$\krj\ (in Rayleigh-Jeans temperature units) to define the free-free mask. In addition, we use the Galactic mask (CS-CR75) from the \planck\ 
component separation results \citep{planck2013-p06} to exclude the synchrotron emission from the Galactic plane and the Galactic ``haze'' \citep{planck2012-IX}. We also apply the 
\planck\ point source mask \citep{planck2013-p08}. 

Our mask focuses on the part of the sky where dust is the dominant emission component at HFI frequencies. This choice makes the spectral leakage from free-free and CO line 
emissions to polarization maps negligible. After masking we are left with 39\,\% of the sky at intermediate Galactic latitudes (10\deg $< |b| <$ 60\deg). The same global mask is used 
for both intensity and polarization correlation analysis to compare results over the same sky.

\section{Cross-correlation method}\label{sec:ccanaly}

We use the CC analysis adopted in many studies \citep{Banday:1996, Gorski:1996, Davies:2006, page2007, Ghosh:2012, planck2013-XII} to extract the signal correlated with the 
$353$\,GHz template in intensity and polarization. The only underlying assumption is that the spatial structure in the $353$\,GHz template and in the map under analysis are locally 
correlated. To reduce this assumption, we apply the CC analysis locally over patches of sky of 10\deg\ radius (Sect.~\ref{sec:implementation}). Our choice for the dust template is 
presented in Sect.~\ref{sec:dust_template}. The methodology is introduced  for intensity in Sect.~\ref{sec:cc_em} and for polarization in Sect.~\ref{sec:cc_pol}. The practical 
implementation of the method is outlined in Sect.~\ref{sec:implementation}.

\subsection{$353$\,GHz template}\label{sec:dust_template}

We perform the CC analysis locally in the pixel domain using the \planck\ $353$\,GHz maps of Stokes parameters as representative internal templates for dust emission in intensity 
(\StokesI\ with the ZLE subtracted)  and polarization (\StokesQ\ and \StokesU). Our choice of a \planck\ map as a dust template  addresses some of the issues plaguing alternative 
choices. First, unlike the \hi\ map, the $353$\,GHz map traces the dust in both \hi\ and H$_2$ gas \citep{Reach:1998, planck2011-7.12,planck2013-XVII}. Second, unlike the full-sky 
\citet{Finkbeiner:1999} 94\,GHz (hereafter FDS) map, the $353$\,GHz map does not rely on an extrapolation over a large frequency range, from 100\microns\ to the \planck\ bands. 
The main drawback of  the $353$\,GHz template is that it includes CMB and CIB anisotropies. By introducing the global mask, we work with the sky region where the CIB 
anisotropies are small compared to dust emission. However, the contribution of the CMB to the CC coefficients, most significant at microwave frequencies, needs to be subtracted.

\subsection{Intensity}\label{sec:cc_em}

\subsubsection{Correlation with the $353$\,GHz template}

For the intensity data, the CC coefficient ($\alphaInu$) is obtained by minimizing the $\chiI$\ expression given by,
\begin{equation}
\chiI\ = \ \sum_{k=1}^{N_{\rm pix}} \left [ \nI (k) \ - \ [\alphaInu]_{353}^{\rm 1T} \ \tI (k)  - \ a \right ]^2 ,  \label{eq:4.1}
\end{equation}
where $\nI$ and $\tI$ denote the data and the $353$\,GHz template maps, respectively. This is a linear fit and the solution is computed analytically. Here the CC coefficient is a 
number in \kcmb \ \kcmb$^{-1}$, as both $\nI$ and $\tI$ are expressed in \kcmb\ units. The constant offset, $a$, takes into account the local mean present in the template as well as 
in the data. The sum is over the unmasked pixels, $k$, within a given sky patch. We are insensitive to the residual dipole present at \planck\ frequencies because we perform local 
correlation over 10\deg\ radius patches. The index `$\rm 1T$' represents the $353$\,GHz correlated coefficient at a given frequency $\nu$ that we obtained using one template only. 

The CC coefficient at a given frequency includes the contribution from all the emission components that are correlated with the $353$\,GHz template (Appendix~\ref{sec:math_cc}). 
It can be decomposed into the following terms: 
\begin{align}
[\alphaInu]_{353}^{\rm 1T} = \ & \alphaI (c^1_{353}) + \ \alphaInu (d_{353})    + \ \alphaInu (s_{353}) + \ \alphaInu (f_{353}) + \ \alphaInu (a_{353}) \ , \label{eq:4.2}
\end{align}
where  $c^1_{353}$, $d_{353}$, $s_{353}$, $f_{353}$, and $a_{353}$ refer to the CMB, dust, synchrotron, free-free, and AME signals that are correlated with the $353$\,GHz 
template, respectively. The CMB CC coefficient term is achromatic  because Eq.~\eqref{eq:4.2} is expressed in \kcmb\ units. We neglect the contributions of the three CO lines, point 
sources, and the CIB anisotropies, since these are subdominant within our global mask  (Sect.~\ref{sec:mask}). We also neglect the cross-correlation of the ZLE with the dust 
template. The chance correlations between the emission components we neglect and the dust template contribute to the statistical uncertainties on the dust SED, but do not bias it. 
We checked this with Monte Carlo simulations (Appendix~\ref{sec:simul})  and repeated our analysis on HFI maps with the ZLE subtracted. The correlation terms of the 
synchrotron and AME components are negligible at $\nu \ge 100$\,GHz, as synchrotron and AME both have a steep spectrum that falls off fast at high frequencies. The free-free 
emission is weak outside the Galactic plane at high frequencies and does not contribute significantly to the CC coefficients. The synchrotron, AME and free-free terms only 
become significant at  $\nu < 100$\,GHz inside our global mask.

\subsubsection{Correlation with two and three templates}\label{sec:cc_em_two_temp}

To remove   $\alphaInu (s_{353})$ and  $\alphaInu (f_{353})$ in Eq.~\eqref{eq:4.2}, we cross-correlate the \planck\ and \wmap\ data  with either two or three templates (including the dust template). We use the $353$\,GHz and the $408$\,MHz maps for the fit with two templates, and add the DDD \halpha\ map for the three-template fit.
The $\chiI$ expressions that we minimize for these two cases are
\begin{align}
\chiI&= \ \sum_{k=1}^{N_{\rm pix}} \left[ \nI (k)  -  \ [\alphaInu]_{353}^{\rm 2T} \ \tI (k)    - \  [\alphaInu]_{0.408}^{\rm 2T} \  \hI (k) - \ a \right]^2 \ , \\ 
\chiI&= \ \sum_{k=1}^{N_{\rm pix}}  \left[ \nI (k)  - \  [\alphaInu]_{353}^{\rm 3T} \  \tI (k)   - \  [\alphaInu]_{0.408}^{\rm 3T} \  \hI (k) \right.  \nonumber\\
&\hspace{4.2cm} \left. - \  [\alphaInu]_{\halpha}^{\rm 3T} \ \fI (k) - \ a \right]^2 \ ,
\end{align}
where $\nI$, $\tI$, $\hI$, and $\fI$ denote the data at a frequency $\nu$, the \planck\ $353$\,GHz, Haslam 408\,MHz, and DDD \halpha\ maps, respectively.  For these multiple template fits, the CC coefficients are given by
\begin{align}
[\alphaInu]_{353}^{\rm 2T}&=   \ \alphaI (c^2_{353}) + \ \alphaInu (d_{353})  + \ \alphaInu (f_{353}) + \ \alphaInu (a_{353})  \label{eq:4.3} \\
[\alphaInu]_{353}^{\rm 3T}&=   \ \alphaI (c^3_{353}) +  \ \alphaInu (d_{353})  + \ \alphaInu (a_{353}) \ . \label{eq:4.4}
\end{align}
The indices `$\rm 2T$' and `$\rm 3T$' are used here to distinguish the CC coefficients for the fit with two and three templates, respectively. The use of additional templates  
removes the corresponding terms from the right hand side of these equations.  Equation~\eqref{eq:4.4} is used to derive the mean dust SED in intensity. 
Equations~\eqref{eq:4.2},~\eqref{eq:4.3}, and~\eqref{eq:4.4} may be combined to derive $\alphaInu (s_{353})$ and  $\alphaInu (f_{353})$.

\subsection{Polarization}\label{sec:cc_pol}

For the polarization data, we cross-correlate both the Stokes  \StokesQ\ and \StokesU\  $353$\,GHz templates with the \StokesQ\ and \StokesU\ maps for all \planck\ and \wmap\ 
frequencies.  Ideally in CC analysis, the template is free from noise, but the \planck\ $353$\,GHz polarization templates do contain noise, which may bias the CC coefficients. To 
circumvent this problem, we use two  independent \StokesQ\  and \StokesU\  maps made with the two detector sets DS1 and DS2 at $353$\,GHz  as templates 
(Sect.~\ref{sec:systematics}). The maps made with each of the two detector sets have independent noise and dust BPM. Using two polarization detector sets at $353$\,GHz with 
independent noise realizations reduces the noise bias in the determination of the $353$\,GHz CC coefficients. We use the detector set maps rather than the half-ring maps because 
the removal of glitches induces some noise correlation between the two half-ring maps that affects the data at all multipoles \citep{planck2013-p03,planck2013-p03e}.

The polarization CC coefficient ($\alphaPnu$) is derived by minimizing the $\chiP$ expression given by
\begin{align}
\chiP&= \sum_{i=1}^2 \sum_{k=1}^{N_{\rm pix}}\  \left [ \StokesQ_{\nu} (k)  - \  [\alphaPnu]_{353}^{\rm 1T} \  \StokesQ_{353}^i (k)- \ a \right ]^2  \nonumber \\
&\hspace{2cm} + \ \left [ \StokesU_{\nu} (k)  - \  [\alphaPnu]_{353}^{\rm 1T} \  \StokesU_{353}^i (k) - \ b \right ]^2 \ ,  \label{eq:4.5}
\end{align}
where the index $i$ takes the values 1 and 2, which correspond to the DS1 and DS2 maps at $353$\,GHz.  The summation $k$ is over the unmasked pixels within a 
given sky patch. The constant offsets $a$ and $b$ take into account the local mean present in the template as well as in the data Stokes~\StokesQ\ and \StokesU\ maps, 
respectively. At $353$\,GHz, we cross-correlate the DS1 and DS2 maps of \StokesQ\ and \StokesU\ among themselves, minimizing
\begin{align}
\chiP&= \sum_{\substack{i=1\\ i \neq j}}^2 \sum_{k=1}^{N_{\rm pix}} \ \left [ \StokesQ_{353}^j (k)  - \  [\alphaPt]_{353}^{\rm 1T} \  \StokesQ_{353}^i (k) - \ a\right ]^2  \nonumber \\
&\hspace{1cm} + \ \left [ \StokesU_{353}^j (k)  - \  [\alphaPt]_{353}^{\rm 1T} \  \StokesU_{353}^i (k) - \ b \right ]^2 \ .
\end{align}

The CC coefficients, $\alphaPnu$, comprise the contributions of CMB, dust, synchrotron and possibly AME polarization. The free-free polarization is expected to be negligible theoretically \citep{Rybicki:1979} and has been constrained to a few percent observationally  \citep{Macellari:2011}. The  polarization decomposition is given by
\begin{equation}
[\alphaPnu]^{\rm 1T}_{353}   = \ \alphaP (c^1_{353}) + \ \alphaPnu (d_{353}) + \ \alphaPnu (s_{353}) + \ \alphaPnu (a_{353}) \ .  \label{eq:4.6}
\end{equation}
The polarized CMB CC coefficient, $\alphaP (c^1_{353})$, is achromatic because Eq.~\eqref{eq:4.6} is expressed in \kcmb\ units.  Unlike for intensity, due to the absence of any 
polarized synchrotron template free from Faraday rotation \citep{Gardner:1966}, we cannot perform a fit with two templates  to remove $\alphaPnu (s_{353})$ in Eq.~\eqref{eq:4.6}.

We have performed Monte Carlo simulations at the HFI frequencies in order to estimate the uncertainty on the CC coefficient induced by the noise and other Galactic emission present in the data (see Appendix~\ref{sec:simul}).

\subsection{Implementation}\label{sec:implementation}

Here we describe how we implement the CC method. The \planck, \wmap, \hi\ and DIRBE sky maps are smoothed to a common resolution of 1\deg, taking into account the effective 
beam response of each map, and reduced to a \healpix\ resolution $\Nside=128$.  For the \planck\ and \wmap\ maps, we use the effective beams defined in multipole space that are 
provided in the \planck\ Legacy Archive\footnote{\url{http://archives.esac.esa.int}} (PLA) and  LAMBDA website \citep{planck2013-p03c, planck2013-p02d, Bennett:2012}. The 
Gaussian approximation of the average beam  widths for \planck\ and \wmap\ maps are quoted in Table~\ref{tab:2.1}. For the \hi\ and DIRBE maps, we also use Gaussian 
beams with the widths given in Table~\ref{tab:2.1}. For the polarization data, we use the ``\ensuremath{\tt ismoothing}'' routine of \healpix\ that decomposes the \StokesQ\ and 
\StokesU\ maps  into $E$ and $B$ $a_{\ell m}$s, applies Gaussian smoothing of 1\deg\ in harmonic space (after deconvolving the effective azimuthally
symmetric beam response for each map), and transforms the smoothed $E$ and $B$ $a_{\ell m}$s back into \StokesQ\ and \StokesU\ maps at $\Nside=128$ resolution.

We divide the intermediate Galactic latitudes into sky patches with 10\deg\ radius centred on \healpix\ pixels for $\Nside=8$.  For a much smaller radius we would have too few 
independent sky pixels within a given sky patch to measure the mean dust SED. For a much larger radius we would have too few sky patches to estimate the statistical uncertainty 
on the computation of the mean dust SED. Each sky patch contains roughly 1500 pixels at $\Nside=128$ resolution. We only consider 400 sky patches ($N_{\rm bins} $), which 
have 500 or more unmasked pixels. We then cross-correlate the $353$\,GHz \planck\ internal template with the \wmap\ and \planck\ maps between 23 and $353$\,GHz, locally in 
each sky patch to extract the $353$\,GHz correlated emission in intensity, along with its polarization counterpart. The sky patches used are not strictly independent.  Each sky pixel 
is part of a few sky patches, which is required to  sample properly the spatial variations of the CC coefficients.  The mean number of times each pixel is used in CC coefficients 
($N_{\rm visit}$) is estimated with the following formula:
\begin{equation}
N_{\rm visit}  =  \frac{N_{\rm bins}  \times \langle N_{\rm pixels} \rangle} {0.39 \times N_{\rm total}} \sim 5 \ ,
\end{equation}
where $N_{\rm total} =12 \times N_{\rm side}^2$ is the total number of pixels at 1\deg\ resolution, 0.39 is the fraction of the sky used in our analysis and $\langle N_{\rm pixels} 
\rangle = 1000$ is the average number of pixels per sky patch after masking.

\section{Component separation methodology}\label{sec:comp_sep}

At the highest frequencies ($\nu \ge$ 100\,GHz) within our mask, the two main contributors to the CC coefficient are the CMB and dust emission. In this section, we detail how we 
separate them and estimate the spectral index of the dust emission (\betad) in intensity and polarization.

\subsection{Separation of dust emission for intensity }\label{sec:dust_spectral}

The CC coefficients at $\nu \ge$ 100\,GHz can be written as 
\begin{equation}
[\alphaInu]_{353}^{\rm 3T} = \ \alphaI (c^3_{353}) + \ \alphaInu(d_{353}) \ , \label{eq:5.1}
\end{equation}
where  $c^3_{353}$ and $d_{353}$ are the $353$\,GHz correlated CMB and dust emission, respectively. 
The CMB CC coefficient is achromatic in \kcmb\ units, i.e., in temperature units relative to the CMB blackbody spectrum. To remove the CMB contribution, we work with the 
differences of CC coefficients between two given frequencies. To measure the dust spectral index both in intensity and polarization, we choose to work with colour ratios defined 
between two given frequencies $\nu_2$ and $\nu_1$ as
\begin{align}
R_{\nu_0}^{\rm I}(\nu_2,\nu_1)&=
\dfrac{[\alpha_{\nu_2}^{\rm I}]_{353}^{\rm 3T} -[\alpha_{\nu_0}^{\rm I}]_{353}^{\rm 3T} }{[\alpha_{\nu_1}^{\rm I}]_{353}^{\rm 3T} -[\alpha_{\nu_0}^{\rm I}]_{353}^{\rm 3T} } \label{eq:5.2} \\
& = \dfrac{\alpha_{\nu_2}^{\rm I}(d_{353})-\alpha_{\nu_0}^{\rm I}(d_{353})}{\alpha_{\nu_1}^{\rm I}(d_{353})-\alpha_{\nu_0}^{\rm I}(d_{353})} \nonumber  \ , 
\end{align}
where $\nu_0$ represents the reference CMB frequency which is chosen to be 100\,GHz in the present analysis.  To convert the measured colour ratio into \betad\ we follow earlier 
studies \citep{planck2013-XVII,planck2013-p06b} by approximating the SED of the dust emission with a modified blackbody (MBB, hereafter) spectrum
\citep{planck2011-7.13,planck2011-7.12,planck2013-p06b} given by
\begin{align}
\alphaInu(d_{353}) = F_{\nu} \ C_{\nu}  \ A_{\rm d} \ \nu^{\ \betad} \ B_{\nu} (\Td) \ , \label{eq:5.4}
\end{align}
where \Td\ is the colour temperature and \betad\ is the spectral index of the dust emission. The factor $F_{\nu}$ takes into account the conversion from \MJysr\ (with the photometric 
convention $\nu \StokesI_{\nu}$=constant)  to  \kcmb\ units, while $C_{\nu}$ is the colour correction that depends on the value of \betad\ and \Td. The colour correction is computed 
knowing the bandpass filters at the HFI frequencies \citep{planck2013-p03d} and the spectrum of the dust emission. Using Eq.~\eqref{eq:5.4}, the colour ratio can be written as
a function of \betad\ and \Td:
\begin{align}
R_{\nu_0}^{\rm I}(\nu_2,\nu_1)&= \dfrac{F_{\nu_2} C_{\nu_2}  \nu_2^{\ \betad} B_{\nu_2}(\Td) - F_{\nu_0} C_{\nu_0} 
  \nu_0^{\ \betad} B_{\nu_0}(\Td)}{F_{\nu_1} C_{\nu_1}  \nu_1^{\ \betad}
  B_{\nu_1}(\Td)-F_{\nu_0} C_{\nu_0}  \nu_0^{\ \betad}
  B_{\nu_0}(\Td)} \nonumber \\ &=g(\betad, \Td)  \ . \label{eq:5.5}
\end{align}
In Sect.~\ref{sec:measuring_betaI}, we use the three \planck\ maps, at $100$, $217$, and $353$\,GHz, to compute  \RIT\ and measure the dust spectral index (\betadmmI) at microwave frequencies (or mm wavelengths), for each sky patch. In the next section, we explain how we determine  \Td.

\subsection{Measuring colour temperatures in intensity} \label{sec:Td}

The dust temperatures inferred from an MBB fit of the \planck\ at $\nu \ge 353$\,GHz and the \iras\ 100\microns\ sky maps at 5\arcmin\ resolution \citep{planck2013-p06b} cannot be 
used to compute mean temperatures within each sky patch because the fits are nonlinear.  The two frequencies, \planck\ 857\,GHz and DIRBE 100\microns\ (3000\,GHz), which 
are close to the dust emission peak, are well suited to measure \Td\ for each sky patch.  We use the \planck\ $353$\,GHz map as a template to compute the colour ratio \RIIT\ over 
each sky patch, as described in Eq.~\eqref{eq:5.2}. The superscript ${\rm I}$ on the colour ratio and \betad\ denote intensity. As the CMB signal is negligible at  these frequencies, we 
work directly with the ratio \RIIT, without subtracting the 100\,GHz CC measure. We assume a mean dust spectral index at submm frequencies, \betasubmmI, of 1.50. The choice of \betasubmmI\
value is based on the MBB fit to the  dust emissivities at 100 $\mu$m and the \planck\ 353, 545 and 857 GHz frequencies, for each sky patch. Due to the \betasubmmI -- \Td\ anti-correlation,
the variations of the \betasubmmI\ values  just increases the scatter of the \Td\ values by about 20\,\% as compared to \Td\ values derived using fixed \betasubmmI. However, the \Td\ values from the 
MBB fits are closely correlated with \Td\ values determined using the ratio \RIIT\ and a fixed spectral index. We use the colour ratio \RIIT\ and mean $\betasubmmI=1.50$  to estimate \Td\ values for 
each sky patch by inverting the relation given in Eq.~\eqref{eq:5.5}.

\begin{figure}[h!]
\includegraphics[width=8.8cm]{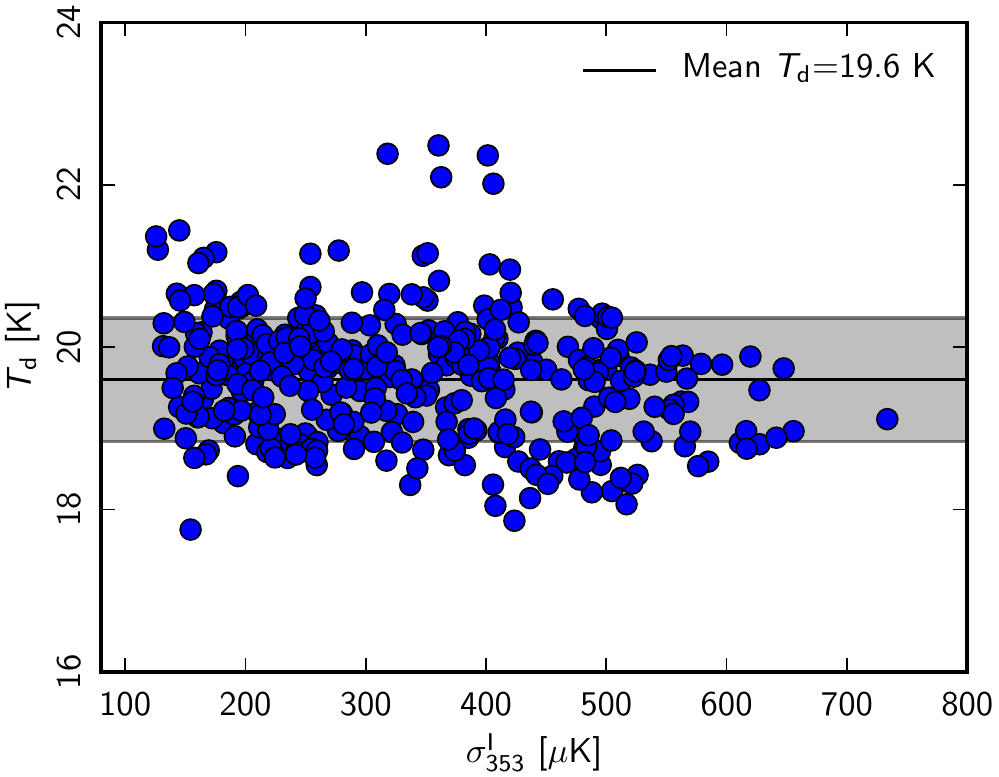} 
\caption{Dust colour temperatures, \Td, computed from \RIIT, are plotted versus  the local dispersion of the $353$\,GHz intensity template, \sigmaI. The mean \Td\ is 
$19.6$\,K, with the  $1\sigma$ dispersion of $0.8$\,K across sky patches (shaded area).}
\label{fig:5.1}
\end{figure}

In Fig.~\ref{fig:5.1}, we plot  the derived \Td\ versus the local brightness dispersion of the \planck\ $353$\,GHz template in intensity (\sigmaI). We point out that \sigmaI\ is not an 
uncertainty in the $353$\,GHz intensity template. The mean value of \Td\ over our mask at intermediate Galactic latitudes is $19.6$\,K. The $1\sigma$ dispersion of \Td\  over the 400 sky patches is $0.8$\,K. This value is slightly smaller that  the mean value at high Galactic latitudes, $20.4 \pm 1.1$\,K for $\betasubmmI = 1.57 \pm 0.11$\footnote{These values are derived from a grey-body fit of dust SED at $\nu \ge 353\,$GHz. The error-bar on the dust temperature is associated with that on the spectral index}, 
we obtained repeating the dust-\hi\ correlation analysis of \citet{planck2013-XVII} on the same full-mission \planck\ data.

The choice of \betasubmmI\ used in this paper is different from the one derived from the analysis of high Galactic latitude data \citep{planck2013-XVII} and the analysis of the whole sky \citep{planck2013-p06b} using public release \planck\ 2013 data. This difference results  
from a change in the photometric calibration by $1.9$\,\%, $-2.2$\,\%, $-3.5$\,\%,  at 353, 545 and $857\,$GHz, 
between the DX11d and the \planck\ 2013 data. The new calibration factors make the mean \betasubmmI\ slightly smaller and \Td\ slightly higher. To estimate uncertainties on \betasubmmI, we run a set of Monte-Carlo simulations that take into account the absolute and relative calibration uncertainties present in the DIRBE and  the \planck\ full-mission  HFI data at $\nu \ge 353$\,GHz. 
We assume that the MBB spectrum is a good fit to the data and apply $1\sigma$ photometric uncertainties of 1\,\%, 7\,\%,  7\,\%, and 13\,\% at 353, 545, 857, and 3000 GHz respectively. To get multiple SED realizations, we vary the MBB spectrum within the photometric uncertainty at each frequency used for the fit, independently of others. Then we perform the MBB SED fit and  find that the $1\sigma$ dispersion on the mean value of \betasubmmI\ is $0.16\,(\mathrm{syst.})$. The new value of $\betasubmmI=1.50$ is well within the range of values and systematic uncertainties quoted in Table 3 of  \citealt{planck2013-p06b} (using 2013 \planck\ data) for the same region of the sky.

\subsection{Separation of CMB emission in intensity}\label{sec:cmb_est}

The CC coefficient, derived in Eq.~\eqref{eq:4.2}, contains the CMB contribution that is achromatic in \kcmb\ units. We determine this CMB contribution assuming that the dust emission is well approximated by a MBB spectrum from 100 to $353$\,GHz. For each sky patch, we use the values of  \betadmmI\ and \Td\ from Sects.~\ref{sec:measuring_betaI} and \ref{sec:Td}. We solve for two parameters, the CMB contribution, $\alpha^{\rm I} (c^3_{353})$, and the dust amplitude, $A_{\rm d}^{\rm \StokesI}$, by minimizing
\begin{equation}
\chi_{\rm s}^2 = \sum_{\nu}\ \left ( \frac{[\alphaInu]_{353}^{\rm 3T} - \alpha^{\rm I} (c^3_{353}) - \ F_{\nu}\  C_{\nu} \ A_{\rm d}^{\rm \StokesI} \ \nu^{\ \betadmmI} \ B_{\nu} (\Td) }{\sigma_{\alphaInu}} \right )^2  \ , 
\end{equation}
where $\sigma_{\alphaInu}$ is the uncertainty on the CC coefficient, determined using the Monte Carlo simulations (Appendix~\ref{sec:simul}). The joint spectral fit of 
$\alpha^{\rm I} (c^3_{353})$, \betadmmI, \Td, and $A_{\rm d}^{\rm \StokesI}$ leads to a degeneracy between the fitted parameters. To avoid this problem, we fix the values of 
\betadmmI\ and \Td\ for each sky patch based on the colour ratios, independent of the value of $\alpha^{\rm I} (c^3_{353})$. The CMB contributions are subtracted from the CC 
coefficients at all  frequencies, including the LFI and \wmap\ data not used in the fit. After CMB subtraction, the CC coefficient ($\tildealphaInu$) for the $353$\,GHz template is
\begin{equation}
[\tildealphaInu]_{353}^{\rm 3T}   =  [\alphaInu]_{353}^{\rm 3T}  - \alpha^{\rm I} (c^3_{353}) =  \ \alphaInu (d_{353})  \ .\label{eq:6.1}
\end{equation}
We perform the same exercise on the one- and two-template fits to derive the CMB subtracted CC coefficients.

\subsection{Separation of dust emission for polarization}\label{sec:5.4}

As for our analysis of the intensity data, we write the 353 GHz correlated polarized CC coefficients at $\nu \ge 100$\,GHz  as 
\begin{equation}
[\alphaPnu]_{353}^{\rm 1T} = \ \alphaP (c^1_{353}) + \ \alphaPnu(d_{353}) \ , \label{eq:5.4.1}
\end{equation}
where  $c^1_{353}$ and $d_{353}$ are the CMB and dust polarized emission correlated with the $353$ polarization templates. The 
contributions from synchrotron and AME  to the polarized CC coefficients are assumed to be negligible at HFI frequencies. 
Like for intensity in Eq.~\eqref{eq:5.5}, we compute  \RIP\ combining  the three polarized CC coefficients at  100, 217 and $353\,$GHz.
We assume that the temperature of the dust grains contributing to the polarization is the same as that determined for the 
dust emission in intensity (Sect.~\ref{sec:Td}), and derive \betadmmP\ at microwave frequencies. 

To separate the contribution of dust and the CMB to the polarized CC coefficients, we follow the method described in Sect.~\ref{sec:cmb_est}, and 
rely on the Monte Carlo simulations described in Appendix~\ref{sec:simul} to estimate uncertainties. The CMB contribution is 
subtracted at all frequencies, including the LFI and \wmap\ data.

\section{Dust spectral index for intensity}\label{sec:betaI}

Here we estimate the dust spectral index  \betadmmI\ at microwave frequencies ($\nu \le 353$\,GHz) and  mm wavelengths. We present the results of the data analysis and estimate the uncertainties, including possible systematic effects. 

\subsection{Measuring \betadmmI\ }\label{sec:measuring_betaI}

\begin{figure}[h!]
\begin{tabular}{c}
\includegraphics[width=8.8cm]{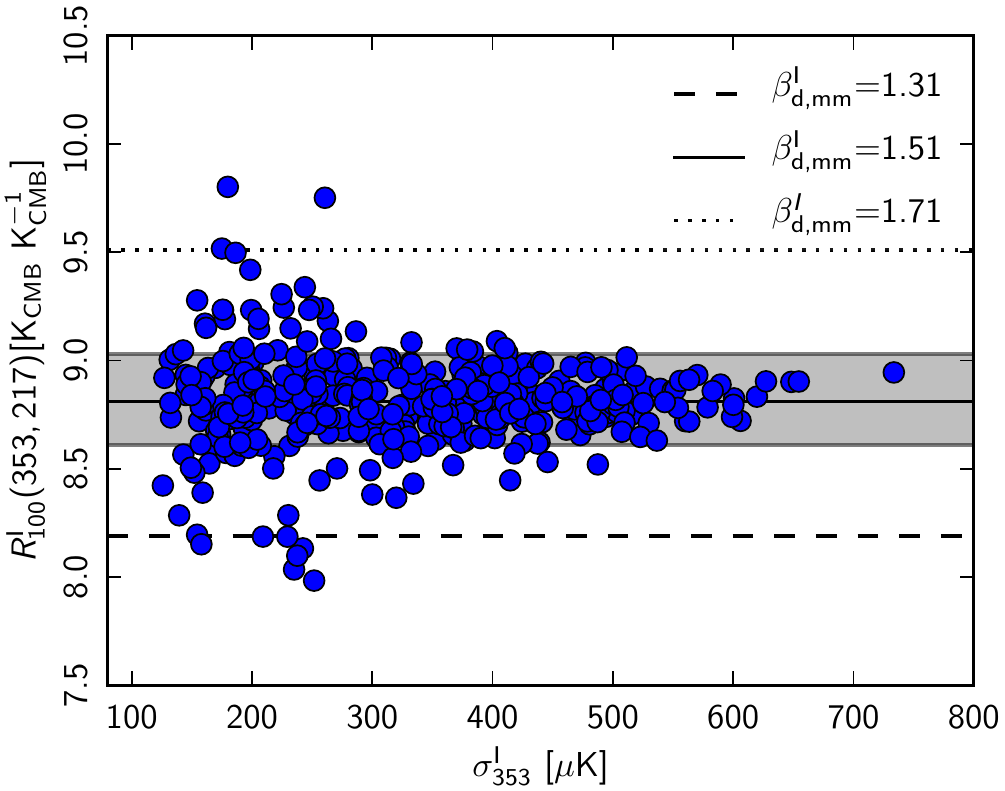}  \\
\includegraphics[width=8.8cm]{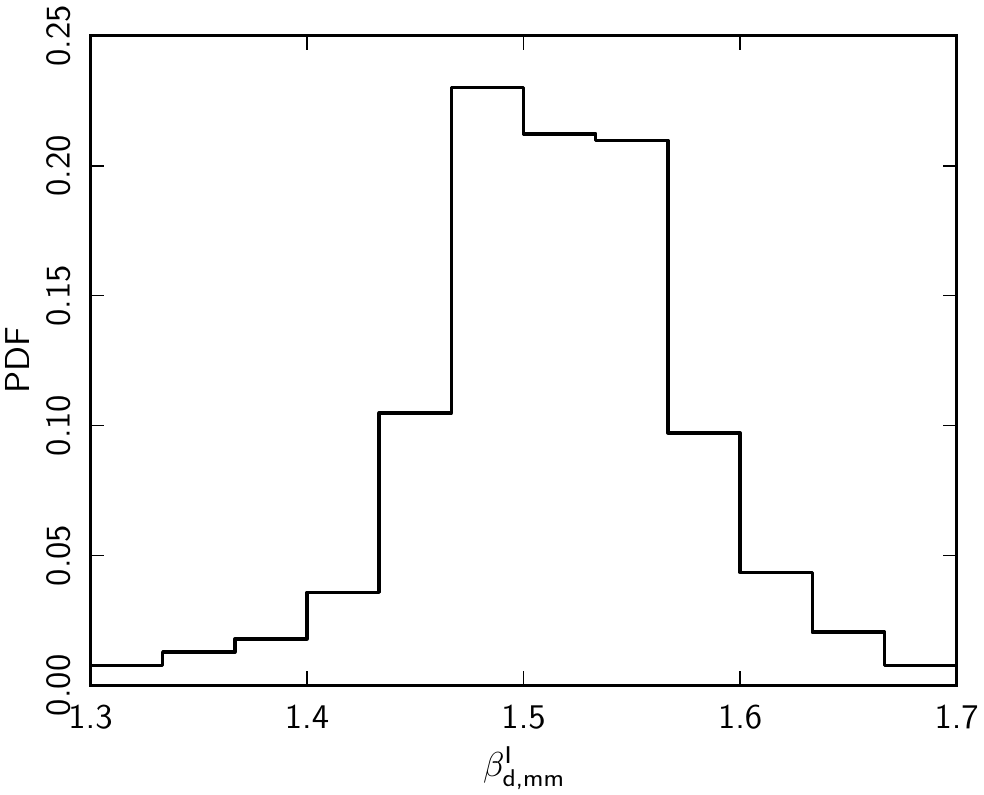}
\end{tabular}
\caption{\emph{Top}: Colour ratio  \RIT\ versus dispersion of the $353$\,GHz template \sigmaI\ for each sky patch. The $1\sigma$ dispersion of the \RIT\ values is shown as the 
shaded grey area. The mean ratio corresponds to  a spectral index of 1.51 (solid line) for  a mean  $\Td=19.6$\,K. \emph{Bottom}: probability distribution function (PDF) of the 
\betadmmI\  values derived from \RIT\ using the specific value of \Td\ for each sky patch. The measured $1\sigma$ dispersion of \betadmmI\  is  0.07.}
\label{fig:6.1}
\end{figure}

We use the three full mission \planck\ maps, at $100$, $217$, and $353$\,GHz, to derive a mean \betadmmI\ using the three-template fit, assuming an MBB spectrum for the dust emission (Sect.~\ref{sec:dust_spectral}). 
The 217 and $353$\,GHz maps have the highest signal-to-noise ratio for dust emission at microwave frequencies, whereas the $100$\,GHz map is used as a reference frequency to 
subtract the CMB contribution at the CC level. We estimate \RIT\ for each sky patch using the relation given by Eq.~\eqref{eq:5.2}.  The values of \RIT\ are plotted in top panel of Fig.~
\ref{fig:6.1} as a function of \sigmaI, which allows us to identify the statistical noise and systematic effects due to uncertainties on the CC coefficients. Our Monte Carlo simulations 
(Appendix~\ref{sec:simul}) show that the uncertainties on \RIT\ scale approximately as the inverse square-root of \sigmaI, and that the scatter in the measured \RIT\ for sky patches 
with low \sigmaI is due to data noise.

For each sky patch, we derive \betadmmI\  from \RIT\  by inverting Eq.~\eqref{eq:5.5} for the values of  \Td\ derived in Sect.~\ref{sec:Td}. The histogram of \betadmmI\ for all sky patches 
is presented in the bottom panel of Fig.~\ref{fig:6.1}.  The mean value of \betadmmI\  from the 400 sky patches is 1.514 (round-off to 1.51) with  $1\sigma$ dispersion of 0.065 (round-off to 0.07). The 
statistical uncertainty on the mean \betadmmI\  is 0.01, which is computed 
from the $1\sigma$ deviation divided by the square root of the number of independent sky patches (400/$N_{\rm visit}$) used. This estimate of the statistical error bar on  \betadmmI\ 
takes into account the uncertainties associated with  the  chance correlation between the dust template and emission components (CO lines, point sources, the CIB anisotropies and 
the ZLE) not fitted with templates.  It also includes uncertainties on the subtraction of the CMB contribution.

\begin{table}[tmb]
\begingroup
\newdimen\tblskip \tblskip=5pt
\caption{\label{tab:6.1} Dust spectral indices for intensity derived applying the three-template fit on distinct subsets of the \planck\  data (Sect.~\ref{sec:systematics}). Here the index ``Full" refers to the full mission \planck\ 2014 data, which is used in Sect.~\ref{sec:measuring_betaI} to produce Fig.~\ref{fig:6.1}. The scatter of the seven measurements for the subsets of the \planck\ data is within the $1\sigma$ statistical uncertainty on the mean \betadmmI.}
\nointerlineskip
\vskip -3mm
\setbox\tablebox=\vbox{
   \newdimen\digitwidth 
   \setbox0=\hbox{\rm 0} 
   \digitwidth=\wd0 
   \catcode`*=\active 
   \def*{\kern\digitwidth}

   \newdimen\signwidth 
   \setbox0=\hbox{+} 
   \signwidth=\wd0 
   \catcode`!=\active 
   \def!{\kern\signwidth}

\halign{
\hbox to 1.3in{#\leaderfil}\tabskip 2.0em&
\hfil #\hfil\tabskip=0pt\cr
\noalign{\doubleline}
Data sets& \betadmmI\cr 
\noalign{\vskip 4pt\hrule\vskip 6pt}
      Full &  1.514\cr
      YR1& 1.514\cr
      YR2 & 1.519\cr
      HR1& 1.515\cr 
      HR2& 1.518\cr
      DS1& 1.514\cr
      DS2& 1.520\cr
\noalign{\vskip 5pt\hrule\vskip 3pt}}}
\endPlancktablewide
\par
\endgroup
\end{table}

\begin{figure}[h!]
\includegraphics[width=8.8cm]{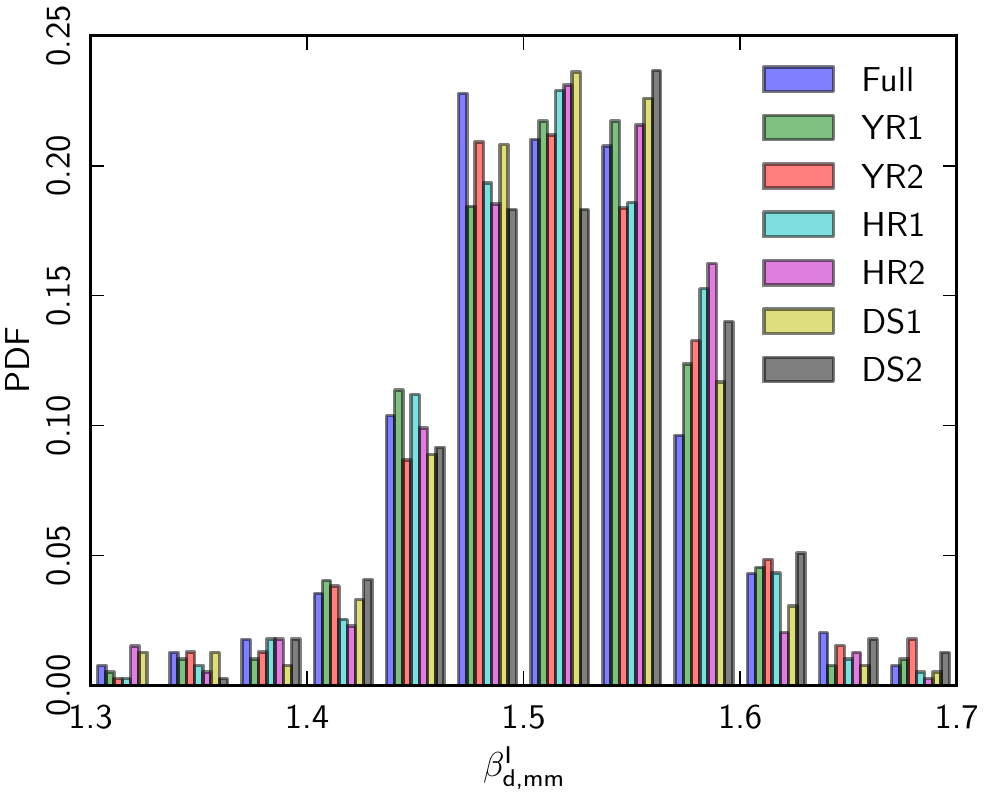}
\caption{Same plot as in the bottom panel of Fig.~\ref{fig:6.1}, including our results for the subsets of the \planck\ data listed in Table~\ref{tab:6.1}.
The bin per bin measurements of \betadmmI\ using subsets of the \planck\ data are compatible with the one obtained using the full mission data (Sect.~\ref{sec:measuring_betaI}).}
\label{fig:6.4}
\end{figure}

\subsection{Uncertainties on \betadmmI }\label{sec:err_betaI}

We use the full mission \planck\ intensity maps as a reference data for the mean dust spectral index value.
To assess the systematic uncertainties on the mean spectral index, we repeat our CC analysis on maps made with subsets of the \planck\ data (Sect.~\ref{sec:systematics}), keeping 
the same ZLE-subtracted \planck\ $353$\,GHz map as a template. For each set of maps, we compute the mean \betadmmI\ from \RIT\ values. Table~\ref{tab:6.1} lists the 
\betadmmI\ values derived from the three-template fit applied to each data sub-set. The six measurements of \betadmmI\ from various data splits  
are within the $1\sigma$ statistical uncertainties on the mean intensity dust spectral index.  We find a mean dust spectral index  $\betadmmI=1.51\pm0.01\,(\mathrm{stat.})$. This spectral index is very close to the mean index of 1.50 at sub-mm wavelengths we 
derived from MBB fits to the \planck\ data at $\nu \ge 353$\,GHz in Sect.~\ref{sec:Td}.

\subsection{Dependence of \betadmmI\ on the choice of \betasubmmI }

\begin{figure}[h!]
\includegraphics[width=8.8cm]{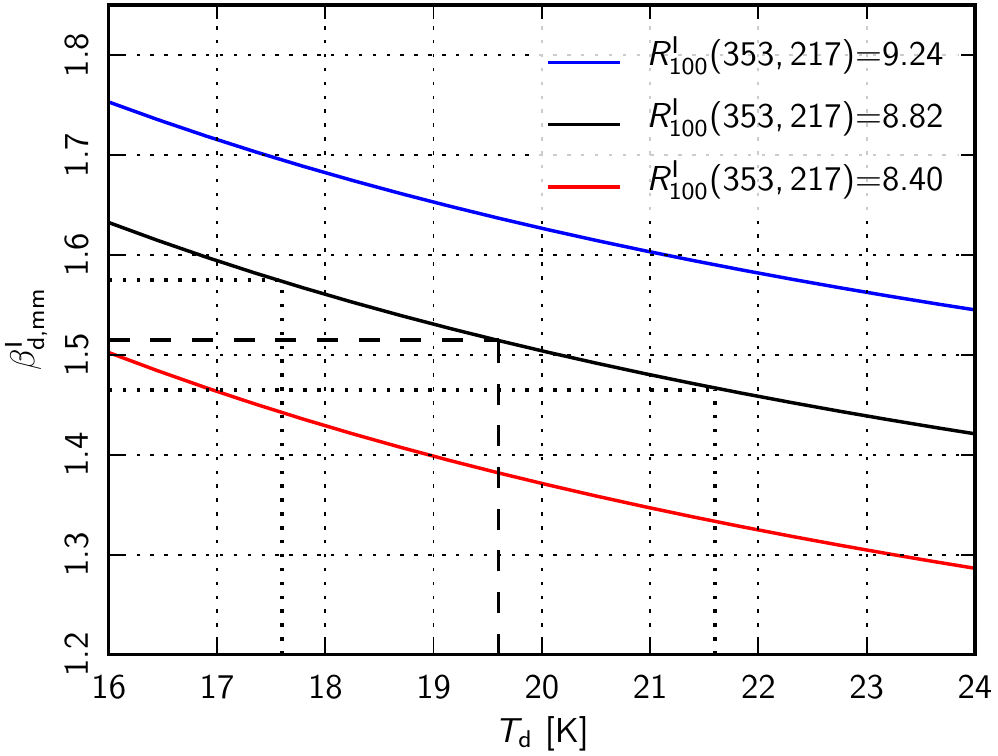}  
\caption{Variation of \betadmmI\ with \Td\ for constant values of \RIT. The dashed line corresponds to a mean $\Td=19.6$\,K and $\betadmmI=1.51$ for the best-fit value of $\RIT=8.82$. The two dotted line lines correspond to a change in \Td\ of $\pm \,2$\,K about its mean value, resulting in change of \betadmmI\ value of $\pm\,0.05$.}
\label{fig:6.2}
\end{figure}

In Fig.~\ref{fig:6.2} we plot  \RIT\ as a function of \betadmmI\ and \Td. The central black line corresponds to the median value of \RIT\ obtained using the CC analysis. 
Figure~\ref{fig:6.2} shows that varying \Td\ by $\pm$\,2\,K, a $\pm$\,2.5$\sigma$ deviation from the mean value, 
changes \betadmmI\ by $\pm$\,0.05. To estimate \betadmmI, we 
use \Td, which in turn depends on \betasubmmI. We repeat our analysis with two different starting values of \betasubmmI, which are within 1$\sigma$ systematics uncertainties 
derived in Sect.~\ref{sec:Td}. For the values of $\betasubmmI=1.34$ and 1.66, we find $\betadmmI=1.50$ and 1.53, respectively. This is due to the fact that in Rayleigh-Jeans limit, the effect of \Td\  is low,
and the shape of the spectrum is dominated by \betadmmI. Our determination of mean \betadmmI\ is robust and independent of the initial choice of \betasubmmI\ used for the analysis.

\subsection{Alternative approach of measuring \betadmmI\ }

\begin{figure}[h!]
\includegraphics[width=8.8cm]{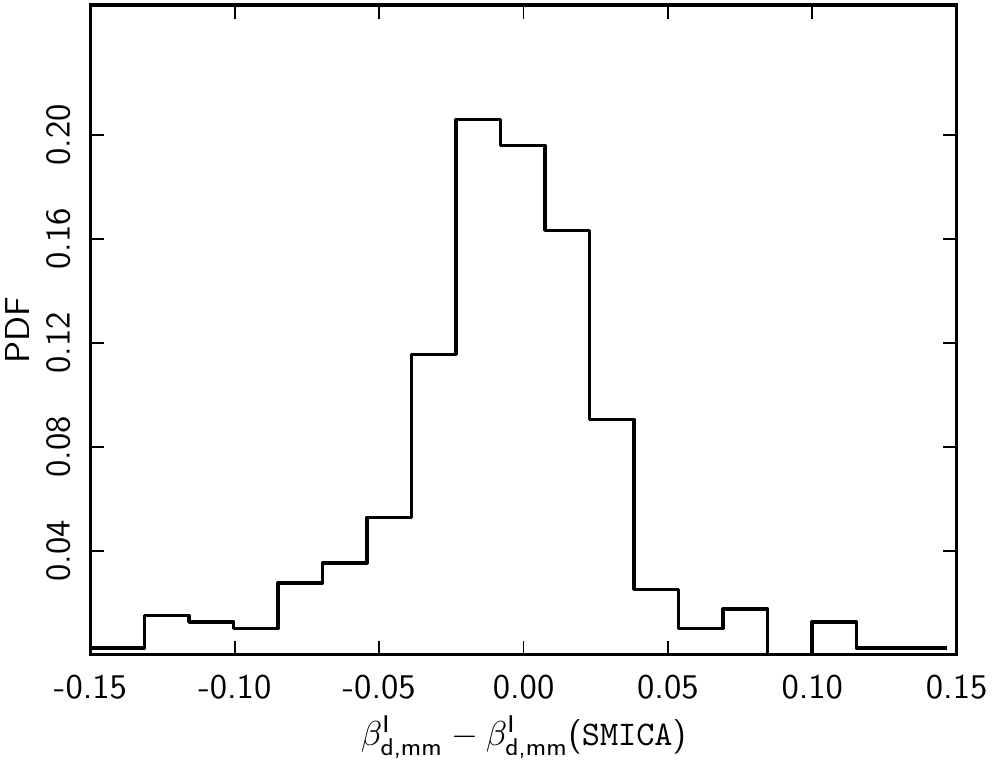} 
\caption{Histogram of the difference between the spectral indices \betadmmI\ from Eq.~(\ref{eq:5.5}) and \betadmmI\ (\smica) from Eq.~(\ref{eq:smica}), where we applied
the CC analysis to \planck\ maps corrected for CMB anisotropies with the \smica\ CMB  map.  }
\label{fig:6.3}
\end{figure}

To derive the dust spectral index from \RIT, we assume an MBB spectrum for the dust emission between 100 and $353$\,GHz (Sect.~\ref{sec:dust_spectral}). To validate this 
assumption, we repeat our CC analysis with \planck\ maps corrected for CMB anisotropies using the CMB map from the spectral matching independent component analysis (\smica, 
\citealt{planck2013-p06}). We infer \betadmmI\ (\smica) directly from the ratio between the 353 and $217$\,GHz CC coefficients without subtracting the $100$\,GHz CC coefficient, 
i.e.,
\begin{equation} 
R^{\rm I}_{\smica} (353,217) = \dfrac{[\alpha'_{353}]_{(353-\smica)}^{\rm 1T}  }{[\alpha'_{217}]_{(353-\smica)}^{\rm 1T} }  \label{eq:smica} \ ,
\end{equation}
where $\alpha'$ refers to the CC coefficients computed with maps corrected for CMB anisotropies using the \smica\ map.  The histogram of the difference between the 
two sets of spectral indices \betadmmI\ and \betadmmI\ (\smica) is presented in Fig.~\ref{fig:6.3}. The mean difference between the two estimates is zero.

\subsection{Comparison with other studies}

Our  determination of the spectral index \betadmmI\ of the dust emission in intensity at intermediate Galactic latitudes may be compared with the results from similar analyses of the 
\planck\ data. In \citet{planck2013-XVII}, the CC analysis has been applied to the \planck\ data at high Galactic latitudes ($b < -30$\deg) using an \hi\ map as a dust template free 
from CIB and CMB anisotropies. This is a suitable template to derive the spectral dependence of dust emission at high Galactic latitudes. The same methodology of colour ratios has 
been used in that work. The mean dust spectral index, $\betadmmI=1.53\pm0.03$, from  \citet{planck2013-XVII} agrees with the mean value we find in this paper. In an 
analysis of the diffuse emission in the Galactic plane, the spectral index of dust at millimetre wavelengths is found to increase from $\betadmmI=1.54$, for lines of sight where the 
medium is mostly atomic, to $\betadmmI=1.66$, where the medium is predominantly molecular \citep{planck2013-XIV}. The three studies indicate that the spectral index \betadmmI\ 
is remarkably similar over the diffuse ISM observed at high and intermediate Galactic latitudes, and in the Galactic plane.

\section{Spectral energy distribution of dust intensity}\label{sec:em_SED} 

In this section, we derive the mean SED of dust emission for intensity with its uncertainties from our CC analysis. The detailed spectral modelling of the dust SED is discussed in Sect.~\ref{sec:em_model}.

\subsection{Mean dust SED}\label{sec:em_mean_SED}

\begin{figure}[h!]
\begin{tabular}{c}
\includegraphics[width=8.8cm]{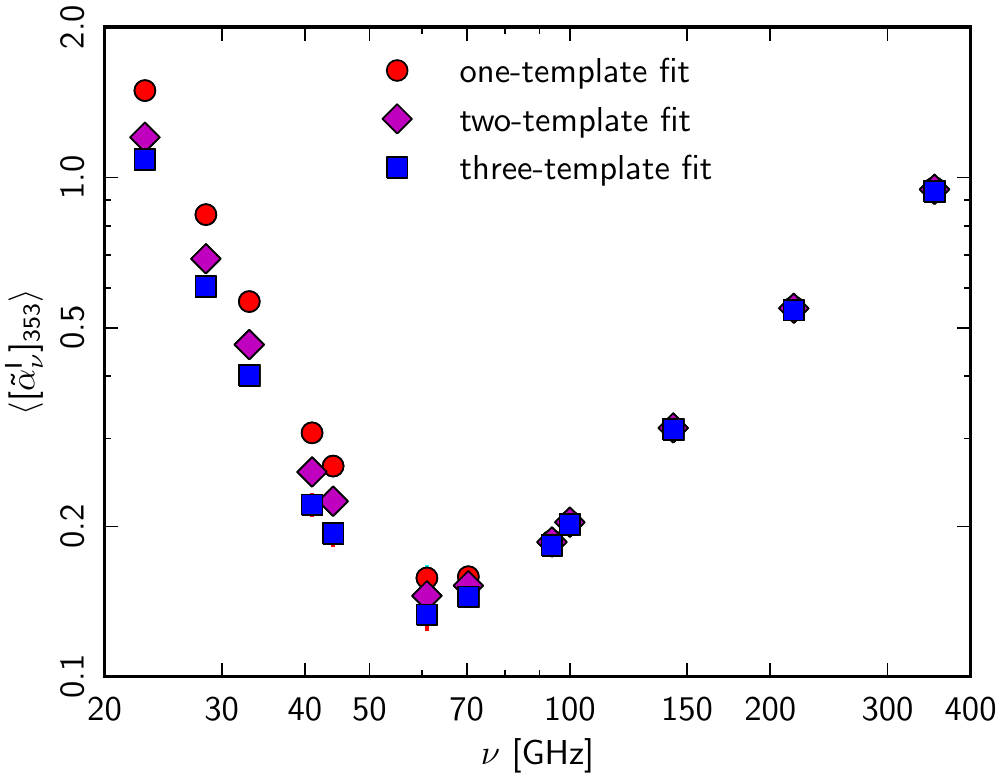}  \\
\includegraphics[width=8.8cm]{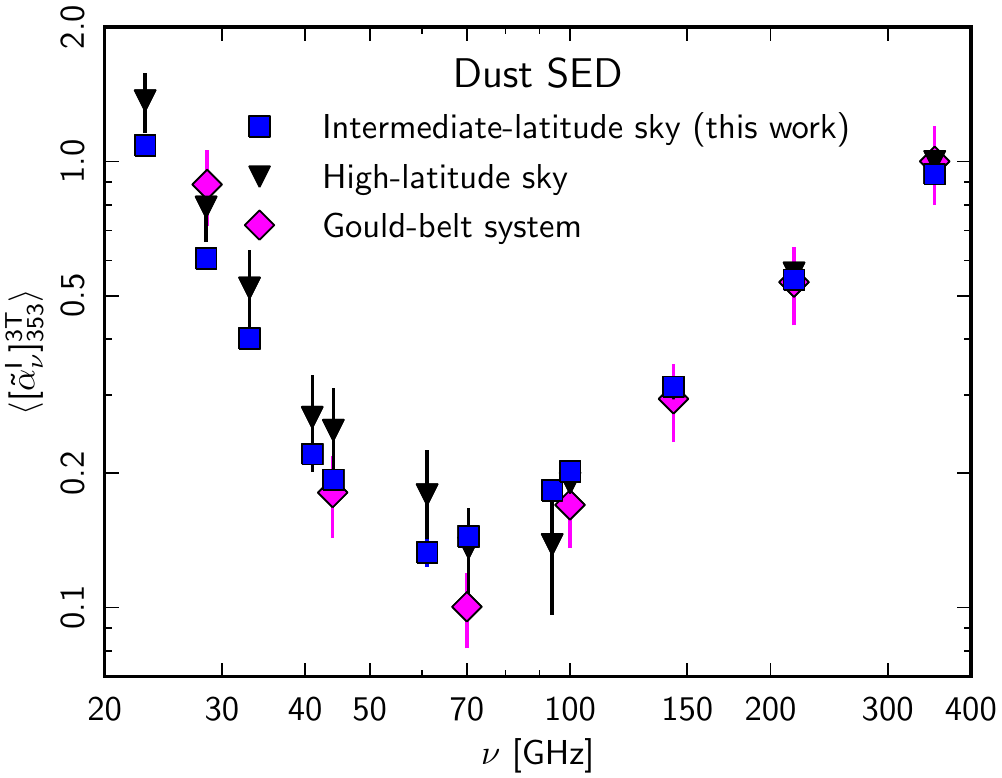}
\end{tabular}
\caption{\emph{Top}: The three SEDs in \krj\ units, normalized to 1 at $353$\,GHz, obtained by averaging, CMB corrected, CC coefficients 
from the fits with one (cyan circles), two (blue circles)  and three templates (red circles). The uncertainties at each frequency are estimated using the subsets of \planck\ and \wmap\ data. 
\emph{Bottom}: Our SED from the  three-template fit (red circles) is compared with the dust SED 
in  \citet{planck2013-XII} for the Gould Belt system (squares) and that 
at high latitude sky (inverted triangles) obtained applying the dust-\hi\ correlation analysis in \citet{planck2013-XVII}  to 
the full \planck\ mission data.}
\label{fig:7.1}
\end{figure}

We use the full mission \planck\ maps for the spectral modelling of the dust SED in intensity.
The mean SED is obtained by averaging the CC coefficients after CMB subtraction (see Sect.~\ref{sec:cmb_est}) over all sky patches. The mean SED is expressed in \krj\  units, normalized to 1 at 353 GHz. The three SEDs obtained from the fits with one, 
two, and three templates are shown in Fig.~\ref{fig:7.1}. The SED values obtained with the three templates fit are listed in Table~\ref{tab:11.1}. The three SEDs are identical at the 
highest frequencies.  They differ at $\nu < 100$\,GHz due to the non-zero correlation between the $353$\,GHz dust template with synchrotron and free-free emission. At 23\,GHz, 
after CMB correction the CC coefficient from the fit with three templates is lower by 9 and 35\,\% from those derived from the fits with two and one template, respectively.  
The 9\,\% difference accounts for the free-free emission correlated with dust and the 35\,\% difference for the combination of both synchrotron and free-free emission correlated with 
dust. At these low frequencies, the fit with three templates provides the best separation of the dust from the synchrotron and free-free emission. It is this SED that we call the dust 
SED hereafter.

Our dust SED is similar to that measured for the Gould Belt system \citep{planck2013-XII} and at high Galactic latitudes  \citep{planck2013-XVII}. It shows the thermal dust emission at $\nu \ge 70\,$GHz and is dominated by AME at lower frequencies. At \wmap\ frequencies, our dust SED is similar to that obtained using the FDS 94\,GHz map as a dust 
template in \citet{Davies:2006} and \citet{Ghosh:2012}. The mean dust SED derived using the three-template fit depends on the correction of the DDD \halpha\ map for dust 
scattering and extinction \citep{Dickinson:2003,Witt:2010, Brandt:2012,Bennett:2012}. In Appendix~\ref{sec:dust_scattering}, we study the impact of both assumptions on the mean 
dust SED, which are within a few percent at frequencies below 70\,GHz and have no impact on frequencies above 100\,GHz.

The total uncertainty on the dust SED includes the inter-calibration uncertainties on the data, the statistical uncertainties estimated from the variations of 
the CC coefficients across the sky patches, and the uncertainties due to the CMB subtraction, as discussed for \betadmmI\  in Sect.~\ref{sec:measuring_betaI}. The inter-calibration 
uncertainties ($c_{\nu}$) for \planck\ and \wmap\ data are given in Table~\ref{tab:11.1}. The statistical uncertainties are computed from the $1\sigma$ dispersion of the CC coefficients 
over the 400 sky patches divided by the square root 
of the number of independent sky patches (400/$N_{\rm visit}$). All three types of uncertainty, listed in Table~\ref{tab:11.1}, are added together in quadrature to compute the total 
uncertainty on the mean dust SED. They are shown in Fig.~\ref{fig:7.1}, but most do not appear because they are smaller than the size of the symbols.


\begin{table*}[tmb]
\begingroup
\newdimen\tblskip \tblskip=5pt
\caption{\label{tab:11.1} Mean microwave SEDs obtained from the fit with one, two, and three templates  using the CC analysis.}
\nointerlineskip
\vskip -3mm
\footnotesize
\tiny
\setbox\tablebox=\vbox{
   \newdimen\digitwidth 
   \setbox0=\hbox{\rm 0} 
   \digitwidth=\wd0 
   \catcode`*=\active 
   \def*{\kern\digitwidth}
   \newdimen\signwidth 
   \setbox0=\hbox{+} 
   \signwidth=\wd0 
   \catcode`!=\active 
   \def!{\kern\signwidth}
\halign{
\hbox to 0.8 in{#\leaderfil}\tabskip 1.7em&
\hfil #\hfil&
\hfil #\hfil&
\hfil #\hfil&
\hfil #\hfil&
\hfil #\hfil&
\hfil #\hfil&
\hfil #\hfil&
\hfil #\hfil&
\hfil #\hfil&
\hfil #\hfil&
\hfil #\hfil&
\hfil #\hfil\tabskip=0pt\cr
\noalign{\doubleline}
\omit&\multispan{12}\hfil Experiment\hfil\cr
\omit&\multispan{12}\hfil Frequency [GHz]\hfil\cr
\noalign{\vskip -3pt}
\omit&\multispan{12}\hrulefill\cr
\noalign{\vskip 2pt}
\omit Quantity\hfil& \wmap& \planck& \wmap& \wmap& \planck& \wmap& \planck& \wmap& \planck& \planck& \planck& \planck\cr
\omit& 23& 28.4& 33& 41& 44.1& 61& 70.4& 94& 100& 143& 217& 353\cr
\noalign{\vskip 4pt\hrule\vskip 3pt}
$\langle[\tildealphaInu]_{353}^{\rm 3T}\rangle$&  1.1202  & 0.5813  & 0.3955  & 0.2223  & 0.1857  & 0.1335  & 0.1361  & 0.1745  & 0.2108  & 0.3058  & 0.5837  & 1.0000\cr
$\sigma_{\rm stat}$& 0.0319  & 0.0175  & 0.0126  & 0.0073  & 0.0056  & 0.0038  & 0.0023  & 0.0038  & 0.0023  & 0.0029  & 0.0045  & 0.0073\cr
$c_{\nu}$ [\%]&  1.0  &  1.0  &  1.0  &  1.0  &  1.0  &  1.0  &  0.5  &  1.0  &  0.5  &  0.5  &  0.5  &  1.0\cr
$\sigma_{\rm cmb}$& 0.0086  & 0.0083  & 0.0085  & 0.0083  & 0.0081  & 0.0079  & 0.0037  & 0.0070  & 0.0035  & 0.0026  & 0.0015  & 0.0007\cr
 $\sigma_{\rm tot}$& 0.0346  & 0.0201  & 0.0156  & 0.0113  & 0.0100  & 0.0089  & 0.0044  & 0.0081  & 0.0043  & 0.0041  & 0.0055  & 0.0119\cr
$\sigma_{\rm tot}$&  32.3  &  29.0  &  25.3  &  19.7  &  18.6  &  15.1  &  31.0  &  21.6  &  49.3  &  74.3  & 106.9  &  83.9\cr
$C$& 1.0732  & 1.0000  & 1.0270  & 1.0480  & 1.0000  & 1.0450  & 0.9810  & 0.9927  & 1.0877  & 1.0191  & 1.1203  & 1.1114\cr
$U$& 0.9864  & 0.9487  & 0.9723  & 0.9577  & 0.9328  & 0.9091  & 0.8484  & 0.7998  & 0.7942  & 0.5921  & 0.3343  & 0.0751\cr
\noalign{\vskip 4pt\hrule\vskip 3pt}
$\langle[\tildealphaInu]_{353}^{\rm 1T}\rangle$& 1.5120  & 0.7952  & 0.5469  & 0.3046  & 0.2494  & 0.1556  & 0.1469  & 0.1755  & 0.2100  & 0.3037  & 0.5817  & 1.0000\cr
$\sigma_{\rm tot}^{\rm 1T}$ &0.0565  & 0.0304  & 0.0219  & 0.0139  & 0.0118  & 0.0089  & 0.0053  & 0.0077  & 0.0047  & 0.0042  & 0.0050  & 0.0113\cr
$\langle[\tildealphaInu]_{353}^{\rm 2T}\rangle$& 1.2274  & 0.6530  & 0.4515  & 0.2561  & 0.2135  & 0.1443  & 0.1419  & 0.1756  & 0.2108  & 0.3050  & 0.5823  & 1.0000\cr
 $\sigma_{\rm tot}^{\rm 2T}$& 0.0369  & 0.0218  & 0.0168  & 0.0116  & 0.0102  & 0.0084  & 0.0045  & 0.0075  & 0.0042  & 0.0039  & 0.0051  & 0.0115\cr
 \noalign{\vskip 5pt\hrule\vskip 3pt}}}
\endPlancktablewide
\ \ \ $\langle[\tilde{\alpha}_{\nu}^{\StokesI}]_{353}^{\rm 3T}\rangle\equiv$ mean dust SED in  \krj\  units, normalized to 1 at $353$\,GHz, 
from the fit  with three templates.  The values are not colour corrected.\par
\ \ \ $\sigma_{\rm stat}\equiv$ statistical uncertainty on the mean dust SED.\par
\ \ \ $c_{\nu}\equiv$ uncertainties on the inter-calibration [\%] between \planck\ and \wmap\ frequencies \citep{planck2013-p01,Bennett:2012}.  \par
\ \ \ $\sigma_{\rm cmb}\equiv$ uncertainty on the mean dust SED introduced by the CMB subtraction multiplied by the inter-calibration factor $c_{\nu}$. \par
\ \ \ $\sigma_{\rm tot}\equiv$ total uncertainty  on the mean dust SED. \par
\ \ \ S/N $\equiv$ signal-to-noise ratio on the mean dust SED. \par
\ \ \  $C\equiv$ colour-correction factors computed with a linear combination of the power-law model and the MBB parameters listed in Table~\ref{tab:7.1}. \par
\ \ \ $U\equiv$ unit conversion factors from thermodynamic (\kcmb) to Rayleigh-Jeans (\krj) temperature. \par
\ \ \ $\langle[\tilde{\alpha}_{\nu}^{\StokesI}]_{353}^{\rm 1T}\rangle\equiv$ mean intensity SED in  \krj\ units, normalized to 1 at $353$\,GHz, derived from the correlation of the maps with the one-template fit.  The values are not colour corrected.\par
\ \ \ $\sigma_{\rm tot}^{\rm 1T} \equiv$ total uncertainty  on the mean intensity SED with the one-template fit. \par
\ \ \ $\langle[\tilde{\alpha}_{\nu}^{\StokesI}]_{353}^{\rm 2T}\rangle\equiv$ mean intensity SED in  \krj\ units, normalized to 1 at $353$\,GHz, derived from the correlation of the maps with the two-template fit.  The values are not colour corrected.\par
\ \ \ $\sigma_{\rm tot}^{\rm 2T} \equiv$ total uncertainty  on the mean intensity SED with the two-template fit. \par
\endgroup
\end{table*}

\subsection{Parametric modelling for intensity dust SED}\label{sec:em_model}

A spectral fit of the dust SED is required to separate the thermal dust emission from the AME.  We continue to use an MBB spectrum for the thermal dust emission, and consider two models with different spectra for the AME. 

\begin{figure*}
\begin{tabular}{cc}
\includegraphics[width=8.7cm]{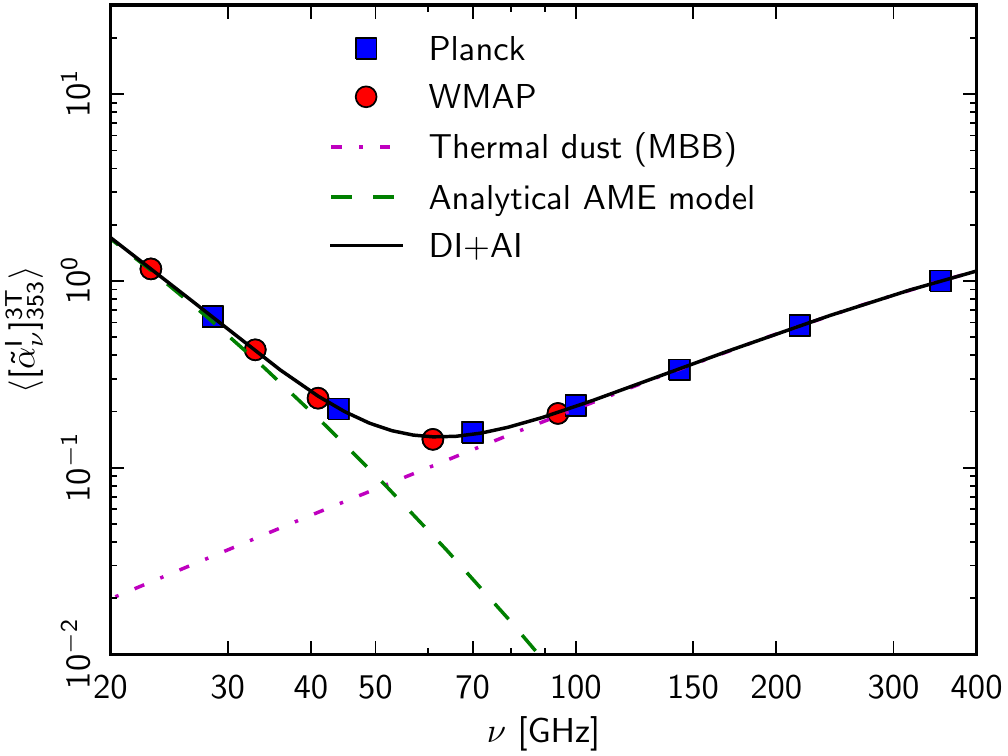}&
\includegraphics[width=8.7cm]{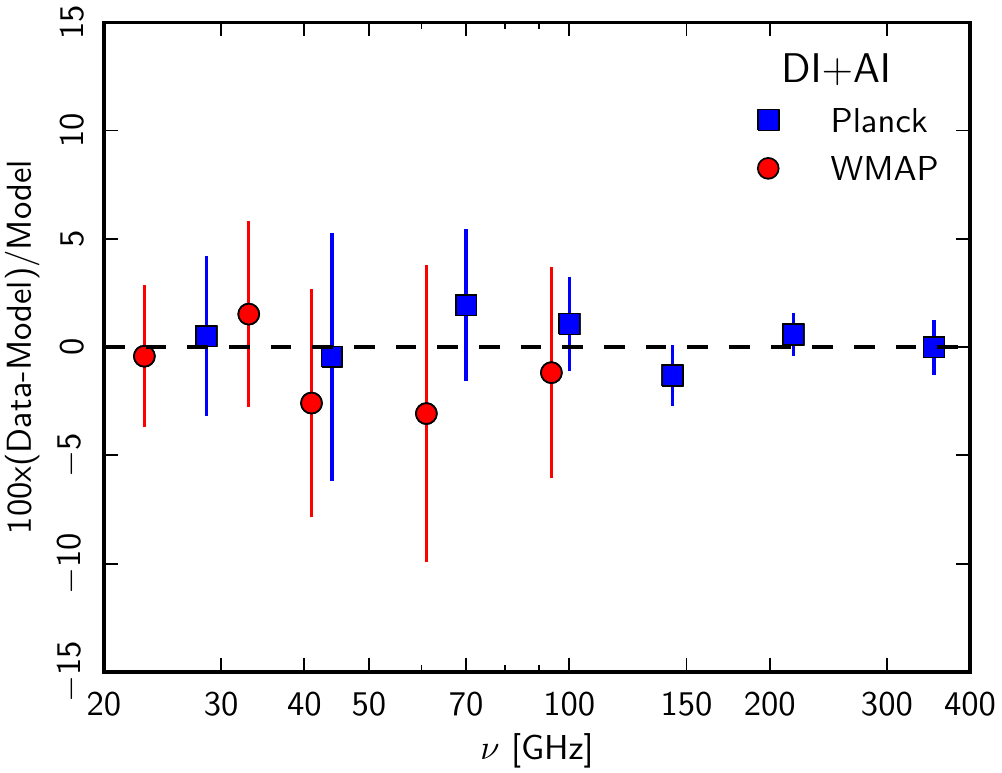} \\
\includegraphics[width=8.7cm]{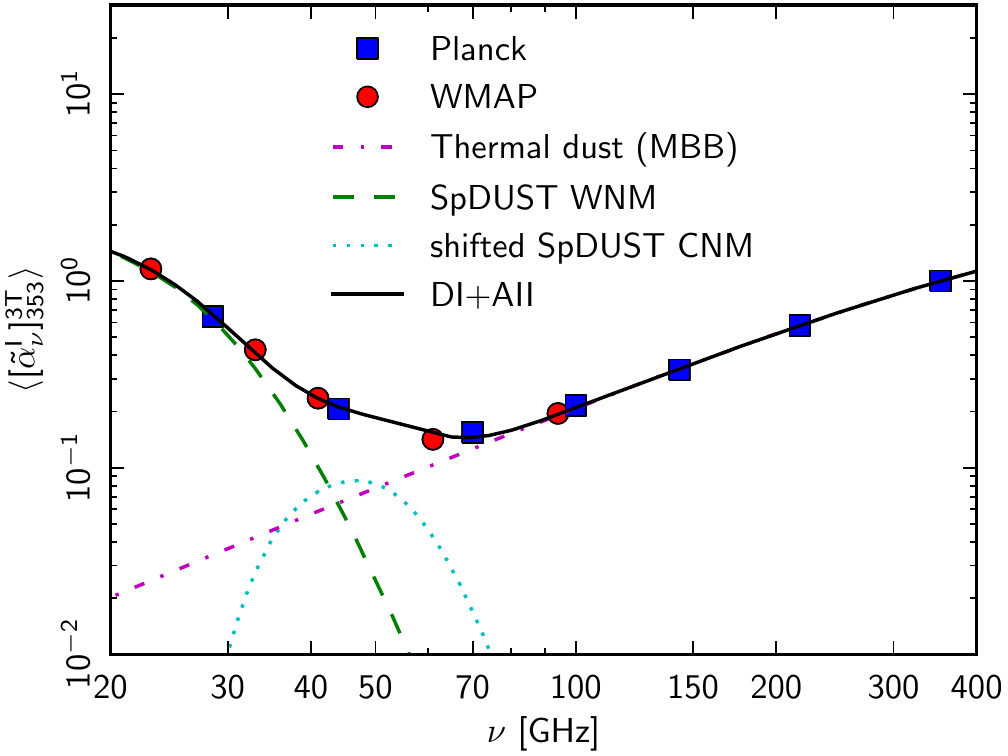}&
\includegraphics[width=8.7cm]{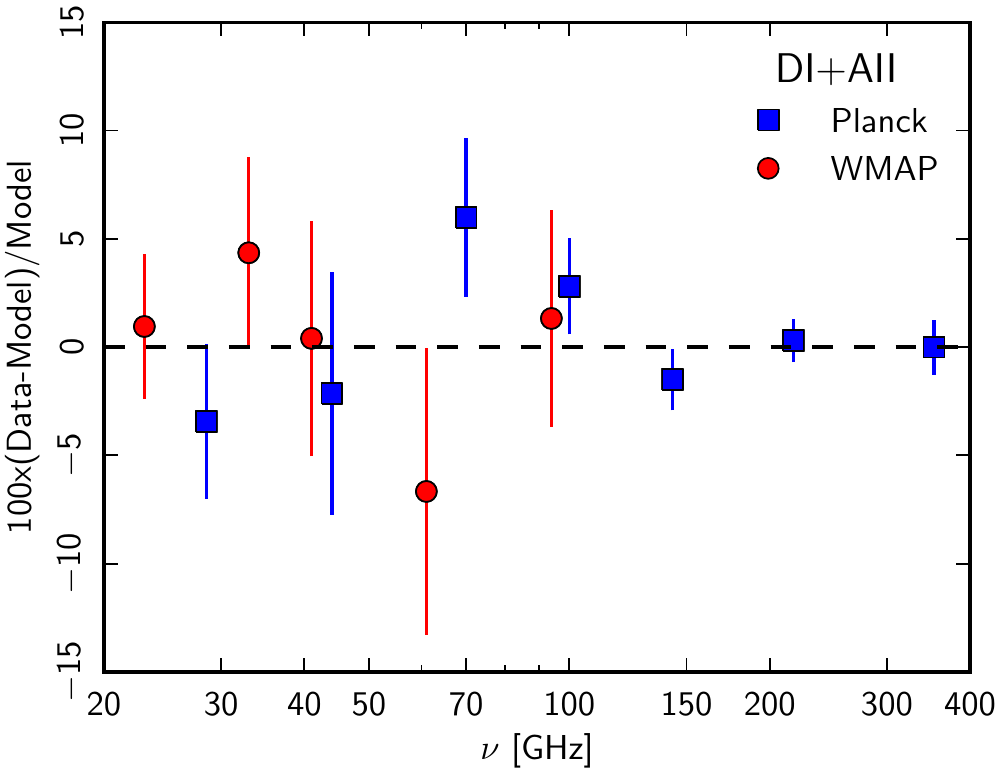} 
\end{tabular}
\caption{Mean dust SED in \krj\ units, normalized to 1 at $353$\,GHz, with different spectral fits and the respective residuals. The two parametric model fits are DI+AI (\emph{top left}), and DI+AII (\emph{bottom left}), as presented in Sect.~\ref{sec:em_model}. \emph{Right}: residuals after removing the best fit model listed in Table~\ref{tab:7.1}, from the mean dust SED. The two spectral models provide good fit to the data, with residuals compatible with zero.}
\label{fig:7.2}
\end{figure*}

\begin{itemize}

\item Model DI+AI: In this first approach, we use the analytical model of the AME (AI) introduced by \citet{bonaldi:2007}, which is a parabola in the log $I_{\nu}$ -- log $\nu$ plane, parameterized by the peak frequency ($\nu_{\rm p}$) and the slope $-m_{60}$ at 60\,GHz. The AME model ($M_{\rm a}$) normalized at 23\,GHz (in \krj\ units) is given by:
\begin{align}
& \log~M_{\rm a} = - \left(\frac{m_{60}\log~\nu_{\rm p}}{\log(\nu_{\rm p}/60\,{\rm GHz})} + 2 \right)\log\left(\frac{\nu}{23\,{\rm GHz}}\right) \nonumber \\
& + \frac{m_{60}}{2\log(\nu_{\rm p}/60\,{\rm GHz})} \left [ (\log(\nu/1\,{\rm GHz}))^2 - (\log~23)^2 \right ]\ .
\end{align}
This model is a good fit to spectra of dipole emission from small spinning dust particles computed with the \spdust\ code \citep{Ali-Haimoud:2009,Silsbee:2011}.  The second 
component is the MBB spectrum of the thermal dust emission (DI) with free parameter \betadmmI. The total model is written in \krj\ units 
normalized to 1 at the frequency $\nuref=353\,{\rm GHz}$:
\begin{equation}
\langle[\tildealphaInu]_{353}^{\rm 3T}\rangle = \ \AaI\ M_{\rm a} +\  \left(\frac{\nu}{\nuref}\right)^{\betadmmI - 2} \frac{B_{\nu} (\Td)}{B_{\nuref} (\Td)} \label{eq:7.2} \ , 
\end{equation}
where $\AaI$ is the amplitude of AME and \betadmmI\ is the spectral index of thermal dust emission at microwave frequencies.  We fix 
$\Td=19.6$\,K from Sect.~\ref{sec:Td}. The four free parameters of the model are \AaI, $m_{60}$, $\nu_{\rm p}$ and \betadmmI. \\

\item Model DI+AII: In this second approach, the AME component (AII) is a linear combination of two spinning dust components arising from the typical cold neutral medium (CNM) and 
warm neutral medium (WNM). In our analysis, we use the predicted \spdust\ (v2) spectra \citep{Ali-Haimoud:2009,Silsbee:2011} of the CNM and WNM spinning dust components. 
Following the work of \citet{Hoang:2011} and \citet{Ghosh:2012}, we shift both the WNM and CNM spectra in frequency space to fit the observed dust SED. The same DI model of 
the thermal dust emission is considered for this model. In this case, the spectral model is given in \krj\ units, normalized to 1 at the frequency $\nuref$, by:
\begin{align}
\langle[\tildealphaInu]_{353}^{\rm 3T}\rangle& = \ A_{\rm WNM}^{\rm I} \ D_{\rm WNM} (\nu - \Delta \nu_{\rm WNM}) \nonumber \\
& + \ A_{\rm CNM}^{\rm I} \ D_{\rm CNM} (\nu - \Delta \nu_{\rm CNM}) \nonumber \\
& + \ \left(\frac{\nu}{\nuref}\right)^{\betadmmI - 2} \frac{B_{\nu} (\Td)}{B_{\nuref} (\Td)}  \label{eq:7.3} \ ,
\end{align}
where $A_{\rm WNM}^{\rm I}$ is the amplitude of WNM spectrum normalized at 23\,GHz, $A_{\rm CNM}^{\rm I}$ is the amplitude of CNM spectrum normalized at 41\,GHz, 
$D_{\rm WNM}$ is the \spdust\ WNM spectrum, $D_{\rm CNM}$ is the \spdust\ CNM spectrum, $\Delta \nu_{\rm WNM}$ is the shift in the WNM spectrum, $\Delta \nu_{\rm CNM}$ is 
the shift in the CNM spectrum,  $\nuref=353\,{\rm GHz}$ is the reference frequency, \betadmmI\ is the spectral index of the thermal dust emission.  We fix $\betadmmI=1.51$ from Sect.~\ref{sec:err_betaI} 
and $\Td=19.6$\,K from Sect.~\ref{sec:Td}. The four free parameters of the model are the WNM amplitude, the WNM frequency shift, the CNM amplitude and the CNM frequency shift.

\end{itemize}

The fits of the dust SED with models DI+AI and DI+AII are shown in Fig.~\ref{fig:7.2}.  The best fit model parameters are listed in Table~\ref{tab:7.1}.

\begin{table}[tmb]
\begingroup
\newdimen\tblskip \tblskip=5pt
\caption{\label{tab:7.1} Results of the spectral fits to the mean dust SED in intensity using  \planck\ and \wmap\ maps. }
\nointerlineskip
\vskip -3mm
\footnotesize
\setbox\tablebox=\vbox{
   \newdimen\digitwidth 
   \setbox0=\hbox{\rm 0} 
   \digitwidth=\wd0 
   \catcode`*=\active 
   \def*{\kern\digitwidth}
   \newdimen\signwidth 
   \setbox0=\hbox{+} 
   \signwidth=\wd0 
   \catcode`!=\active 
   \def!{\kern\signwidth}
\halign{
\hbox to 1.5cm{#\leaderfil}\tabskip 2.4em&
\hfil #\hfil\tabskip=3em&
\hbox to 1.5cm{#\leaderfil}\tabskip 2.4em&
\hfil #\hfil\tabskip=0pt\cr
\noalign{\doubleline}
\omit Parameters\hfil& DI+AI& \omit Parameters\hfil& DI+AII\cr
\noalign{\vskip 4pt\hrule\vskip 6pt}
 $\betadmmI$                   &  $1.52\pm0.01$&  $A_{\rm WNM}^{\rm I}$& $1.12\pm0.04$\cr
 $\AaI$      & $1.14\pm0.04$& $\Delta \nu_{\rm WNM}$ [GHz]& $-1.7\pm0.8$ \cr
 $\nu_{\rm p}$ [GHz]     & $9.5\pm6.9$& $A_{\rm CNM}^{\rm I}$& $0.07\pm0.01$ \cr
 $m_{60}$& $1.81\pm0.38$& $\Delta \nu_{\rm CNM}$ [GHz]& $22.2\pm1.6$\cr
 $\chi^2/N_{\rm dof}     $& 2.4/8        & $\chi^2/N_{\rm dof}$& 8.7/8\cr
\noalign{\vskip 5pt\hrule\vskip 3pt}}}
\endPlancktablewide
The parameters listed in this table are described in Eqs.~\eqref{eq:7.2} and \ref{eq:7.3} where the dust SED is expressed in \krj\ units and normalized to 1 at $353$\,GHz. The fixed model parameter is $\Td\ =19.6$\,K for two intensity models. \par
\endgroup
\end{table}

\section{Dust spectral index for polarization}\label{sec:betaP}

We now move to the analysis of the polarization data. Like in Sect.~\ref{sec:betaI}, we estimate the polarized dust spectral index (\betadmmP) at microwave frequencies ($\nu \le 353$\,GHz). We present the results of our data analysis and tests of its robustness against systematic uncertainties. 

\subsection{Measuring \betadmmP\ }\label{sec:mean_betap}

\begin{figure}[h!]
\begin{tabular}{c}
\includegraphics[width=8.8cm]{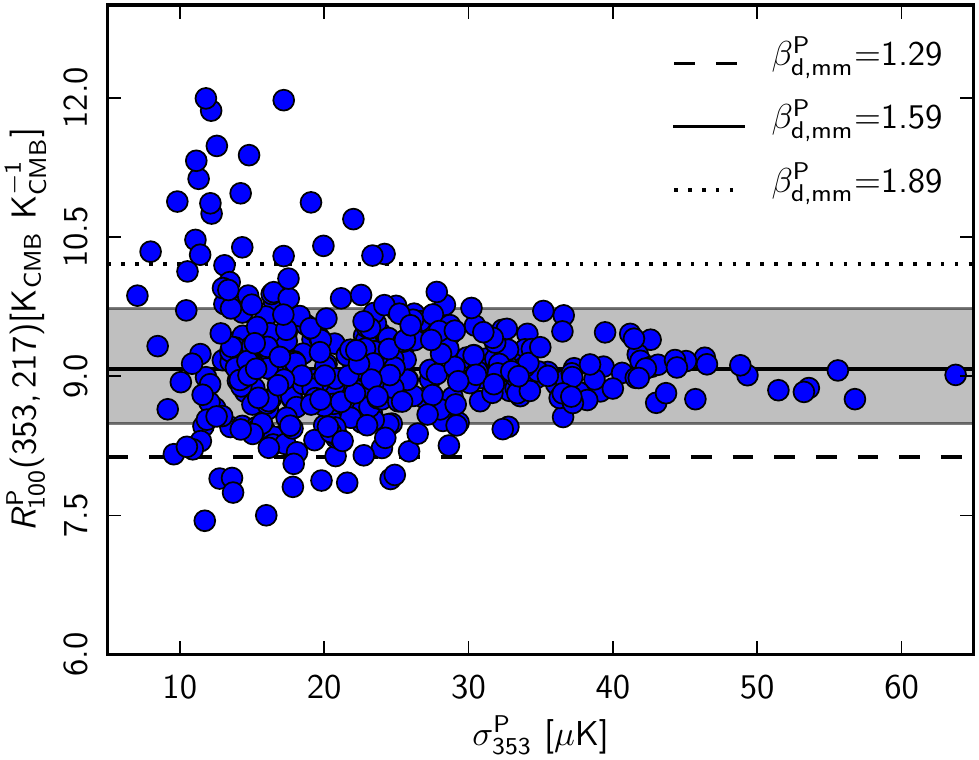} \\
\includegraphics[width=8.8cm]{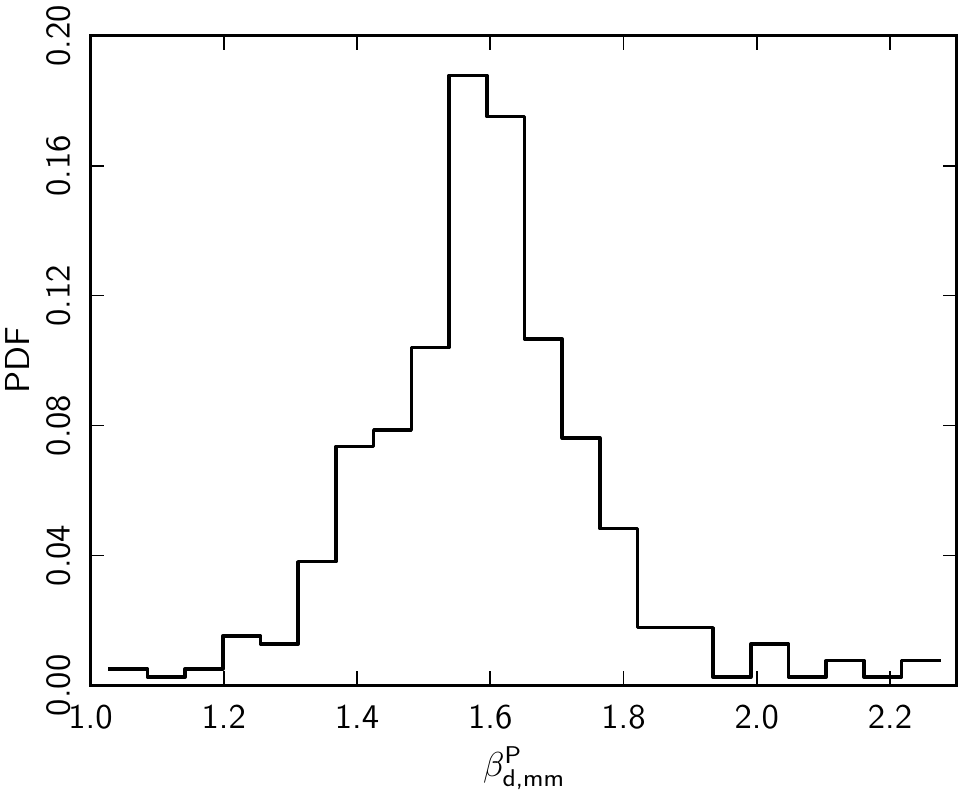} 
\end{tabular}
\caption{\emph{Top}: Colour ratio \RIP\ against the local dispersion of the polarization template at $353$\,GHz. \emph{Bottom}: Histogram of the \betadmmP\ values 
inferred from \RIP\ for all sky patches. The mean value of the spectral index for polarization is $1.59\pm0.02$, which is different from that for intensity $1.51\pm0.01$ (Fig.~\ref{fig:6.1}). } 
\label{fig:8.1}
\end{figure}

Here we use the full mission \planck\ polarization maps, keeping the polarized detector 
set maps (DS1 and DS2) at $353$\,GHz as fixed templates, to derive a mean \betadmmP\ using the one-template fit.
Using the polarization CC coefficients, $[\alphaPnu]_{353}^{\rm 1T}$, we compute \RIP\ for each sky patch using Eq.~\eqref{eq:5.2}. Fig.~\ref{fig:8.1} shows the values of \RIP\ versus 
the local dispersion of the polarized map at $353$\,GHz (\sigmaP) for all the sky patches. To compute \sigmaP, we use a 1\deg\ smoothed map of $\polint_{353}$ derived in 
\citet{planck2014-XIX}.  We derive \betadmmP\ for each sky patch from \RIP, taking into account the local estimate of \Td\ derived from \RIIT\ (Sect.~\ref{sec:Td}). We assume that the 
temperature of the dust grains contributing to the polarization is the same as that determined for the dust emission in intensity. This is not necessarily true if the polarization is 
associated with specific dust grains, e.g., the silicates versus carbon dust \citep{Martin:2007,Draine_Fraisse:2009}. This should be kept in mind in thinking of physical interpretations. 
Here we use the spectral indices \betadmmP\ and \betadmmI\ as a mathematical way to quantify the difference between the dust SED for intensity and polarization.

The scatter on the \RIP\ values increases  for $\sigmaP<$\,20\,$\mu$K due to data noise. The histogram of \betadmmP\ from the 400 sky 
patches is presented in Fig.~\ref{fig:8.1}. The distribution of \betadmmP\ has a mean value of 1.592 (round-off to 1.59), with a $1\sigma$ dispersion of 0.174 (round-off to 0.17). This dispersion is the same if we use
the mean dust temperature of $19.6\,$K for all sky patches. The 
statistical uncertainty on the mean \betadmmP\ is computed from the $1\sigma$ dispersion divided by the square root of the number of independent sky patches 
($400/N_{\rm visit}$) used, which is 0.02.

The mean value of the dust spectral index for polarization is different from that for intensity, $1.51\pm0.01$ (Sect.~\ref{sec:betaI}) over the same sky area. 
In the next section, we check whether the difference of spectral indices in intensity and polarization is a robust result against systematics present in the polarization data.

\begin{table}[tmb]
\begingroup
\newdimen\tblskip \tblskip=5pt
\caption{\label{tab:8.1} Polarized dust spectral indices derived using multiple subsets and templates of the \planck\ data. The full mission \planck\ polarization data along with the DS1 and DS2 templates (first entry in the Table below) is used in Sect.~\ref{sec:mean_betap} to produce Fig.~\ref{fig:8.1}. The scatter of the 20 measurements is consistent with the $1\sigma$ statistical uncertainty on the mean value of \betadmmP.} 
\nointerlineskip
\vskip -3mm
\setbox\tablebox=\vbox{
   \newdimen\digitwidth 
   \setbox0=\hbox{\rm 0} 
   \digitwidth=\wd0 
   \catcode`*=\active 
   \def*{\kern\digitwidth}
   \newdimen\signwidth 
   \setbox0=\hbox{+} 
   \signwidth=\wd0 
   \catcode`!=\active 
   \def!{\kern\signwidth}
\halign{
\hbox to 2.5cm{#\leaderfil}\tabskip 2.0em&
#\hfil&
\hfil #\hfil\tabskip=0pt\cr
\noalign{\doubleline}
\omit Templates\hfil& Data sets& \betadmmP \cr
\noalign{\vskip 4pt\hrule\vskip 6pt}
\noalign{\vskip 5pt}
&  Full & 1.592\cr
& HR1& 1.595\cr
 DS1 and DS2& HR2& 1.595\cr
& YR1 & 1.619\cr
& YR2 & 1.592\cr
\noalign{\vskip 5pt}
\noalign{\hrule}
\noalign{\vskip 5pt}
& Full& 1.602 \cr
&  HR1& 1.603 \cr
YR1 and YR2 & HR2 & 1.606 \cr
& DS1& 1.564 \cr
& DS2& 1.627 \cr
\noalign{\vskip 5pt}
\noalign{\hrule}
\noalign{\vskip 5pt}
& Full& 1.613 \cr
& DS1& 1.579 \cr
HR1 and HR2& DS2& 1.639 \cr
& YR1 & 1.639 \cr
& YR2 & 1.614 \cr
\noalign{\vskip 5pt}
\noalign{\hrule}
\noalign{\vskip 5pt}
& Full& 1.578 \cr
& DS1& 1.560 \cr
S1+S3 and S2+S4& DS2& 1.590 \cr
& HR1 & 1.581 \cr
& HR2 & 1.588 \cr
\noalign{\vskip 5pt\hrule\vskip 3pt}}}
\endPlancktable
\endgroup
\end{table}

\subsection{Uncertainties in \betadmmP\ }\label{sec:err_betaP}

For the mean polarized dust spectral index, we use the results from full mission \planck\ polarization maps with the two detector set maps as fixed templates (Sect.~\ref{sec:mean_betap}).
To estimate the systematic uncertainty for the mean \betadmmP, we apply the CC analysis on multiple subsets of the \planck\ data, 
including the combination of yearly maps (YR1 and YR2), 
the full mission half-ring maps (HR1 and HR2), the combination of odd surveys (S1+S3) and even surveys (S2+S4), and the detector set maps (DS1 and DS2) (see Sect.~\ref{sec:systematics} for more details).  We use 
these subsets of the data as maps and templates at $353$\,GHz.  Table~\ref{tab:8.1} lists the derived mean \betadmmP, for all the sky patches  from each 
combination of the data subsets. 
The dispersion of the  \betadmmP\ values in Table~\ref{tab:8.1}, 0.02,  is consistent with
the $1\sigma$ dispersion on the mean polarization spectra index from statistical uncertainties estimated in Sect.~\ref{sec:mean_betap}, making it difficult to separate the contributions from the statistical noise and the data systematics. Therefore, we use the $1\sigma$ dispersion from the subsets of the \planck\ data, as listed in Table~\ref{tab:8.1}, as a combine statistical and systematic uncertainties on the mean value of \betadmmP. Thus, we find $\betadmmP=1.59\pm0.02~(\text{stat.+ syst.})$. The small difference, 0.08,  between  \betadmmP\ and \betadmmI\  has a $3.6 \, \sigma$ significance, taking into account the total uncertainty on both \betadmmP\ and \betadmmI.

\section{Spectral energy distribution of dust polarization}\label{sec:pol_SED}

We now derive the mean SED for the dust polarization and extend to polarization the parametric modelling already made on the dust SED in intensity (Sect.~\ref{sec:em_model}).

\subsection{Mean polarized SED}\label{sec:pol_mean_SED}

We use the full mission \planck\ maps for the spectral modelling of the dust SED in polarization.
Like in Sect.~\ref{sec:em_mean_SED} for dust emission in intensity, the mean SED for polarization is obtained by averaging the polarization CC coefficients 
after CMB subtraction (see Sect.~\ref{sec:cmb_est}) over all sky patches and is expressed in \krj\ units. The polarization SED is derived from the one-template fit, keeping the templates fixed to the polarized detector 
set maps (DS1 and DS2) at $353$\,GHz (see Sect.~\ref{sec:cc_pol} for more details). 
We compute the mean polarization SED and its uncertainties in a similar manner to that discussed in Sect.~\ref{sec:em_SED}. The mean polarized SED and associated uncertainties are listed in 
Table~\ref{tab:11.2}, and is shown in Fig.~\ref{fig:9.1}.

The polarization SED first decreases with decreasing frequency, then turns up below $60$\,GHz. This is the first time that such a behavior has been observed for polarized emission 
correlated with dust polarization, though it has been seen before for the total sky polarization \citep{Bennett:2012}.


\begin{table*}[tmb]
\begingroup
\newdimen\tblskip \tblskip=5pt
\caption{\label{tab:11.2} Mean microwave SED for polarization computed using the CC analysis.}
\nointerlineskip
\vskip -3mm
\tiny
\setbox\tablebox=\vbox{
   \newdimen\digitwidth 
   \setbox0=\hbox{\rm 0} 
   \digitwidth=\wd0 
   \catcode`*=\active 
   \def*{\kern\digitwidth}
   \newdimen\signwidth 
   \setbox0=\hbox{+} 
   \signwidth=\wd0 
   \catcode`!=\active 
   \def!{\kern\signwidth}
\halign{
\hbox to 0.8 in{#\leaderfil}\tabskip 1.7em&
\hfil #\hfil&
\hfil #\hfil&
\hfil #\hfil&
\hfil #\hfil&
\hfil #\hfil&
\hfil #\hfil&
\hfil #\hfil&
\hfil #\hfil&
\hfil #\hfil&
\hfil #\hfil&
\hfil #\hfil&
\hfil #\hfil\tabskip=0pt\cr
\noalign{\doubleline}
\omit&\multispan{12}\hfil Frequency [GHz]\hfil\cr
\omit&\multispan{12}\hfil Experiment\hfil\cr
\noalign{\vskip -3pt}
\omit&\multispan{12}\hrulefill\cr
\noalign{\vskip 2pt}
\omit Quantity\hfil& \wmap& \planck& \wmap& \wmap& \planck& \wmap& \planck& \wmap& \planck& \planck& \planck& \planck\cr
\omit& 23& 28.4& 33& 41& 44.1& 61& 70.4& 94& 100& 143& 217& 353\cr
\noalign{\vskip 4pt\hrule\vskip 3pt}
$\langle[\tildealphaPnu]_{353}^{\rm 1T}\rangle$& 0.9481  & 0.4038  & 0.3351  & 0.1793  & 0.1525  & 0.1179  & 0.1129  & 0.1852  & 0.1900  & 0.3029  & 0.5624  & 1.0000\cr
$\sigma_{\rm stat}$&  0.1201  & 0.0538  & 0.0402  & 0.0292  & 0.0190  & 0.0198  & 0.0118  & 0.0261  & 0.0050  & 0.0048  & 0.0062  & 0.0068\cr
$c_{\nu}$ [\%]&   1.0  &  1.0  &  1.0  &  1.0  &  1.0  &  1.0  &  0.5  &  1.0  &  0.5  &  0.5  &  0.5  &  1.0\cr
$\sigma_{\rm cmb}$& 0.0006  & 0.0006  & 0.0006  & 0.0006  & 0.0006  & 0.0005  & 0.0003  & 0.0005  & 0.0002  & 0.0002  & 0.0001  & 0.0000\cr
 $\sigma_{\rm tot}$& 0.1204  & 0.0539  & 0.0403  & 0.0293  & 0.0190  & 0.0199  & 0.0118  & 0.0262  & 0.0051  & 0.0050  & 0.0067  & 0.0114\cr
S/N&   7.9  &   7.5  &   8.3  &   6.1  &   8.0  &   5.9  &   9.6  &   7.1  &  37.1  &  60.4  &  83.6  &  87.7\cr
\noalign{\vskip 5pt\hrule\vskip 3pt}}}
\endPlancktablewide
\ \ \ $\langle[\tilde{\alpha}_{\nu}^{P}]_{353}^{\rm 1T}\rangle\equiv$ \vtop{\hsize=165mm\noindent Mean polarization SED in \krj\  units, normalized to 1 at $353$\,GHz, from the correlation with the $353$\,GHz templates.  The values are not colour corrected.}\par
\ \ \ $\sigma_{\rm stat}\equiv$ Statistical uncertainty on the mean polarization SED.\par
\ \ \ $c_{\nu}\equiv$ Uncertainties on the inter-calibration [\%] between \planck\ and \wmap\ frequencies \citep{planck2013-p01,Bennett:2012}.  \par
\ \ \ $\sigma_{\rm cmb}\equiv$ Uncertainty on the mean polarized SED introduced by the CMB-subtraction multiplied by the inter-calibration factor $c_{\nu}$.\par
\ \ \ $\sigma_{\rm tot}\equiv$ Total uncertainty on the mean polarized SED.\par
\endgroup
\end{table*}

\subsection{Low frequency rise of the polarization SED}\label{sec:low_freq}

In this section, we show that a synchrotron component correlated with dust is the most likely interpretation for the low frequency rise of the polarization SED .
 
\subsubsection{Synchrotron polarization correlated with dust}\label{sec:pol_synch}

The polarized dust and synchrotron emissions may be written as
\begin{align}
[\StokesQ^{\rm s}_{\nu},\StokesU^{\rm s}_{\nu}]&= \ \polfrac_{\rm s} \ \StokesI^{\rm s}_{\nu} \ [\cos 2\polang_{\rm s},\sin 2\polang_{\rm s}] \ , \\
[\StokesQ^{\rm d}_{353},\StokesU^{\rm d}_{353}]& = \ \polfrac_{\rm d} \ \StokesI^{\rm d}_{353} \ [\cos 2\polang_{\rm d},\sin 2\polang_{\rm d}] \ ,
\end{align}
where $\polfrac_{\rm s}$ and $\polfrac_{\rm d}$ are the polarization fractions, and $\polang_{\rm s}$ and $\polang_{\rm d}$ are the polarization angles, for synchrotron and dust, respectively. After correlation with the $353$\,GHz \StokesI, \StokesQ\ and \StokesU\ templates, we have
\begin{equation}
\langle \alphaPnu (s_{353}) \rangle \ \le \frac{\polfrac_{\rm s}}{\polfrac_{\rm d}} \ \langle \alphaInu (s_{353})\rangle \label{eq:8.2} \ ,
\end{equation}
where $\langle \alphaInu (s_{353})\rangle$ and $\langle \alphaPnu (s_{353})\rangle$ are the mean SEDs of the synchrotron emission correlated with dust in intensity and polarization. 
The upper limit in Eq.~\eqref{eq:8.2} is obtained when the synchrotron and dust polarization angles are identical, which is not what is observed comparing the \planck\ $353$\,GHz and 
\wmap\ $23$\,GHz polarization data \citep{planck2014-XIX}. Both emission processes trace the same large-scale Galactic magnetic field (GMF), but they give different weights to 
different parts of the line of sight. The CC analysis only keeps the synchrotron emission that arises from the same volume of interstellar space as the dust emission. For example, it is 
expected to filter out the synchrotron emission from the Galactic halo, where there is little dust. Thus, to validate our interpretation of the low frequency rise of the polarization SED 
with synchrotron and no AME polarization, we need to show that the upper limit in Eq.~\eqref{eq:8.2} holds.  

\begin{figure}[h!]
\includegraphics[width=8.8cm]{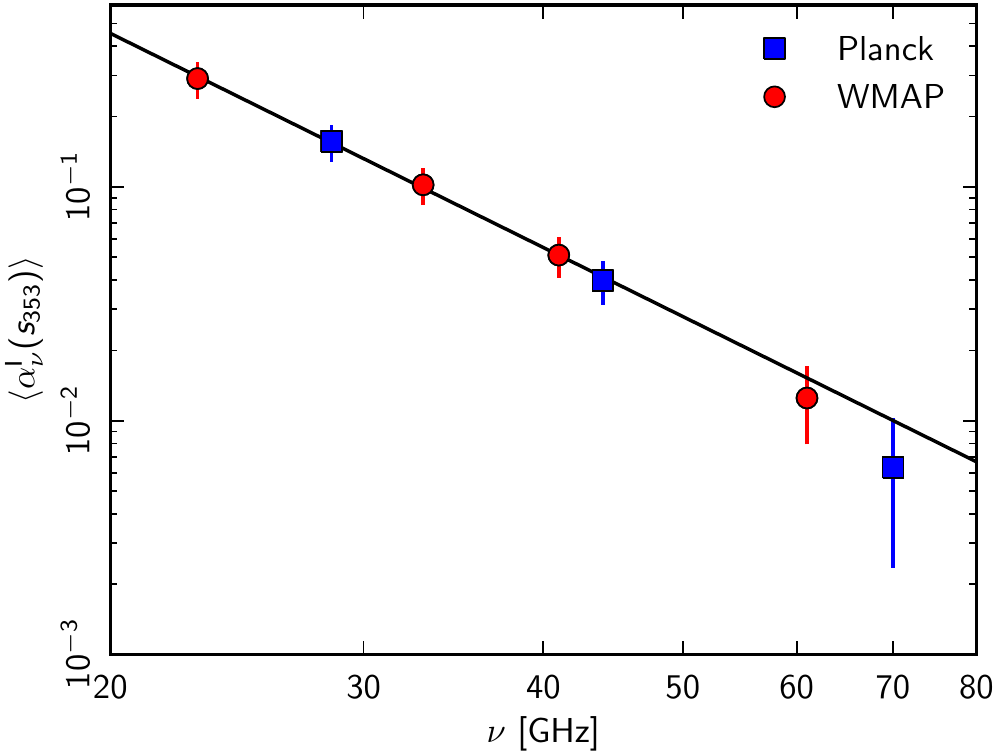} 
\caption{Spectral energy  distribution of the dust-correlated synchrotron emission in intensity. The SED is the ratio between the synchrotron emission at frequency $\nu$ and the dust 
emission at $353$\,GHz emission in units of \krj. The mean spectral index for the intensity is $\beta_{\rm s}^{\rm I}=-3.04\pm0.36$. }
\label{fig:9.2}
\end{figure}

The intrinsic polarization of synchrotron emission is about 75\,\% for typical relativistic electron spectra \citep{Rybicki:1979,Longair:1994}, whereas the analysis of \planck\ 
polarization maps indicates that the intrinsic polarization of dust at $353$\,GHz can reach about 20\,\%  \citep{planck2014-XIX,planck2014-XX}.
To compute the synchrotron SED in intensity, $\langle \alphaInu (s_{353})\rangle$, we combine the $353$\,GHz correlated CC coefficients, corrected for the CMB (Sect.~
\ref{sec:cmb_est}), obtained from the fits with one and two templates (Sect.~\ref{sec:cc_em}). The SED of the synchrotron emission correlated with dust is then obtained by taking the 
difference between the CC coefficients in Eqs.~\eqref{eq:4.2} and~\eqref{eq:4.3}, and  averaging over all sky patches. It is shown in Fig.~\ref{fig:9.2}. 
We fit this synchrotron SED with a power-law (PL) model.  The normalized 
amplitude of synchrotron emission at 23\,GHz is $A_{\rm s}^{\rm I}=0.30$. This is the ratio between the 23 and 353\,GHz emission in units of \krj. The mean synchrotron spectral 
index derived from the fit is $\beta_{\rm s}^{\rm I}=-3.04\pm0.36$. The uncertainty on $\beta_{\rm s}$ is overestimated as the uncertainties on the synchrotron SED are highly correlated across all \wmap\ and \planck\ frequencies. However, this is not critical for our study because we do not use the uncertainty on $\beta_{\rm s}$  in the paper.  The derived mean $\beta_{\rm s}^{\rm I}$ of the dust-correlated synchrotron emission is consistent with the spectral 
index of 408-MHz-correlated synchrotron emission obtained using \wmap\ data \citep{MAMD:2008, Dickinson:2009, Gold2010,Ghosh:2012}, and the spectral index of the polarized synchrotron emission \citep{Fuskeland:2014}.

\begin{figure*}
\begin{tabular}{cc}
\includegraphics[width=8.8cm]{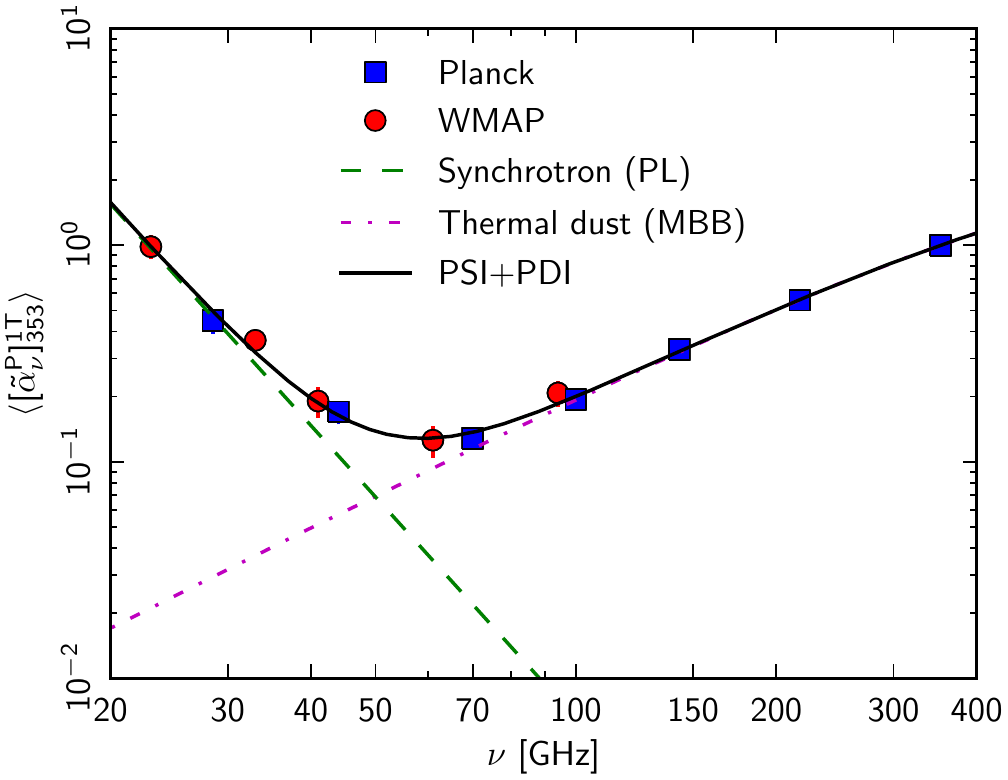}&
\includegraphics[width=8.8cm]{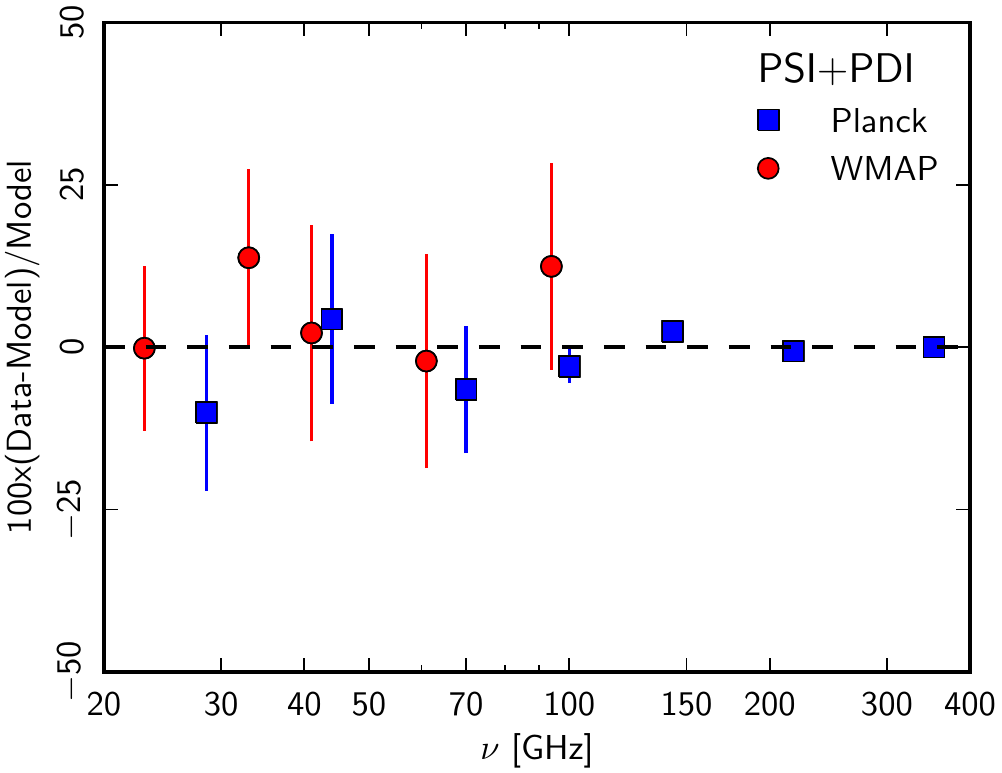} \\
\includegraphics[width=8.8cm]{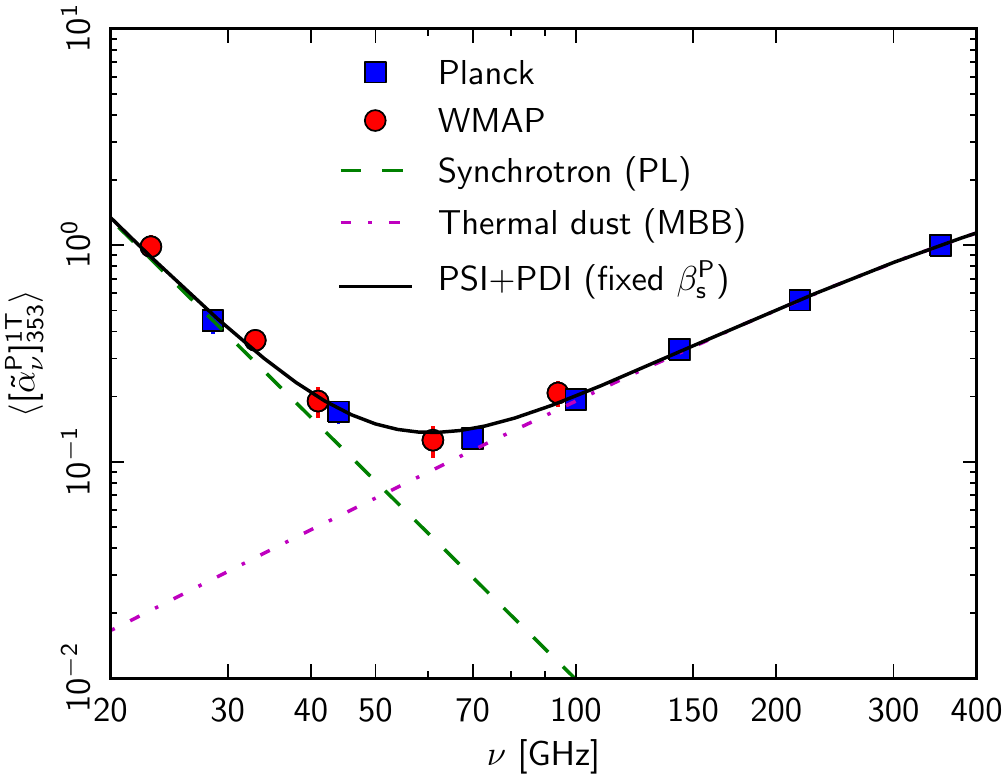}&
\includegraphics[width=8.8cm]{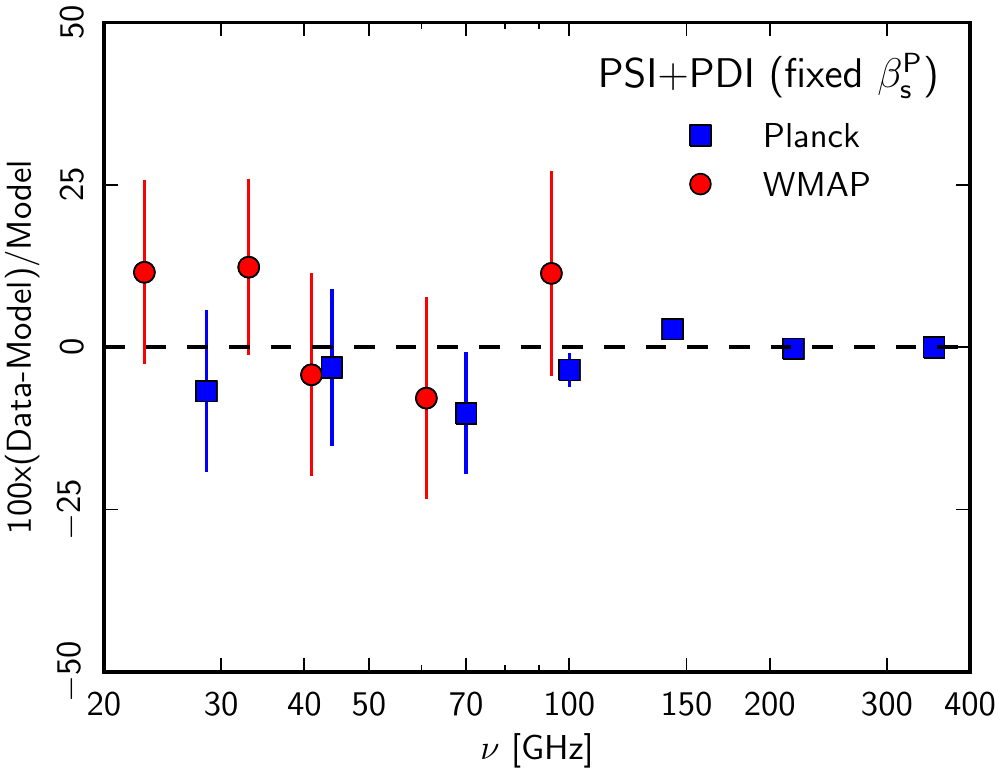} \\
\end{tabular}
\caption{Mean polarized SED in \krj\ units, normalized to 1 at $353$\,GHz correlated with the Stokes \StokesQ\ and \StokesU\ $353$\,GHz maps.  The polarized spectral model with and without the 
constraint on $\beta_{\rm s}^{\rm P}$ match the observed data points. } 
\label{fig:9.1}
\end{figure*}

Using $\langle\alpha^{\rm I}_{23}(s_{353})\rangle$, we find the theoretical upper limit on $\langle\alpha^{\rm P}_{23}(s_{353})\rangle$ to be
\begin{equation}
\langle \alpha^{\rm P}_{23}(s_{353})\rangle \le \left ( \frac{0.75}{0.20} \right ) \times 0.30 = 1.1 \ . 
\end{equation}
The measured value of $\langle[\tildealphaPnu]_{353}^{\rm 1T}\rangle$ is 0.95 (shown in Fig.~\ref{fig:9.1}), which is within the upper limit. The difference between the measured and 
theoretical upper limit in Eq.~\eqref{eq:8.2} can be explained by the fact that polarization angles traced by synchrotron and dust emission are not perfectly aligned. We point out that 
that this statement refers to the synchrotron emission correlated with the dust template in intensity, which is not one to one correlated with dust in polarization.

\subsubsection{Upper limit on AME polarization}

We can set an upper limit on the polarization fraction of AME by assuming that the synchrotron and dust polarization are totally uncorrelated. Within this hypothesis, the low 
frequency rise of the polarization SED is entirely due to polarized AME. Since both AME and thermal dust emission are associated with interstellar matter, it is reasonable to assume 
that the polarization angles are the same for AME and dust. We obtain the $353$\,GHz correlated AME polarization at $23$\,GHz as, 
\begin{align}
\langle [\tilde{\alpha}_{23}^{\rm P}]_{353}^{\rm 1T}\rangle = \langle \alpha_{23}^{\rm P}(a_{353}) \rangle& = \frac{\polfrac_{\rm a}}{\polfrac_{\rm d}} \langle \alpha_{23}^{\rm I}(a_{353}) \rangle \ , \nonumber \\
\text{i.e.,} \ \ 0.95& = \frac{\polfrac_{\rm a}}{\polfrac_{\rm d}} \langle \alpha_{23}^{\rm I}(a_{353}) \rangle \ , \nonumber\\
\text{so} \ \ \polfrac_{\rm a}&= \frac{\polfrac_{\rm d} \times 0.95} {\langle \alpha_{23}^{\rm I}(a_{353}) \rangle} \ . \nonumber \\
& = \frac{\polfrac_{\rm d} \times 0.95} {A_{\rm a}^{\rm I}} \ . \hspace{0.4cm}  
\end{align}
We use the mean AME amplitude, $A_{\rm a}^{\rm I}=1.14$, from model DI+AI and Table~\ref{tab:7.1}, together with $\polfrac_{\rm d}=20$\,\% \citep{planck2014-XIX,planck2014-XX}, 
to derive an upper limit  on the intrinsic polarization fraction of AME of about $16$\,\%. This is much higher than upper limits reported from the analysis of compact sources 
\citep{Dickinson:2011, Lopez-Caraballo:2011,Rubino:2012} and theoretical predictions \citep{Lazarian:2000, Hoang:2013}. Thus AME is unlikely to be the sole explanation for the 
low frequency rise of the polarization SED, even if we cannot exclude some contribution from AME.

\subsection{Parametric modelling for polarized dust SED}\label{sec:pol_model}

In this section we present 	a spectral model that fits the observed polarization SED. We model the polarization SED with a combination of polarized synchrotron and dust 
components. This model does not include AME. We account for the rise of the SED towards the lowest frequencies with the synchrotron component.  For the synchrotron component 
we use the PL model with two parameters: the amplitude; and the spectral index.  The PL model of synchrotron emission is related to the power-law energy distribution of the 
cosmic-ray electron spectrum \citep{Abdo:2009,Ackermann:2010, Ackermann:2012}. 
The model is the superposition of a power-law synchrotron spectrum and the MBB for the thermal dust 
emission. We refer to this model as PSI+PDI.  It is described by the equation:
\begin{align}
\langle [\tildealphaPnu]_{353}^{\rm 1T}\rangle& = \ \AsP\  \left( \frac{\nu}{\nub}\right)^{\beta_{\rm s}^{\rm P}} + \left(\frac{\nu}{\nuref}\right)^{\betadmmP - 2} \frac{B_{\nu} (\Td)}{B_{\nuref} (\Td)}  \ , \label{eq:9.3} 
\end{align}
\normalsize
where $\AsP$ is the amplitude of polarized synchrotron components in \krj\ units, and \betadmmP\ is the polarized dust spectral index. The polarized dust SED, expressed in \krj\ units, is normalized to 1 at 353 GHz. Like for the two intensity models, we fix $\Td=19.6$\,K. 
We fit three parameters: the synchrotron amplitude, the synchrotron and dust spectral index. 
We also fit this model with an additional constraint that the spectral index of $353$\,GHz correlated synchrotron component is the same for intensity and polarization. By doing so 
we have one less parameter to fit and increase the number of degrees of freedom by one.

The fits to the polarization SED for model with and without the constraint on $\beta_{\rm s}^{\rm P}$ are shown in Fig.~\ref{fig:9.1}. The parameters for the best-fit models are listed in 
Table~\ref{tab:10.1}. The two models provide very similar fits to the observed polarized SED.  If we force the spectral indices of the synchrotron for intensity and polarization to be equal, 
we find an equally good fit to the polarization SED. These results are further discussed in the next section.

\begin{table}[tmb]
\begingroup
\newdimen\tblskip \tblskip=5pt
\caption{\label{tab:10.1} Results of the spectral fits to the mean polarized dust SED obtained using \planck\ and \wmap\ data. }
\nointerlineskip
\vskip -3mm
\setbox\tablebox=\vbox{
   \newdimen\digitwidth 
   \setbox0=\hbox{\rm 0} 
   \digitwidth=\wd0 
   \catcode`*=\active 
   \def*{\kern\digitwidth}
   \newdimen\signwidth 
   \setbox0=\hbox{+} 
   \signwidth=\wd0 
   \catcode`!=\active 
   \def!{\kern\signwidth}
\halign{
\hbox to 1.6 cm{#\leaderfil}\tabskip 1.8em&
\hfil #\hfil&
\hfil #\hfil\tabskip=0pt\cr
\noalign{\doubleline}
Parameters $^{a}$& Unconstrained $\beta_{\rm s}^{\rm P}$& Fixed $\beta_{\rm s}^{\rm P}$\cr
\noalign{\vskip 4pt\hrule\vskip 6pt}
 $\AsP$                                & $0.97\pm0.10$& $0.86\pm0.06$\cr
 $\beta_{\rm s}^{\rm P}$    & $-3.40\pm0.28$& $-3.04$\cr
 $\betadmmP$                    & $1.57\pm 0.01$& $1.58\pm0.01$\cr
 $\chi^2/N_{\rm dof}$         & 6.6/9& 8.6/10\cr
\noalign{\vskip 5pt\hrule\vskip 3pt}}}
\endPlancktable
\tablenote{a} The parameters of the model PSI+PDI are described in Eq.~\eqref{eq:9.3} for fixed $\Td=19.6$\,K, where the dust SED is expressed in \krj\ units and normalized to 1 at $353$\,GHz.   \par
\endgroup
\end{table}

\section{Comparison of the dust SEDs for intensity and polarization}\label{sec:discussion}

We now compare the dust SEDs for intensity and polarization and discuss the frequency dependence of the polarization fraction within the context of existing dust models.

\subsection{Spectral dependence of the polarization fraction}


\begin{figure}[h!]
\begin{tabular}{c}
\includegraphics[width=8.7cm]{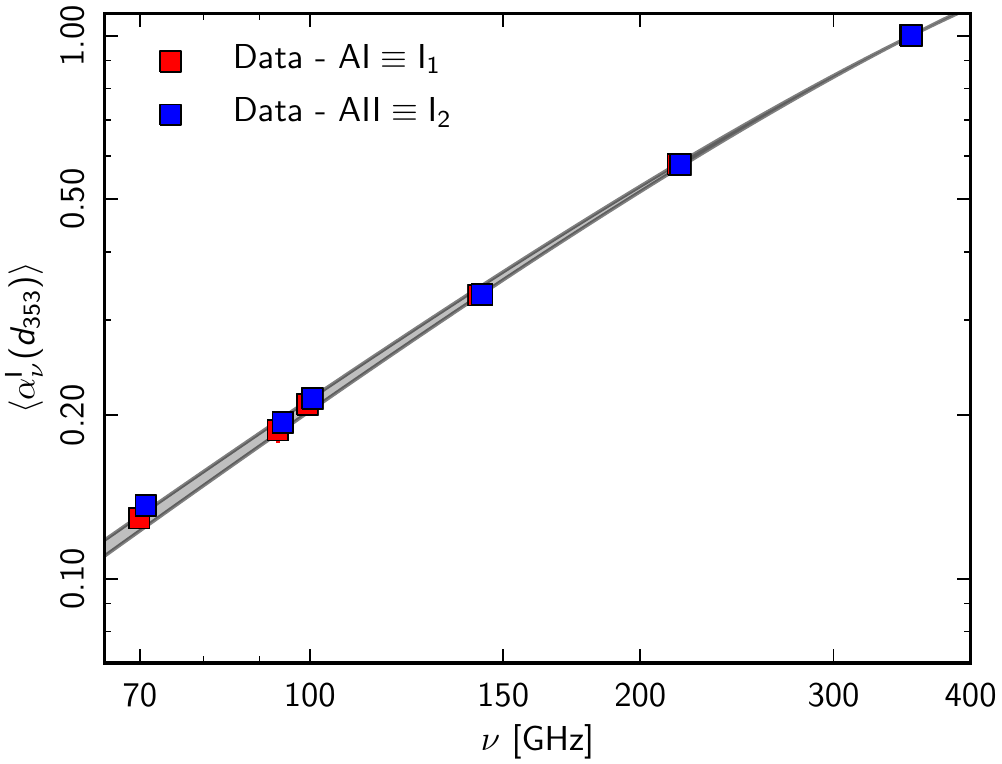} \\
\includegraphics[width=8.7cm]{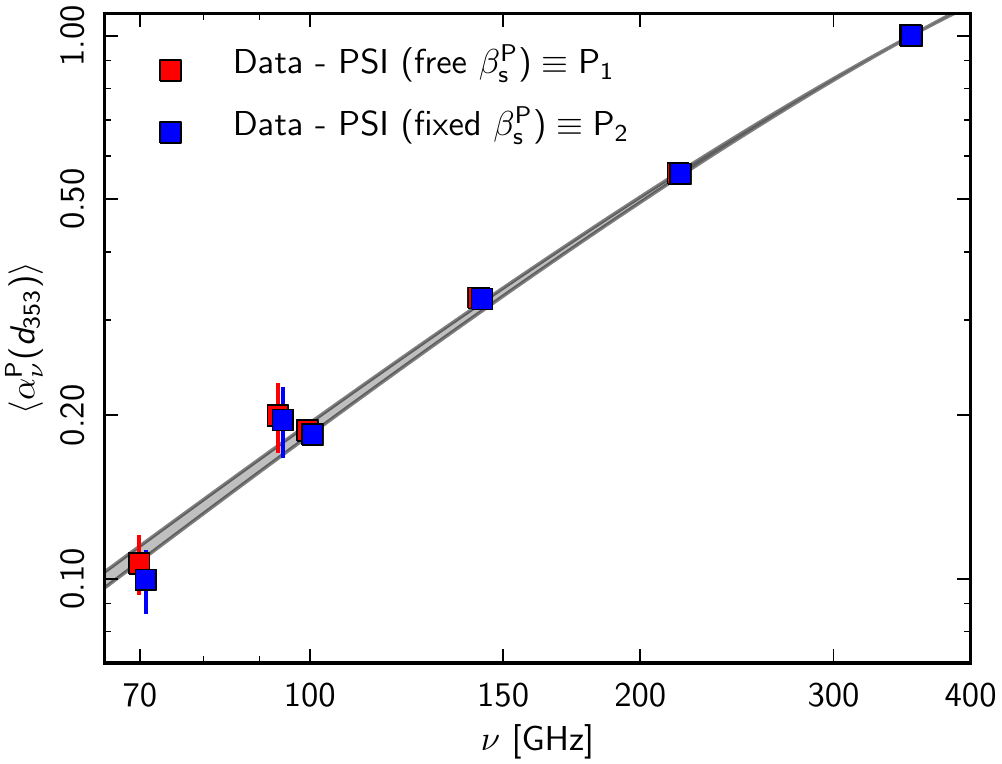} \\
\includegraphics[width=8.7cm]{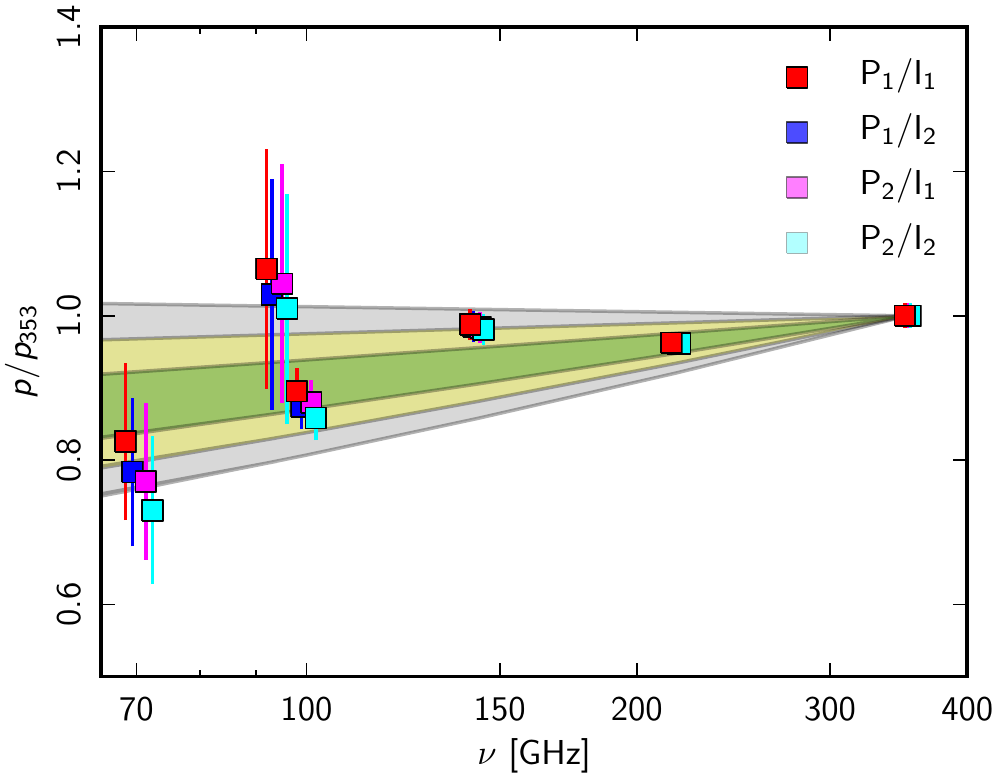} 
\end{tabular}
\caption{Frequency dependence of the dust SED in intensity without AME (\emph{top}), in polarization without synchrotron (\emph{middle}), and the polarization fraction (\emph{bottom}) for the four different combinations of dust models. 
The SEDs are plotted in units of \krj\  and normalized to 1 at $353$\,GHz. The shaded areas represent $1\sigma$ (green), $2\sigma$ (yellow), and $3\sigma$ (grey) statistical uncertainty on the mean normalized polarization fraction. We used the mean spectra from the same 400 sky patches for both the intensity and polarization analysis.}
\label{fig:10.1}
\end{figure}


Figure~\ref{fig:10.1} presents the dust SEDs for
intensity and polarization derived from the spectral decomposition in emission components (Sects.~\ref{sec:em_model} and \ref{sec:pol_model}). 
The SED for intensity, obtained after AME subtraction, is shown in the top panel of Fig.~\ref{fig:10.1} from 70 to $353$\,GHz.  
The two sets of data points computed for each of the two AME models are very close to each other.  
The dust SED for polarization, obtained after subtraction of the synchrotron component,  is shown in the middle panel of Fig.~\ref{fig:10.1}. 
The data points are plotted for the two spectral indices in Table~\ref{tab:10.1}, which differ by 0.4.  The small difference between the corresponding points shows that
the subtraction of the  synchrotron component  has a very small impact on the dust polarization SED even at $70$\,GHz. This indicates that our 
dust polarization SED  is robust with respect to uncertainties on the  spectral index of polarized synchrotron, including a possible steepening of the spectrum with increasing 
frequencies as discussed in \citet{Gold2010}. The polarization fraction, shown in the bottom panel of Fig.~\ref{fig:10.1}, is normalized with respect to the data point at $353$\,GHz. 
The uncertainties on \polfrac\ include the uncertainties from both \polint\ and \StokesI. The data suggest that there is a small decrease in \polfrac\ 
by $21 \pm 6\%$ from 353 to $70$\,GHz.

\subsection{The contribution of carbon dust and silicates to polarization}

The results of this work set new observational constraints on dust models including silicate and carbon grains with possibly different 
polarization properties \citep{Hildebrand:1999,Martin:2007}. 
We discuss the spectral dependence of \polfrac\ using the models from \citet{Draine_Li:2007}, \citet{Compeigne:2011}, and \citet{Jones:2013}. 
In these three models, the thermal dust emission is the electric dipole emission from two types of grains, silicates and carbon grains, with distinct optical properties and thereby 
temperatures.  The first two models use the same optical properties for silicates, but distinct properties for carbon grains; \citet{Draine_Li:2007} use the optical properties of graphite, 
while \citet{Compeigne:2011} use results from laboratory measurements of amorphous carbon. The spectral index for carbon grains is 2 in the \citet{Draine_Li:2007} model and 1.6 
in \citet{Compeigne:2011}. Over microwave frequencies, the opacity of silicates scales as $\nu^{1.6}$ in both models. \citet{Jones:2013} use optical properties of amorphous carbon 
grains, which depend on the hydrogen fraction and degree of aromatization \citep{Jones:2012}. The spectral index of the carbon dust at $353$\,GHz varies between 1.2 and 2.3, 
depending on the nature of the carbon grains (see Fig.~14 of \citealt{Jones:2013} for more details).

SEDs from the first two models have been compared to the \planck\ dust SED in intensity in \citet{planck2013-XVII} and 
\citet{planck2014-XXIX}. The differences between the model and the data are within 
5--15\,\% at $\nu < 353$\,GHz.  \citet{Draine_Fraisse:2009} have used the \citet{Draine_Li:2007} dust model to compute spectra for dust polarization. They predict a systematic 
increase of the polarization fraction \polfrac\ at microwave frequencies when only silicates contribute to dust polarization. This prediction is not what we report in this paper.
However, model predictions for the spectral dependency of $p$ are related to the difference in spectral index between carbon and silicate grains, which is not known. Thus, a 
difference between carbon and silicate polarization may be the correct physical interpretation of the spectral dependence of \polfrac, even if the data do not match the 
\citet{Draine_Fraisse:2009} model. 

Calculations of the polarized SED for the \citet{Compeigne:2011} and \citet{Jones:2013} models are  needed to assess quantitatively this  
interpretation. In the \citet{Compeigne:2011} model, the spectra from silicates and carbon grains are very similar at long wavelengths and we do not expect \polfrac\ to depend on 
wavelength when only silicates contribute to the polarization.  In the \citet{Jones:2013} model the contribution from carbon dust grains could be dominant at $\nu < 353$\,GHz.

\subsection{Microwave dust emission}

The dust SED in polarization and the spectral dependence of \polfrac\ allows us to discuss  two possible interpretations of the dust microwave emission.

\citet{Meny:2007} introduced a physical description of FIR/microwave dust emission, where the microwave dust opacity of amorphous grains is dominated by low energy 
transitions associated with disorder in the structure of the solids on atomic scales. This contribution is modelled by transitions in two-level systems (TLS). The TLS model is 
supported by experimental results on silicates \citep{Agladze:1996, Boudet:2005, Coupeaud:2011}, which indicate that the opacity of amorphous silicate grains flattens towards long 
wavelengths. 
The TLS model has been used to model dust emission spectra by \citet{Paradis:2011}.  
It was also proposed as a possible interpretation of the flattening  
of the dust SED in intensity  from FIR to mm wavelengths, which was reported  in two earlier Planck papers on 
the diffuse dust emission in the Galactic plane in \citet{planck2013-XIV} and at high Galactic latitudes in \citet{planck2013-XVII}. Our analysis based on new \Planck\ data does not confirm this flattening, but does not 
dismiss it either because the error-bars on the submm spectral index from calibration uncertainties remain significant: 0.16 ($1\sigma$) from Sect.~\ref{sec:Td}. 
Within the TLS model, a flattening of the dust SED in intensity would be due to a decrease in the spectral index of silicate grains at long wavelengths. Since silicate grains are polarized, the flattening of the dust SED should also be seen in polarization.  
This expectation is not supported by the results of our data analysis, namely the  difference in spectral indices of the thermal dust emission in intensity and polarization
($\betadmmI$ and $\betadmmP$) reported in Sect.~\ref{sec:err_betaP}.  

MDE has been introduced by \citet{Draine_Hensley:2012a} to explain the flattening of the dust SED at sub-mm wavelengths in the Small Magellanic Cloud \citep{planck2011-6.4b}.  MDE could also contribute to the long wavelength emission of Galactic dust. Model spectra of MDE are presented in
\citet{Draine_Hensley:2012b} for Galactic dust. The contribution of MDE could be significant at frequencies  smaller than a few hundred GHz and increasing towards smaller frequencies. 
If the magnetic particles are inclusions randomly-oriented within interstellar grains, their emission is polarized in a direction perpendicular to that of the dipolar electric emission \citep{Draine_Hensley:2012b}.  In this case, we expect this emission component to reduce the polarization of the dust emission. This could possibly account for the observed decrease in \polfrac\ from 353 to $70$\,GHz. 

A  fit of the SEDs in polarization and intensity with the models of \citet{Paradis:2011} and \citet{Draine_Hensley:2012b} would be necessary to test these two interpretations quantitatively.

\section{Conclusion}\label{sec:conclusion}

We have characterized the frequency dependence of dust emission in intensity and polarization by analysing \planck\ data over 39\,\% of the sky at intermediate Galactic latitudes.
We use the \planck\ $353$\,GHz \StokesI, \StokesQ, and \StokesU\ maps as templates for dust emission in intensity and polarization. We cross-correlate them with the \planck\ and 
\wmap\ data, at 12 frequencies from 23 to $353$\,GHz. The main results of the data analysis are as follows.

\begin{itemize}

\item The mean spectral index of the dust emission measured between 100 and $353$\,GHz is $\betadmmI = 1.51\pm0.01$. This value agrees with that reported by 
\citet{planck2013-XVII} for the high Galactic latitude sky and \citet{planck2013-XIV} for diffuse emission in the Galactic plane. 
The microwave spectral index \betadmmI\ is close to
that at submm wavelengths, which is  derived from fits to the \planck\ data at $\nu \ge 353$\,GHz with the full-mission \planck\ data. 

\item We determine the mean dust SED in intensity from 23 to $353$\,GHz. We separate the dust and AME contributions to the SED. The dust contribution is well fit by combining the 
modified blackbody spectrum with a spectral index of 1.51 with the mean temperature of $19.6$\,K. The two parametric models we use for the AME yield the same SED for the dust for frequencies  $\nu \ge 70\,$GHz.

\item The mean spectral index for dust polarization, measured between 100 and $353$\,GHz, is $\betadmmP = 1.59\pm0.02$, 
assuming the temperature of aligned dust grains 
contributing to the polarization is the same as that determined from the dust emission in intensity.  
We show that the small difference with $\betadmmI $, the spectral index measured in a similar way for dust 
intensity,  is a robust result against systematic uncertainties estimated comparing results of our data analysis obtained 
on various subsets of the \Planck\ data. 

\item We determine the SED of the dust-correlated polarized emission from $23$ to $353$\,GHz. This SED decreases with decreasing frequency and turns up below $60$\,GHz, 
very much like the dust SED in intensity due to AME. We show that the low frequency rise of the polarization SED may be explained by synchrotron polarization correlated with dust. 

\item We use a parametric model to separate the synchrotron and dust polarization and to characterize the spectral dependence of the dust polarization fraction.  The polarization 
fraction \polfrac\ of the dust emission decreases by $(21\pm6)$\,\% from 353 to $70$\,GHz.  We discuss this result within the context of existing 
dust models. It  could indicate differences in polarization efficiency among components of interstellar dust (e.g., carbon and silicate grains).  

\end{itemize}

Our observational results provide inputs to quantify and optimize the separation between Galactic and CMB polarization \citep{planck2014-XXX}.  Our CC analysis yields a spectral decomposition of the 
diffuse emission into its main components: thermal emission from dust; free-free; synchrotron; and AME.  This spectral decomposition may be combined with power spectra of the 
templates, as discussed in Appendix~\ref{sec:power_spectra}, to quantify the Galactic emission in intensity as a function of the observed frequency and multipole (see Figs.~27 and 
28 in \citet{planck2013-p01}).  For polarization, we are currently missing a synchrotron polarization template that is free from Faraday rotation.  Polarized synchrotron templates, 
which are expected from C-BASS \citep{King:2010} at 5\,GHz and QUIJOTE \citep{Hoyland:2012} between 10 and 20\,GHz, will be important for future correlation analyses.

\begin{acknowledgements}

The \planck~Collaboration acknowledges the support of: ESA; CNES and CNRS/INSU-IN2P3-INP (France); ASI, CNR, and INAF (Italy); NASA and DoE (USA); STFC and UKSA 
(UK); CSIC, MICINN and JA (Spain); Tekes, AoF and CSC (Finland); DLR and MPG (Germany); CSA (Canada); DTU Space (Denmark); SER/SSO (Switzerland); RCN (Norway); SFI 
(Ireland); FCT/MCTES (Portugal); and DEISA (EU). A detailed description of the \planck~Collaboration and a list of its members can be found at 
\url{http://www.rssd.esa.int/index.php?project=PLANCK\&page=Planck\_Collaboration}. The research leading to these results has received funding from the European Research 
Council under the European Union's Seventh Framework Programme (FP7/2007-2013) / ERC grant agreement n$^\circ$ 267934. We acknowledge the use of the Legacy Archive for 
Microwave Background Data Analysis (LAMBDA), part of the High Energy Astrophysics Science Archive Center (HEASARC). HEASARC/LAMBDA is a service of the Astrophysics 
Science Division at the NASA Goddard Space Flight Center. Some of the results in this paper have been derived using the \healpix\ package.

\end{acknowledgements}

\bibliographystyle{aat}
\def\eprinttmppp@#1arXiv:@{#1}
\providecommand{\arxivlink[1]}{\href{http://arxiv.org/abs/#1}{arXiv:#1}}
\def\eprinttmp@#1arXiv:#2 [#3]#4@{\ifthenelse{\equal{#3}{x}}{\ifthenelse{
\equal{#1}{}}{\arxivlink{\eprinttmppp@#2@}}{\arxivlink{#1}}}{\arxivlink{#2}
  [#3]}}
\providecommand{\eprintlink}[1]{\eprinttmp@#1arXiv: [x]@}
\providecommand{\eprint}[1]{\eprintlink{#1}}
\providecommand{\adsurl}[1]{\href{#1}{ADS}}


\begin{thebibliography}{105}
\expandafter\ifx\csname natexlab\endcsname\relax\def\natexlab#1{#1}\fi

\bibitem[{{Abdo} {et~al.}(2009){Abdo}, {Ackermann}, {Ajello}, {Atwood},
  {Axelsson}, {Baldini}, {Ballet}, {Barbiellini}, {Bastieri}, {Battelino},
  {Baughman}, {Bechtol}, {Bellazzini}, {Berenji}, {Blandford}, {Bloom},
  {Bogaert}, {Bonamente}, {Borgland}, {Bregeon}, {Brez}, {Brigida}, {Bruel},
  {Burnett}, {Caliandro}, {Cameron}, {Caraveo}, {Carlson}, {Casandjian},
  {Cecchi}, {Charles}, {Chekhtman}, {Cheung}, {Chiang}, {Ciprini}, {Claus},
  {Cohen-Tanugi}, {Cominsky}, {Conrad}, {Cutini}, {Dermer}, {de Angelis}, {de
  Palma}, {Digel}, {di Bernardo}, {Do Couto E Silva}, {Drell}, {Dubois},
  {Dumora}, {Edmonds}, {Farnier}, {Favuzzi}, {Focke}, {Frailis}, {Fukazawa},
  {Funk}, {Fusco}, {Gaggero}, {Gargano}, {Gasparrini}, {Gehrels}, {Germani},
  {Giebels}, {Giglietto}, {Giordano}, {Glanzman}, {Godfrey}, {Grasso},
  {Grenier}, {Grondin}, {Grove}, {Guillemot}, {Guiriec}, {Hanabata}, {Harding},
  {Hartman}, {Hayashida}, {Hays}, {Hughes}, {J{\'o}hannesson}, {Johnson},
  {Johnson}, {Johnson}, {Kamae}, {Katagiri}, {Kataoka}, {Kawai}, {Kerr},
  {Kn{\"o}dlseder}, {Kocevski}, {Kuehn}, {Kuss}, {Lande}, {Latronico},
  {Lemoine-Goumard}, {Longo}, {Loparco}, {Lott}, {Lovellette}, {Lubrano},
  {Madejski}, {Makeev}, {Massai}, {Mazziotta}, {McConville}, {McEnery},
  {Meurer}, {Michelson}, {Mitthumsiri}, {Mizuno}, {Moiseev}, {Monte},
  {Monzani}, {Moretti}, {Morselli}, {Moskalenko}, {Murgia}, {Nolan}, {Norris},
  {Nuss}, {Ohsugi}, {Omodei}, {Orlando}, {Ormes}, {Ozaki}, {Paneque},
  {Panetta}, {Parent}, {Pelassa}, {Pepe}, {Pesce-Rollins}, {Piron}, {Pohl},
  {Porter}, {Profumo}, {Rain{\`o}}, {Rando}, {Razzano}, {Reimer}, {Reimer},
  {Reposeur}, {Ritz}, {Rochester}, {Rodriguez}, {Romani}, {Roth}, {Ryde},
  {Sadrozinski}, {Sanchez}, {Sander}, {Saz Parkinson}, {Scargle}, {Schalk},
  {Sellerholm}, {Sgr{\`o}}, {Smith}, {Smith}, {Spandre}, {Spinelli}, {Starck},
  {Stephens}, {Strickman}, {Strong}, {Suson}, {Tajima}, {Takahashi},
  {Takahashi}, {Tanaka}, {Thayer}, {Thayer}, {Thompson}, {Tibaldo}, {Tibolla},
  {Torres}, {Tosti}, {Tramacere}, {Uchiyama}, {Usher}, {van Etten},
  {Vasileiou}, {Vilchez}, {Vitale}, {Waite}, {Wallace}, {Wang}, {Winer},
  {Wood}, {Ylinen}, \& {Ziegler}}]{Abdo:2009}
{Abdo}, A.~A., {Ackermann}, M., {Ajello}, M., {et~al.}, {Measurement of the
  Cosmic Ray e$^{+}$+e$^{-}$ Spectrum from 20GeV to 1TeV with the Fermi Large
  Area Telescope}. 2009, Physical Review Letters, 102, 181101,
  \eprint{0905.0025}

\bibitem[{{Ackermann} {et~al.}(2012){Ackermann}, {Ajello}, {Allafort},
  {Atwood}, {Baldini}, {Barbiellini}, {Bastieri}, {Bechtol}, {Bellazzini},
  {Berenji}, {Blandford}, {Bloom}, {Bonamente}, {Borgland}, {Bouvier},
  {Bregeon}, {Brigida}, {Bruel}, {Buehler}, {Buson}, {Caliandro}, {Cameron},
  {Caraveo}, {Casandjian}, {Cecchi}, {Charles}, {Chekhtman}, {Cheung},
  {Chiang}, {Ciprini}, {Claus}, {Cohen-Tanugi}, {Conrad}, {Cutini}, {de
  Angelis}, {de Palma}, {Dermer}, {Digel}, {Do Couto E Silva}, {Drell},
  {Drlica-Wagner}, {Favuzzi}, {Fegan}, {Ferrara}, {Focke}, {Fortin},
  {Fukazawa}, {Funk}, {Fusco}, {Gargano}, {Gasparrini}, {Germani}, {Giglietto},
  {Giommi}, {Giordano}, {Giroletti}, {Glanzman}, {Godfrey}, {Grenier}, {Grove},
  {Guiriec}, {Gustafsson}, {Hadasch}, {Harding}, {Hayashida}, {Hughes},
  {J{\'o}hannesson}, {Johnson}, {Kamae}, {Katagiri}, {Kataoka},
  {Kn{\"o}dlseder}, {Kuss}, {Lande}, {Latronico}, {Lemoine-Goumard}, {Llena
  Garde}, {Longo}, {Loparco}, {Lovellette}, {Lubrano}, {Madejski}, {Mazziotta},
  {McEnery}, {Michelson}, {Mitthumsiri}, {Mizuno}, {Moiseev}, {Monte},
  {Monzani}, {Morselli}, {Moskalenko}, {Murgia}, {Nakamori}, {Nolan}, {Norris},
  {Nuss}, {Ohno}, {Ohsugi}, {Okumura}, {Omodei}, {Orlando}, {Ormes}, {Ozaki},
  {Paneque}, {Parent}, {Pesce-Rollins}, {Pierbattista}, {Piron}, {Pivato},
  {Porter}, {Rain{\`o}}, {Rando}, {Razzano}, {Razzaque}, {Reimer}, {Reimer},
  {Reposeur}, {Ritz}, {Romani}, {Roth}, {Sadrozinski}, {Sbarra}, {Schalk},
  {Sgr{\`o}}, {Siskind}, {Spandre}, {Spinelli}, {Strong}, {Takahashi},
  {Takahashi}, {Tanaka}, {Thayer}, {Thayer}, {Tibaldo}, {Tinivella}, {Torres},
  {Tosti}, {Troja}, {Uchiyama}, {Usher}, {Vandenbroucke}, {Vasileiou},
  {Vianello}, {Vitale}, {Waite}, {Winer}, {Wood}, {Wood}, {Yang}, \&
  {Zimmer}}]{Ackermann:2012}
{Ackermann}, M., {Ajello}, M., {Allafort}, A., {et~al.}, {Measurement of
  Separate Cosmic-Ray Electron and Positron Spectra with the Fermi Large Area
  Telescope}. 2012, Physical Review Letters, 108, 011103, \eprint{1109.0521}

\bibitem[{{Ackermann} {et~al.}(2010){Ackermann}, {Ajello}, {Atwood}, {Baldini},
  {Ballet}, {Barbiellini}, {Bastieri}, {Baughman}, {Bechtol}, {Bellardi},
  {Bellazzini}, {Belli}, {Berenji}, {Blandford}, {Bloom}, {Bogart},
  {Bonamente}, {Borgland}, {Brandt}, {Bregeon}, {Brez}, {Brigida}, {Bruel},
  {Buehler}, {Burnett}, {Busetto}, {Buson}, {Caliandro}, {Cameron}, {Caraveo},
  {Carlson}, {Carrigan}, {Casandjian}, {Ceccanti}, {Cecchi}, {{\c C}elik},
  {Charles}, {Chekhtman}, {Cheung}, {Chiang}, {Cillis}, {Ciprini}, {Claus},
  {Cohen-Tanugi}, {Conrad}, {Corbet}, {Deklotz}, {Dermer}, {de Angelis}, {de
  Palma}, {Digel}, {di Bernardo}, {Do Couto E Silva}, {Drell}, {Drlica-Wagner},
  {Dubois}, {Fabiani}, {Favuzzi}, {Fegan}, {Fortin}, {Fukazawa}, {Funk},
  {Fusco}, {Gaggero}, {Gargano}, {Gasparrini}, {Gehrels}, {Germani},
  {Giglietto}, {Giommi}, {Giordano}, {Giroletti}, {Glanzman}, {Godfrey},
  {Grasso}, {Grenier}, {Grondin}, {Grove}, {Guiriec}, {Gustafsson}, {Hadasch},
  {Harding}, {Hayashida}, {Hays}, {Horan}, {Hughes}, {J{\'o}hannesson},
  {Johnson}, {Johnson}, {Johnson}, {Kamae}, {Katagiri}, {Kataoka}, {Kerr},
  {Kn{\"o}dlseder}, {Kuss}, {Lande}, {Latronico}, {Lemoine-Goumard}, {Llena
  Garde}, {Longo}, {Loparco}, {Lott}, {Lovellette}, {Lubrano}, {Makeev},
  {Mazziotta}, {McEnery}, {Mehault}, {Michelson}, {Minuti}, {Mitthumsiri},
  {Mizuno}, {Moiseev}, {Monte}, {Monzani}, {Moretti}, {Morselli}, {Moskalenko},
  {Murgia}, {Nakamori}, {Naumann-Godo}, {Nolan}, {Norris}, {Nuss}, {Ohsugi},
  {Okumura}, {Omodei}, {Orlando}, {Ormes}, {Ozaki}, {Paneque}, {Panetta},
  {Parent}, {Pelassa}, {Pepe}, {Pesce-Rollins}, {Petrosian}, {Pinchera},
  {Piron}, {Porter}, {Profumo}, {Rain{\`o}}, {Rando}, {Rapposelli}, {Razzano},
  {Reimer}, {Reimer}, {Reposeur}, {Ripken}, {Ritz}, {Rochester}, {Romani},
  {Roth}, {Sadrozinski}, {Saggini}, {Sanchez}, {Sander}, {Sgr{\`o}}, {Siskind},
  {Smith}, {Spandre}, {Spinelli}, {Stawarz}, {Stephens}, {Strickman}, {Strong},
  {Suson}, {Tajima}, {Takahashi}, {Takahashi}, {Tanaka}, {Thayer}, {Thayer},
  {Thompson}, {Tibaldo}, {Tibolla}, {Torres}, {Tosti}, {Tramacere}, {Turri},
  {Uchiyama}, {Usher}, {Vandenbroucke}, {Vasileiou}, {Vilchez}, {Vitale},
  {Waite}, {Wallace}, {Wang}, {Winer}, {Wood}, {Yang}, {Ylinen}, \&
  {Ziegler}}]{Ackermann:2010}
{Ackermann}, M., {Ajello}, M., {Atwood}, W.~B., {et~al.}, {Fermi LAT
  observations of cosmic-ray electrons from 7 GeV to 1 TeV}. 2010, \prd, 82,
  092004, \eprint{1008.3999}

\bibitem[{{Agladze} {et~al.}(1996){Agladze}, {Sievers}, {Jones}, {Burlitch}, \&
  {Beckwith}}]{Agladze:1996}
{Agladze}, N.~I., {Sievers}, A.~J., {Jones}, S.~A., {Burlitch}, J.~M., \&
  {Beckwith}, S.~V.~W., {Laboratory Results on Millimeter-Wave Absorption in
  Silicate Grain Materials at Cryogenic Temperatures}. 1996, \apj, 462, 1026

\bibitem[{{Ali-Ha{\"\i}moud} {et~al.}(2009){Ali-Ha{\"\i}moud}, {Hirata}, \&
  {Dickinson}}]{Ali-Haimoud:2009}
{Ali-Ha{\"\i}moud}, Y., {Hirata}, C.~M., \& {Dickinson}, C., {A refined model
  for spinning dust radiation}. 2009, \mnras, 395, 1055, \eprint{0812.2904}

\bibitem[{{Banday} {et~al.}(2003){Banday}, {Dickinson}, {Davies}, {Davis}, \&
  {G{\'o}rski}}]{Banday:2003}
{Banday}, A.~J., {Dickinson}, C., {Davies}, R.~D., {Davis}, R.~J., \&
  {G{\'o}rski}, K.~M., {Reappraising foreground contamination in the COBE-DMR
  data}. 2003, \mnras, 345, 897, \eprint{arXiv:astro-ph/0302181}

\bibitem[{{Banday} {et~al.}(1996){Banday}, {Gorski}, {Bennett}, {Hinshaw},
  {Kogut}, \& {Smoot}}]{Banday:1996}
{Banday}, A.~J., {Gorski}, K.~M., {Bennett}, C.~L., {et~al.}, {Noncosmological
  Signal Contributions to the COBE DMR 4 Year Sky Maps}. 1996, \apjl, 468, L85,
  \eprint{arXiv:astro-ph/9601064}

\bibitem[{{Bennett} {et~al.}(2013){Bennett}, {Larson}, {Weiland}, {Jarosik},
  {Hinshaw}, {Odegard}, {Smith}, {Hill}, {Gold}, {Halpern}, {Komatsu}, {Nolta},
  {Page}, {Spergel}, {Wollack}, {Dunkley}, {Kogut}, {Limon}, {Meyer}, {Tucker},
  \& {Wright}}]{Bennett:2012}
{Bennett}, C.~L., {Larson}, D., {Weiland}, J.~L., {et~al.}, {Nine-year
  Wilkinson Microwave Anisotropy Probe (WMAP) Observations: Final Maps and
  Results}. 2013, \apjs, 208, 20, \eprint{1212.5225}

\bibitem[{{Bonaldi} {et~al.}(2007){Bonaldi}, {Ricciardi}, {Leach}, {Stivoli},
  {Baccigalupi}, \& {de Zotti}}]{bonaldi:2007}
{Bonaldi}, A., {Ricciardi}, S., {Leach}, S., {et~al.}, {WMAP 3-yr data with
  Correlated Component Analysis: anomalous emission and impact of component
  separation on the CMB power spectrum}. 2007, \mnras, 382, 1791,
  \eprint{0707.0469}

\bibitem[{{Boudet} {et~al.}(2005){Boudet}, {Mutschke}, {Nayral}, {J{\"a}ger},
  {Bernard}, {Henning}, \& {Meny}}]{Boudet:2005}
{Boudet}, N., {Mutschke}, H., {Nayral}, C., {et~al.}, {Temperature Dependence
  of the Submillimeter Absorption Coefficient of Amorphous Silicate Grains}.
  2005, \apj, 633, 272

\bibitem[{{Brandt} \& {Draine}(2012)}]{Brandt:2012}
{Brandt}, T.~D. \& {Draine}, B.~T., {The Spectrum of the Diffuse Galactic
  Light: The Milky Way in Scattered Light}. 2012, \apj, 744, 129,
  \eprint{1109.4175}

\bibitem[{{Chiar} {et~al.}(2006){Chiar}, {Adamson}, {Whittet}, {Chrysostomou},
  {Hough}, {Kerr}, {Mason}, {Roche}, \& {Wright}}]{Chiar:2006}
{Chiar}, J.~E., {Adamson}, A.~J., {Whittet}, D.~C.~B., {et~al.},
  {Spectropolarimetry of the 3.4 {$\mu$}m Feature in the Diffuse ISM toward the
  Galactic Center Quintuplet Cluster}. 2006, \apj, 651, 268,
  \eprint{arXiv:astro-ph/0607245}

\bibitem[{{Chon} {et~al.}(2004){Chon}, {Challinor}, {Prunet}, {Hivon}, \&
  {Szapudi}}]{Chon:2004}
{Chon}, G., {Challinor}, A., {Prunet}, S., {Hivon}, E., \& {Szapudi}, I., {Fast
  estimation of polarization power spectra using correlation functions}. 2004,
  \mnras, 350, 914, \eprint{astro-ph/0303414}

\bibitem[{{Compi{\`e}gne} {et~al.}(2011){Compi{\`e}gne}, {Verstraete}, {Jones},
  {Bernard}, {Boulanger}, {Flagey}, {Le Bourlot}, {Paradis}, \&
  {Ysard}}]{Compeigne:2011}
{Compi{\`e}gne}, M., {Verstraete}, L., {Jones}, A., {et~al.}, {The global dust
  SED: tracing the nature and evolution of dust with DustEM}. 2011, \aap, 525,
  A103, \eprint{1010.2769}

\bibitem[{{Coupeaud} {et~al.}(2011){Coupeaud}, {Demyk}, {Meny}, {Nayral},
  {Delpech}, {Leroux}, {Depecker}, {Creff}, {Brubach}, \&
  {Roy}}]{Coupeaud:2011}
{Coupeaud}, A., {Demyk}, K., {Meny}, C., {et~al.}, {Low-temperature FIR and
  submillimetre mass absorption coefficient of interstellar silicate dust
  analogues}. 2011, \aap, 535, A124, \eprint{1109.2758}

\bibitem[{{Davies} {et~al.}(2006){Davies}, {Dickinson}, {Banday}, {Jaffe},
  {G{\'o}rski}, \& {Davis}}]{Davies:2006}
{Davies}, R.~D., {Dickinson}, C., {Banday}, A.~J., {et~al.}, {A determination
  of the spectra of Galactic components observed by the Wilkinson Microwave
  Anisotropy Probe}. 2006, \mnras, 370, 1125, \eprint{arXiv:astro-ph/0511384}

\bibitem[{{de Oliveira-Costa} {et~al.}(1999){de Oliveira-Costa}, {Tegmark},
  {Gutierrez}, {Jones}, {Davies}, {Lasenby}, {Rebolo}, \&
  {Watson}}]{Oliveira-Costa:1999}
{de Oliveira-Costa}, A., {Tegmark}, M., {Gutierrez}, C.~M., {et~al.},
  {Cross-Correlation of Tenerife Data with Galactic Templates-Evidence for
  Spinning Dust?} 1999, \apjl, 527, L9, \eprint{arXiv:astro-ph/9904296}

\bibitem[{{Dickinson} {et~al.}(2003){Dickinson}, {Davies}, \&
  {Davis}}]{Dickinson:2003}
{Dickinson}, C., {Davies}, R.~D., \& {Davis}, R.~J., {Towards a free-free
  template for CMB foregrounds}. 2003, \mnras, 341, 369,
  \eprint{arXiv:astro-ph/0302024}

\bibitem[{{Dickinson} {et~al.}(2009){Dickinson}, {Eriksen}, {Banday}, {Jewell},
  {G{\'o}rski}, {Huey}, {Lawrence}, {O'Dwyer}, \& {Wandelt}}]{Dickinson:2009}
{Dickinson}, C., {Eriksen}, H.~K., {Banday}, A.~J., {et~al.}, {Bayesian
  Component Separation and Cosmic Microwave Background Estimation for the
  Five-Year WMAP Temperature Data}. 2009, \apj, 705, 1607, \eprint{0903.4311}

\bibitem[{{Dickinson} {et~al.}(2011){Dickinson}, {Peel}, \&
  {Vidal}}]{Dickinson:2011}
{Dickinson}, C., {Peel}, M., \& {Vidal}, M., {New constraints on the
  polarization of anomalous microwave emission in nearby molecular clouds}.
  2011, \mnras, 418, L35, \eprint{1108.0308}

\bibitem[{{Dobler} \& {Finkbeiner}(2008)}]{Dobler:2008a}
{Dobler}, G. \& {Finkbeiner}, D.~P., {Extended Anomalous Foreground Emission in
  the WMAP Three-Year Data}. 2008, \apj, 680, 1222, \eprint{0712.1038}

\bibitem[{{Draine} \& {Fraisse}(2009)}]{Draine_Fraisse:2009}
{Draine}, B.~T. \& {Fraisse}, A.~A., {Polarized Far-Infrared and Submillimeter
  Emission from Interstellar Dust}. 2009, \apj, 696, 1, \eprint{0809.2094}

\bibitem[{{Draine} \& {Hensley}(2012)}]{Draine_Hensley:2012a}
{Draine}, B.~T. \& {Hensley}, B., {The Submillimeter and Millimeter Excess of
  the Small Magellanic Cloud: Magnetic Dipole Emission from Magnetic
  Nanoparticles?} 2012, \apj, 757, 103, \eprint{1205.6810}

\bibitem[{{Draine} \& {Hensley}(2013)}]{Draine_Hensley:2012b}
{Draine}, B.~T. \& {Hensley}, B., {Magnetic Nanoparticles in the Interstellar
  Medium: Emission Spectrum and Polarization}. 2013, \apj, 765, 159,
  \eprint{1205.7021}

\bibitem[{{Draine} \& {Lazarian}(1998)}]{Draine_Lazarian:1998}
{Draine}, B.~T. \& {Lazarian}, A., {Electric Dipole Radiation from Spinning
  Dust Grains}. 1998, \apj, 508, 157, \eprint{arXiv:astro-ph/9802239}

\bibitem[{{Draine} \& {Lazarian}(1999)}]{Draine_Lazarian:1999}
{Draine}, B.~T. \& {Lazarian}, A. 1999, in Bulletin of the American
  Astronomical Society, Vol.~31, American Astronomical Society Meeting
  Abstracts \#194, 890

\bibitem[{{Draine} \& {Li}(2007)}]{Draine_Li:2007}
{Draine}, B.~T. \& {Li}, A., {Infrared Emission from Interstellar Dust. IV. The
  Silicate-Graphite-PAH Model in the Post-Spitzer Era}. 2007, \apj, 657, 810,
  \eprint{arXiv:astro-ph/0608003}

\bibitem[{{Erickson}(1957)}]{Erickson1957}
{Erickson}, W.~C., {A Mechanism of Non-Thermal Radio-Noise Origin.} 1957, \apj,
  126, 480

\bibitem[{{Finkbeiner} {et~al.}(1999){Finkbeiner}, {Davis}, \&
  {Schlegel}}]{Finkbeiner:1999}
{Finkbeiner}, D.~P., {Davis}, M., \& {Schlegel}, D.~J., {Extrapolation of
  Galactic Dust Emission at 100 Microns to Cosmic Microwave Background
  Radiation Frequencies Using FIRAS}. 1999, \apj, 524, 867,
  \eprint{arXiv:astro-ph/9905128}

\bibitem[{{Fuskeland} {et~al.}(2014){Fuskeland}, {Wehus}, {Eriksen}, \&
  {N{\ae}ss}}]{Fuskeland:2014}
{Fuskeland}, U., {Wehus}, I.~K., {Eriksen}, H.~K., \& {N{\ae}ss}, S.~K.,
  {Spatial Variations in the Spectral Index of Polarized Synchrotron Emission
  in the 9yr WMAP Sky Maps}. 2014, \apj, 790, 104, \eprint{1404.5323}

\bibitem[{{Gardner} \& {Whiteoak}(1966)}]{Gardner:1966}
{Gardner}, F.~F. \& {Whiteoak}, J.~B., {The Polarization of Cosmic Radio
  Waves}. 1966, \araa, 4, 245

\bibitem[{{Ghosh} {et~al.}(2012){Ghosh}, {Banday}, {Jaffe}, {Dickinson},
  {Davies}, {Davis}, \& {Gorski}}]{Ghosh:2012}
{Ghosh}, T., {Banday}, A.~J., {Jaffe}, T., {et~al.}, {Foreground analysis using
  cross-correlations of external templates on the 7-year Wilkinson Microwave
  Anisotropy Probe data}. 2012, \mnras, 422, 3617, \eprint{1112.0509}

\bibitem[{{Gold} {et~al.}(2011){Gold}, {Odegard}, {Weiland}, {Hill}, {Kogut},
  {Bennett}, {Hinshaw}, {Chen}, {Dunkley}, {Halpern}, {Jarosik}, {Komatsu},
  {Larson}, {Limon}, {Meyer}, {Nolta}, {Page}, {Smith}, {Spergel}, {Tucker},
  {Wollack}, \& {Wright}}]{Gold2010}
{Gold}, B., {Odegard}, N., {Weiland}, J.~L., {et~al.}, {Seven-year Wilkinson
  Microwave Anisotropy Probe (WMAP) Observations: Galactic Foreground
  Emission}. 2011, \apjs, 192, 15, \eprint{1001.4555}

\bibitem[{{Gorski} {et~al.}(1996){Gorski}, {Banday}, {Bennett}, {Hinshaw},
  {Kogut}, {Smoot}, \& {Wright}}]{Gorski:1996}
{Gorski}, K.~M., {Banday}, A.~J., {Bennett}, C.~L., {et~al.}, {Power Spectrum
  of Primordial Inhomogeneity Determined from the Four-Year COBE DMR Sky Maps}.
  1996, \apjl, 464, L11, \eprint{arXiv:astro-ph/9601063}

\bibitem[{{G{\'o}rski} {et~al.}(2005){G{\'o}rski}, {Hivon}, {Banday},
  {Wandelt}, {Hansen}, {Reinecke}, \& {Bartelmann}}]{Gorski:2005}
{G{\'o}rski}, K.~M., {Hivon}, E., {Banday}, A.~J., {et~al.}, {HEALPix: A
  Framework for High-Resolution Discretization and Fast Analysis of Data
  Distributed on the Sphere}. 2005, \apj, 622, 759,
  \eprint{arXiv:astro-ph/0409513}

\bibitem[{{Haslam} {et~al.}(1982){Haslam}, {Salter}, {Stoffel}, \&
  {Wilson}}]{Haslam:1982}
{Haslam}, C.~G.~T., {Salter}, C.~J., {Stoffel}, H., \& {Wilson}, W.~E., {A 408
  MHz all-sky continuum survey. II - The atlas of contour maps}. 1982, \aaps,
  47, 1

\bibitem[{{Hauser} {et~al.}(1998){Hauser}, {Arendt}, {Kelsall}, {Dwek},
  {Odegard}, {Weiland}, {Freudenreich}, {Reach}, {Silverberg}, {Moseley},
  {Pei}, {Lubin}, {Mather}, {Shafer}, {Smoot}, {Weiss}, {Wilkinson}, \&
  {Wright}}]{Hauser:1998}
{Hauser}, M.~G., {Arendt}, R.~G., {Kelsall}, T., {et~al.}, {The COBE Diffuse
  Infrared Background Experiment Search for the Cosmic Infrared Background. I.
  Limits and Detections}. 1998, \apj, 508, 25, \eprint{arXiv:astro-ph/9806167}

\bibitem[{{Hildebrand} {et~al.}(1999){Hildebrand}, {Dotson}, {Dowell},
  {Schleuning}, \& {Vaillancourt}}]{Hildebrand:1999}
{Hildebrand}, R.~H., {Dotson}, J.~L., {Dowell}, C.~D., {Schleuning}, D.~A., \&
  {Vaillancourt}, J.~E., {The Far-Infrared Polarization Spectrum: First Results
  and Analysis}. 1999, \apj, 516, 834

\bibitem[{{Hoang} {et~al.}(2011){Hoang}, {Lazarian}, \& {Draine}}]{Hoang:2011}
{Hoang}, T., {Lazarian}, A., \& {Draine}, B.~T., {Spinning Dust Emission:
  Effects of Irregular Grain Shape, Transient Heating, and Comparison with
  Wilkinson Microwave Anisotropy Probe Results}. 2011, \apj, 741, 87,
  \eprint{1105.2302}

\bibitem[{{Hoang} {et~al.}(2013){Hoang}, {Lazarian}, \& {Martin}}]{Hoang:2013}
{Hoang}, T., {Lazarian}, A., \& {Martin}, P.~G., {Constraint on the
  Polarization of Electric Dipole Emission from Spinning Dust}. 2013, \apj,
  779, 152, \eprint{1305.0276}

\bibitem[{{Hoyland} {et~al.}(2012){Hoyland}, {Aguiar-Gonz{\'a}lez}, {Aja},
  {Ari{\~n}o}, {Artal}, {Barreiro}, {Blackhurst}, {Cagigas}, {Cano de Diego},
  {Casas}, {Davis}, {Dickinson}, {Arriaga}, {Fernandez-Cobos}, {de la Fuente},
  {G{\'e}nova-Santos}, {G{\'o}mez}, {Gomez}, {G{\'o}mez-Re{\~n}asco},
  {Grainge}, {Harper}, {Herran}, {Herreros}, {Herrera}, {Hobson}, {Lasenby},
  {Lopez-Caniego}, {L{\'o}pez-Caraballo}, {Maffei}, {Martinez-Gonzalez},
  {McCulloch}, {Melhuish}, {Mediavilla}, {Murga}, {Ortiz}, {Piccirillo},
  {Pisano}, {Rebolo-L{\'o}pez}, {Rubi{\~n}o-Martin}, {Ruiz}, {Sanchez de la
  Rosa}, {Sanquirce}, {Vega-Moreno}, {Vielva}, {Viera-Curbelo}, {Villa},
  {Vizcarg{\"u}enaga}, \& {Watson}}]{Hoyland:2012}
{Hoyland}, R.~J., {Aguiar-Gonz{\'a}lez}, M., {Aja}, B., {et~al.} 2012, in
  Society of Photo-Optical Instrumentation Engineers (SPIE) Conference Series,
  Vol. 8452, Society of Photo-Optical Instrumentation Engineers (SPIE)
  Conference Series

\bibitem[{{Jones}(2012)}]{Jones:2012}
{Jones}, A.~P., {Variations on a theme - the evolution of hydrocarbon solids.
  III. Size-dependent properties - the optEC$_{(s)}$(a) model}. 2012, \aap,
  542, A98

\bibitem[{{Jones} {et~al.}(2013){Jones}, {Fanciullo}, {K{\"o}hler},
  {Verstraete}, {Guillet}, {Bocchio}, \& {Ysard}}]{Jones:2013}
{Jones}, A.~P., {Fanciullo}, L., {K{\"o}hler}, M., {et~al.}, {The evolution of
  amorphous hydrocarbons in the ISM: dust modelling from a new vantage point}.
  2013, \aap, 558, A62

\bibitem[{{Kalberla} {et~al.}(2005){Kalberla}, {Burton}, {Hartmann}, {Arnal},
  {Bajaja}, {Morras}, \& {P{\"o}ppel}}]{Kalberla:2005}
{Kalberla}, P.~M.~W., {Burton}, W.~B., {Hartmann}, D., {et~al.}, {The
  Leiden/Argentine/Bonn (LAB) Survey of Galactic HI. Final data release of the
  combined LDS and IAR surveys with improved stray-radiation corrections}.
  2005, \aap, 440, 775, \eprint{arXiv:astro-ph/0504140}

\bibitem[{{King} {et~al.}(2010){King}, {Copley}, {Davies}, {Davis},
  {Dickinson}, {Hafez}, {Holler}, {John}, {Jonas}, {Jones}, {Leahy},
  {Muchovej}, {Pearson}, {Readhead}, {Stevenson}, \& {Taylor}}]{King:2010}
{King}, O.~G., {Copley}, C., {Davies}, R., {et~al.} 2010, in Society of
  Photo-Optical Instrumentation Engineers (SPIE) Conference Series, Vol. 7741,
  Society of Photo-Optical Instrumentation Engineers (SPIE) Conference Series

\bibitem[{{Kogut} {et~al.}(1996){Kogut}, {Banday}, {Bennett}, {Gorski},
  {Hinshaw}, {Smoot}, \& {Wright}}]{Kogut:1996}
{Kogut}, A., {Banday}, A.~J., {Bennett}, C.~L., {et~al.}, {Microwave Emission
  at High Galactic Latitudes in the Four-Year DMR Sky Maps}. 1996, \apjl, 464,
  L5, \eprint{arXiv:astro-ph/9601060}

\bibitem[{{Kogut} {et~al.}(2007){Kogut}, {Dunkley}, {Bennett}, {Dor{\'e}},
  {Gold}, {Halpern}, {Hinshaw}, {Jarosik}, {Komatsu}, {Nolta}, {Odegard},
  {Page}, {Spergel}, {Tucker}, {Weiland}, {Wollack}, \& {Wright}}]{kogut2007}
{Kogut}, A., {Dunkley}, J., {Bennett}, C.~L., {et~al.}, {Three-Year Wilkinson
  Microwave Anisotropy Probe (WMAP) Observations: Foreground Polarization}.
  2007, \apj, 665, 355, \eprint{0704.3991}

\bibitem[{{Lagache}(2003)}]{Lagache:2003a}
{Lagache}, G., {The large-scale anomalous microwave emission revisited by
  WMAP.} 2003, \aap, 405, 813, \eprint{arXiv:astro-ph/0303335}

\bibitem[{{Lazarian} \& {Draine}(2000)}]{Lazarian:2000}
{Lazarian}, A. \& {Draine}, B.~T., {Resonance Paramagnetic Relaxation and
  Alignment of Small Grains}. 2000, \apjl, 536, L15,
  \eprint{arXiv:astro-ph/0003312}

\bibitem[{{Lehtinen} {et~al.}(2010){Lehtinen}, {Juvela}, \&
  {Mattila}}]{Lehtinen:2010}
{Lehtinen}, K., {Juvela}, M., \& {Mattila}, K., {Scattered H{$\alpha$} emission
  from a large translucent cloud G294-24}. 2010, \aap, 517, A79

\bibitem[{{Leitch} {et~al.}(1997){Leitch}, {Readhead}, {Pearson}, \&
  {Myers}}]{Leitch:1997}
{Leitch}, E.~M., {Readhead}, A.~C.~S., {Pearson}, T.~J., \& {Myers}, S.~T., {An
  Anomalous Component of Galactic Emission}. 1997, \apjl, 486, L23,
  \eprint{arXiv:astro-ph/9705241}

\bibitem[{{Liu} {et~al.}(2014){Liu}, {Mertsch}, \& {Sarkar}}]{Liu:2014}
{Liu}, H., {Mertsch}, P., \& {Sarkar}, S., {Fingerprints of Galactic Loop I on
  the Cosmic Microwave Background}. 2014, \apjl, 789, L29, \eprint{1404.1899}

\bibitem[{{Longair}(1994)}]{Longair:1994}
{Longair}, M.~S. 1994, {High energy astrophysics. Volume 2. Stars, the Galaxy
  and the interstellar medium.} (Cambridge University Press)

\bibitem[{{L{\'o}pez-Caraballo} {et~al.}(2011){L{\'o}pez-Caraballo},
  {Rubi{\~n}o-Mart{\'{\i}}n}, {Rebolo}, \&
  {G{\'e}nova-Santos}}]{Lopez-Caraballo:2011}
{L{\'o}pez-Caraballo}, C.~H., {Rubi{\~n}o-Mart{\'{\i}}n}, J.~A., {Rebolo}, R.,
  \& {G{\'e}nova-Santos}, R., {Constraints on the Polarization of the Anomalous
  Microwave Emission in the Perseus Molecular Complex from Seven-year WMAP
  Data}. 2011, \apj, 729, 25, \eprint{1011.1242}

\bibitem[{{Macellari} {et~al.}(2011){Macellari}, {Pierpaoli}, {Dickinson}, \&
  {Vaillancourt}}]{Macellari:2011}
{Macellari}, N., {Pierpaoli}, E., {Dickinson}, C., \& {Vaillancourt}, J.~E.,
  {Galactic foreground contributions to the 5-year Wilkinson Microwave
  Anisotropy Probe maps}. 2011, \mnras, 418, 888

\bibitem[{{Martin}(2007)}]{Martin:2007}
{Martin}, P.~G. 2007, in EAS Publications Series, Vol.~23, EAS Publications
  Series, ed. M.-A. {Miville-Desch{\^e}nes} \& F.~{Boulanger}, 165--188

\bibitem[{{Mennella} {et~al.}(2011){Mennella}, {Butler}, {Curto}, {Cuttaia},
  {Davis}, {Dick}, {Frailis}, {Galeotta}, {Gregorio}, {Kurki-Suonio},
  {Lawrence}, {Leach}, {Leahy}, {Lowe}, {Maino}, {Mandolesi}, {Maris},
  {Mart{\'{\i}}nez-Gonz{\'a}lez}, {Meinhold}, {Morgante}, {Pearson},
  {Perrotta}, {Polenta}, {Poutanen}, {Sandri}, {Seiffert}, {Suur-Uski},
  {Tavagnacco}, {Terenzi}, {Tomasi}, {Valiviita}, {Villa}, {Watson},
  {Wilkinson}, {Zacchei}, {Zonca}, {Aja}, {Artal}, {Baccigalupi}, {Banday},
  {Barreiro}, {Bartlett}, {Bartolo}, {Battaglia}, {Bennett}, {Bonaldi},
  {Bonavera}, {Borrill}, {Bouchet}, {Burigana}, {Cabella}, {Cappellini},
  {Chen}, {Colombo}, {Cruz}, {Danese}, {D'Arcangelo}, {Davies}, {de Gasperis},
  {de Rosa}, {de Zotti}, {Dickinson}, {Diego}, {Donzelli}, {Efstathiou},
  {En{\ss}lin}, {Eriksen}, {Falvella}, {Finelli}, {Foley}, {Franceschet},
  {Franceschi}, {Gaier}, {G{\'e}nova-Santos}, {George}, {G{\'o}mez},
  {Gonz{\'a}lez-Nuevo}, {G{\'o}rski}, {Gruppuso}, {Hansen}, {Herranz},
  {Herreros}, {Hoyland}, {Hughes}, {Jewell}, {Jukkala}, {Juvela},
  {Kangaslahti}, {Keih{\"a}nen}, {Keskitalo}, {Kilpia}, {Kisner}, {Knoche},
  {Knox}, {Laaninen}, {L{\"a}hteenm{\"a}ki}, {Lamarre}, {Leonardi},
  {Le{\'o}n-Tavares}, {Leutenegger}, {Lilje}, {L{\'o}pez-Caniego}, {Lubin},
  {Malaspina}, {Marinucci}, {Massardi}, {Matarrese}, {Matthai}, {Melchiorri},
  {Mendes}, {Miccolis}, {Migliaccio}, {Mitra}, {Moss}, {Natoli}, {Nesti},
  {N{\o}rgaard-Nielsen}, {Pagano}, {Paladini}, {Paoletti}, {Partridge},
  {Pasian}, {Pettorino}, {Pietrobon}, {Pospieszalski}, {Pr{\'e}zeau}, {Prina},
  {Procopio}, {Puget}, {Quercellini}, {Rachen}, {Rebolo}, {Reinecke},
  {Ricciardi}, {Robbers}, {Rocha}, {Roddis}, {Rubino-Mart{\'{\i}}n},
  {Savelainen}, {Scott}, {Silvestri}, {Simonetto}, {Sjoman}, {Smoot}, {Sozzi},
  {Stringhetti}, {Tauber}, {Tofani}, {Toffolatti}, {Tuovinen}, {T{\"u}rler},
  {Umana}, {Valenziano}, {Varis}, {Vielva}, {Vittorio}, {Wade}, {Watson},
  {White}, \& {Winder}}]{planck2011-1.4}
{Mennella}, A., {Butler}, R.~C., {Curto}, A., {et~al.}, {Planck early results.
  III. First assessment of the Low Frequency Instrument in-flight performance}.
  2011, \aap, 536, A3, \eprint{1101.2038}

\bibitem[{{Meny} {et~al.}(2007){Meny}, {Gromov}, {Boudet}, {Bernard},
  {Paradis}, \& {Nayral}}]{Meny:2007}
{Meny}, C., {Gromov}, V., {Boudet}, N., {et~al.}, {Far-infrared to millimeter
  astrophysical dust emission. I. A model based on physical properties of
  amorphous solids}. 2007, \aap, 468, 171, \eprint{arXiv:astro-ph/0701226}

\bibitem[{{Miville-Desch{\^e}nes} {et~al.}(2008){Miville-Desch{\^e}nes},
  {Ysard}, {Lavabre}, {Ponthieu}, {Mac{\'{\i}}as-P{\'e}rez}, {Aumont}, \&
  {Bernard}}]{MAMD:2008}
{Miville-Desch{\^e}nes}, M.-A., {Ysard}, N., {Lavabre}, A., {et~al.},
  {Separation of anomalous and synchrotron emissions using WMAP polarization
  data}. 2008, \aap, 490, 1093, \eprint{0802.3345}

\bibitem[{{Page} {et~al.}(2007){Page}, {Hinshaw}, {Komatsu}, {Nolta},
  {Spergel}, {Bennett}, {Barnes}, {Bean}, {Dor{\'e}}, {Dunkley}, {Halpern},
  {Hill}, {Jarosik}, {Kogut}, {Limon}, {Meyer}, {Odegard}, {Peiris}, {Tucker},
  {Verde}, {Weiland}, {Wollack}, \& {Wright}}]{page2007}
{Page}, L., {Hinshaw}, G., {Komatsu}, E., {et~al.}, {Three-Year Wilkinson
  Microwave Anisotropy Probe (WMAP) Observations: Polarization Analysis}. 2007,
  \apjs, 170, 335, \eprint{astro-ph/0603450}

\bibitem[{{Paradis} {et~al.}(2011){Paradis}, {Bernard}, {M{\'e}ny}, \&
  {Gromov}}]{Paradis:2011}
{Paradis}, D., {Bernard}, J.-P., {M{\'e}ny}, C., \& {Gromov}, V., {Far-infrared
  to millimeter astrophysical dust emission. II. Comparison of the two-level
  systems (TLS) model with astronomical data}. 2011, \aap, 534, A118,
  \eprint{1107.5179}

\bibitem[{{Planck HFI Core Team}(2011)}]{planck2011-1.5}
{Planck HFI Core Team}, {Planck early results, IV. First assessment of the High
  Frequency Instrument in-flight performance}. 2011, \aap, 536, A4,
  \eprint{1101.2039}

\bibitem[{{\sorthelp{Planck Collaboration 2011A}}{Planck Collaboration
  I}(2011)}]{planck2011-1.1}
{\sorthelp{Planck Collaboration 2011A}}{Planck Collaboration I}, {Planck early
  results. I. The Planck mission}. 2011, \aap, 536, A1, \eprint{1101.2022}

\bibitem[{{\sorthelp{Planck Collaboration 2011Q}}{Planck Collaboration
  XVII}(2011)}]{planck2011-6.4b}
{\sorthelp{Planck Collaboration 2011Q}}{Planck Collaboration XVII}, {Planck
  early results. XVII. Origin of the submillimetre excess dust emission in the
  Magellanic Clouds}. 2011, \aap, 536, A17, \eprint{1101.2046}

\bibitem[{{\sorthelp{Planck Collaboration 2011T}}{Planck Collaboration
  XX}(2011)}]{planck2011-7.2}
{\sorthelp{Planck Collaboration 2011T}}{Planck Collaboration XX}, {Planck early
  results. XX. New light on anomalous microwave emission from spinning dust
  grains}. 2011, \aap, 536, A20, \eprint{1101.2031}

\bibitem[{{\sorthelp{Planck Collaboration 2011X}}{Planck Collaboration
  XXIV}(2011)}]{planck2011-7.12}
{\sorthelp{Planck Collaboration 2011X}}{Planck Collaboration XXIV}, {Planck
  early results. XXIV. Dust in the diffuse interstellar medium and the Galactic
  halo}. 2011, \aap, 536, A24, \eprint{1101.2036}

\bibitem[{{\sorthelp{Planck Collaboration 2011Y}}{Planck Collaboration
  XXV}(2011)}]{planck2011-7.13}
{\sorthelp{Planck Collaboration 2011Y}}{Planck Collaboration XXV}, {Planck
  early results. XXV. Thermal dust in nearby molecular clouds}. 2011, \aap,
  536, A25, \eprint{1101.2037}

\bibitem[{{\sorthelp{Planck Collaboration 2014A}}{Planck Collaboration
  I}(2014)}]{planck2013-p01}
{\sorthelp{Planck Collaboration 2014A}}{Planck Collaboration I},
  {\textit{Planck} 2013 results. I. Overview of products and scientific
  results}. 2014, \aap, 571, A1, \eprint{1303.5062}

\bibitem[{{\sorthelp{Planck Collaboration 2014B}}{Planck Collaboration
  II}(2014)}]{planck2013-p02}
{\sorthelp{Planck Collaboration 2014B}}{Planck Collaboration II},
  {\textit{Planck} 2013 results. II. Low Frequency Instrument data processing}.
  2014, \aap, 571, A2, \eprint{1303.5063}

\bibitem[{{\sorthelp{Planck Collaboration 2014C}}{Planck Collaboration
  III}(2014)}]{planck2013-p02a}
{\sorthelp{Planck Collaboration 2014C}}{Planck Collaboration III},
  {\textit{Planck} 2013 results. III. LFI systematic uncertainties}. 2014,
  \aap, 571, A3, \eprint{1303.5064}

\bibitem[{{\sorthelp{Planck Collaboration 2014D}}{Planck Collaboration
  IV}(2014)}]{planck2013-p02d}
{\sorthelp{Planck Collaboration 2014D}}{Planck Collaboration IV},
  {\textit{Planck} 2013 results. IV. LFI Beams and window functions}. 2014,
  \aap, 571, A4, \eprint{1303.5065}

\bibitem[{{\sorthelp{Planck Collaboration 2014F}}{Planck Collaboration
  VI}(2014)}]{planck2013-p03}
{\sorthelp{Planck Collaboration 2014F}}{Planck Collaboration VI},
  {\textit{Planck} 2013 results. VI. High Frequency Instrument data
  processing}. 2014, \aap, 571, A6, \eprint{1303.5067}

\bibitem[{{\sorthelp{Planck Collaboration 2014G}}{Planck Collaboration
  VII}(2014)}]{planck2013-p03c}
{\sorthelp{Planck Collaboration 2014G}}{Planck Collaboration VII},
  {\textit{Planck} 2013 results. VII. HFI time response and beams}. 2014, \aap,
  571, A7, \eprint{1303.5068}

\bibitem[{{\sorthelp{Planck Collaboration 2014H}}{Planck Collaboration
  VIII}(2014)}]{planck2013-p03f}
{\sorthelp{Planck Collaboration 2014H}}{Planck Collaboration VIII},
  {\textit{Planck} 2013 results. VIII. HFI photometric calibration and
  mapmaking}. 2014, \aap, 571, A8, \eprint{1303.5069}

\bibitem[{{\sorthelp{Planck Collaboration 2014I}}{Planck Collaboration
  IX}(2014)}]{planck2013-p03d}
{\sorthelp{Planck Collaboration 2014I}}{Planck Collaboration IX},
  {\textit{Planck} 2013 results. IX. HFI spectral response}. 2014, \aap, 571,
  A9, \eprint{1303.5070}

\bibitem[{{\sorthelp{Planck Collaboration 2014J}}{Planck Collaboration
  X}(2014)}]{planck2013-p03e}
{\sorthelp{Planck Collaboration 2014J}}{Planck Collaboration X},
  {\textit{Planck} 2013 results. X. HFI energetic particle effects:
  characterization, removal, and simulation}. 2014, \aap, 571, A10,
  \eprint{1303.5071}

\bibitem[{{\sorthelp{Planck Collaboration 2014K}}{Planck Collaboration
  XI}(2014)}]{planck2013-p06b}
{\sorthelp{Planck Collaboration 2014K}}{Planck Collaboration XI},
  {\textit{Planck} 2013 results. XI. All-sky model of thermal dust emission}.
  2014, \aap, 571, A11, \eprint{1312.1300}

\bibitem[{{\sorthelp{Planck Collaboration 2014L}}{Planck Collaboration
  XII}(2014)}]{planck2013-p06}
{\sorthelp{Planck Collaboration 2014L}}{Planck Collaboration XII},
  {\textit{Planck} 2013 results. XII. Diffuse component separation}. 2014,
  \aap, 571, A12, \eprint{1303.5072}

\bibitem[{{\sorthelp{Planck Collaboration 2014M}}{Planck Collaboration
  XIII}(2014)}]{planck2013-p03a}
{\sorthelp{Planck Collaboration 2014M}}{Planck Collaboration XIII},
  {\textit{Planck} 2013 results. XIII. Galactic CO emission}. 2014, \aap, 571,
  A13, \eprint{1303.5073}

\bibitem[{{\sorthelp{Planck Collaboration 2014N}}{Planck Collaboration
  XIV}(2014)}]{planck2013-pip88}
{\sorthelp{Planck Collaboration 2014N}}{Planck Collaboration XIV},
  {\textit{Planck} 2013 results. XIV. Zodiacal emission}. 2014, \aap, 571, A14,
  \eprint{1303.5074}

\bibitem[{{\sorthelp{Planck Collaboration 2014O}}{Planck Collaboration
  XV}(2014)}]{planck2013-p08}
{\sorthelp{Planck Collaboration 2014O}}{Planck Collaboration XV},
  {\textit{Planck} 2013 results. XV. CMB power spectra and likelihood}. 2014,
  \aap, 571, A15, \eprint{1303.5075}

\bibitem[{{\sorthelp{Planck Collaboration 2014P}}{Planck Collaboration
  XVI}(2014)}]{planck2013-p11}
{\sorthelp{Planck Collaboration 2014P}}{Planck Collaboration XVI},
  {\textit{Planck} 2013 results. XVI. Cosmological parameters}. 2014, \aap,
  571, A16, \eprint{1303.5076}

\bibitem[{{\sorthelp{Planck Collaboration 2014ZE}}{Planck Collaboration
  XXX}(2014)}]{planck2013-pip56}
{\sorthelp{Planck Collaboration 2014ZE}}{Planck Collaboration XXX},
  {\textit{Planck} 2013 results. XXX. Cosmic infrared background measurements
  and implications for star formation}. 2014, \aap, 571, A30,
  \eprint{1309.0382}

\bibitem[{{\sorthelp{Planck Collaboration IntI}}{Planck Collaboration Int.
  IX}(2013)}]{planck2012-IX}
{\sorthelp{Planck Collaboration IntI}}{Planck Collaboration Int. IX}, {Planck
  intermediate results. IX. Detection of the Galactic haze with Planck}. 2013,
  \aap, 554, A139, \eprint{1208.5483}

\bibitem[{{\sorthelp{Planck Collaboration IntL}}{Planck Collaboration Int.
  XII}(2013)}]{planck2013-XII}
{\sorthelp{Planck Collaboration IntL}}{Planck Collaboration Int. XII}, {Planck
  intermediate results. XII. Diffuse Galactic components in the Gould Belt
  System}. 2013, \aap, 557, A53, \eprint{1301.5839}

\bibitem[{{\sorthelp{Planck Collaboration IntN}}{Planck Collaboration Int.
  XIV}(2014)}]{planck2013-XIV}
{\sorthelp{Planck Collaboration IntN}}{Planck Collaboration Int. XIV}, {Planck
  intermediate results. XIV. Dust emission at millimetre wavelengths in the
  Galactic plane}. 2014, \aap, 564, A45, \eprint{1307.6815}

\bibitem[{{\sorthelp{Planck Collaboration IntO}}{Planck Collaboration Int.
  XV}(2014)}]{planck2013-XV}
{\sorthelp{Planck Collaboration IntO}}{Planck Collaboration Int. XV}, {Planck
  intermediate results. XV. A study of anomalous microwave emission in Galactic
  clouds}. 2014, \aap, 565, A103, \eprint{1309.1357}

\bibitem[{{\sorthelp{Planck Collaboration IntQ}}{Planck Collaboration Int.
  XVII}(2014)}]{planck2013-XVII}
{\sorthelp{Planck Collaboration IntQ}}{Planck Collaboration Int. XVII}, {Planck
  intermediate results. XVII. Emission of dust in the diffuse interstellar
  medium from the far-infrared to microwave frequencies}. 2014, \aap, 566, A55,
  \eprint{1312.5446}

\bibitem[{{\sorthelp{Planck Collaboration IntS}}{Planck Collaboration Int.
  XIX}(2015)}]{planck2014-XIX}
{\sorthelp{Planck Collaboration IntS}}{Planck Collaboration Int. XIX}, {Planck
  intermediate results. XIX. An overview of the polarized thermal emission from
  Galactic dust}. 2015, \aap, 576, A104, \eprint{1405.0871}

\bibitem[{{\sorthelp{Planck Collaboration IntT}}{Planck Collaboration Int.
  XX}(2015)}]{planck2014-XX}
{\sorthelp{Planck Collaboration IntT}}{Planck Collaboration Int. XX}, {Planck
  intermediate results. XX. Comparison of polarized thermal emission from
  Galactic dust with simulations of MHD turbulence}. 2015, \aap, 576, A105,
  \eprint{1405.0872}

\bibitem[{{\sorthelp{Planck Collaboration IntZD}}{Planck Collaboration Int.
  XXIX}(2014)}]{planck2014-XXIX}
{\sorthelp{Planck Collaboration IntZD}}{Planck Collaboration Int. XXIX},
  {Planck intermediate results. XXIX. All-sky dust modelling with
  \textit{Planck}, IRAS, and WISE observations}. 2014, \aap, submitted,
  \eprint{1409.2495}

\bibitem[{{\sorthelp{Planck Collaboration IntZE}}{Planck Collaboration Int.
  XXX}(2014)}]{planck2014-XXX}
{\sorthelp{Planck Collaboration IntZE}}{Planck Collaboration Int. XXX}, {Planck
  intermediate results. XXX. The angular power spectrum of polarized dust
  emission at intermediate and high Galactic latitudes}. 2014, \aap, in press,
  \eprint{1409.5738}

\bibitem[{{Reach} {et~al.}(1998){Reach}, {Wall}, \& {Odegard}}]{Reach:1998}
{Reach}, W.~T., {Wall}, W.~F., \& {Odegard}, N., {Infrared Excess and Molecular
  Clouds: A Comparison of New Surveys of Far-Infrared and H I 21 Centimeter
  Emission at High Galactic Latitudes}. 1998, \apj, 507, 507,
  \eprint{arXiv:astro-ph/9802169}

\bibitem[{{Rubi{\~n}o-Mart{\'{\i}}n} {et~al.}(2012){Rubi{\~n}o-Mart{\'{\i}}n},
  {L{\'o}pez-Caraballo}, {G{\'e}nova-Santos}, \& {Rebolo}}]{Rubino:2012}
{Rubi{\~n}o-Mart{\'{\i}}n}, J.~A., {L{\'o}pez-Caraballo}, C.~H.,
  {G{\'e}nova-Santos}, R., \& {Rebolo}, R., {Observations of the Polarisation
  of the Anomalous Microwave Emission: A Review}. 2012, Advances in Astronomy,
  2012

\bibitem[{{Rybicki} \& {Lightman}(1979)}]{Rybicki:1979}
{Rybicki}, G.~B. \& {Lightman}, A.~P. 1979, {Radiative processes in
  astrophysics} (Wiley-Interscience)

\bibitem[{{Seon} \& {Witt}(2012)}]{Seon:2012}
{Seon}, K.-I. \& {Witt}, A.~N., {On the Origins of the Diffuse H{$\alpha$}
  Emission: Ionized Gas or Dust-scattered H{$\alpha$} Halos?} 2012, \apj, 758,
  109, \eprint{1208.5645}

\bibitem[{{Silsbee} {et~al.}(2011){Silsbee}, {Ali-Ha{\"\i}moud}, \&
  {Hirata}}]{Silsbee:2011}
{Silsbee}, K., {Ali-Ha{\"\i}moud}, Y., \& {Hirata}, C.~M., {Spinning dust
  emission: the effect of rotation around a non-principal axis}. 2011, \mnras,
  411, 2750, \eprint{1003.4732}

\bibitem[{{Smith} {et~al.}(2000){Smith}, {Wright}, {Aitken}, {Roche}, \&
  {Hough}}]{Smith:2000}
{Smith}, C.~H., {Wright}, C.~M., {Aitken}, D.~K., {Roche}, P.~F., \& {Hough},
  J.~H., {Studies in mid-infrared spectropolarimetry - II. An atlas of
  spectra}. 2000, \mnras, 312, 327

\bibitem[{{Tauber} {et~al.}(2010){Tauber}, {Mandolesi}, {Puget}, {Banos},
  {Bersanelli}, {Bouchet}, {Butler}, {Charra}, {Crone}, {Dodsworth}, \&
  et~al.}]{Tauber:2010}
{Tauber}, J.~A., {Mandolesi}, N., {Puget}, J.-L., {et~al.}, {Planck pre-launch
  status: The Planck mission}. 2010, \aap, 520, A1

\bibitem[{{Vaillancourt}(2002)}]{Vaillancourt:2002}
{Vaillancourt}, J.~E., {Analysis of the Far-Infrared/Submillimeter Polarization
  Spectrum Based on Temperature Maps of Orion}. 2002, \apjs, 142, 53

\bibitem[{{Vaillancourt} {et~al.}(2008){Vaillancourt}, {Dowell}, {Hildebrand},
  {Kirby}, {Krejny}, {Li}, {Novak}, {Houde}, {Shinnaga}, \&
  {Attard}}]{Vaillancourt:2008}
{Vaillancourt}, J.~E., {Dowell}, C.~D., {Hildebrand}, R.~H., {et~al.}, {New
  Results on the Submillimeter Polarization Spectrum of the Orion Molecular
  Cloud}. 2008, \apjl, 679, L25, \eprint{0803.4185}

\bibitem[{{Vaillancourt} \& {Matthews}(2012)}]{Vaillancourt:2012}
{Vaillancourt}, J.~E. \& {Matthews}, B.~C., {Submillimeter Polarization of
  Galactic Clouds: A Comparison of 350 {$\mu$}m and 850 {$\mu$}m Data}. 2012,
  \apjs, 201, 13, \eprint{1204.1378}

\bibitem[{{Witt} {et~al.}(2010){Witt}, {Gold}, {Barnes}, {DeRoo}, {Vijh}, \&
  {Madsen}}]{Witt:2010}
{Witt}, A.~N., {Gold}, B., {Barnes}, III, F.~S., {et~al.}, {On the Origins of
  the High-latitude H{$\alpha$} Background}. 2010, \apj, 724, 1551,
  \eprint{1010.4361}

\bibitem[{{Wolff} {et~al.}(1997){Wolff}, {Clayton}, {Kim}, {Martin}, \&
  {Anderson}}]{Wolff:1997}
{Wolff}, M.~J., {Clayton}, G.~C., {Kim}, S.-H., {Martin}, P.~G., \& {Anderson},
  C.~M., {Ultraviolet Interstellar Linear Polarization. III. Features}. 1997,
  \apj, 478, 395

\bibitem[{{Ysard} {et~al.}(2010){Ysard}, {Miville-Desch{\^e}nes}, \&
  {Verstraete}}]{Ysard:2010}
{Ysard}, N., {Miville-Desch{\^e}nes}, M.~A., \& {Verstraete}, L., {Probing the
  origin of the microwave anomalous foreground}. 2010, \aap, 509, L1,
  \eprint{0906.3360}

\end{thebibliography}

\appendix

\section{Derivation of the CC coefficients} \label{sec:math_cc}

This appendix details how we compute the dust SED using the CC analysis in Sect.~\ref{sec:cc_em}. For simplicity, we present the simplest case for intensity from the fit with one template. 

We minimize the \chiI\ between the data and the $353$\,GHz template maps, as expressed in Eq.~\eqref{eq:4.1}. The CC coefficient is then given as
\begin{equation}
[\alphaInu]_{353}^{\rm 1T} = \frac{\sum_{k=1}^{N_{\rm pix}} \hat{\nI}(k) \ . \ \hatI_{353}(k)}{ \sum_{k=1}^{N_{\rm pix}} \ \hatI_{353}(k)^2} \ , \label{eq:a1.2} 
\end{equation}
where $\hat{\nI}$ and $\hat{I}_{353}$ are the data and $353$\,GHz template with mean values (computed over the $N_{\rm pix}$) subtracted. The observed \planck\ map at a given frequency is written as the sum of the CMB signal, the Galactic signals (synchrotron, free-free, dust and AME) and noise as
\begin{equation}
\nI(k) = I^{\rm c} (k) + I^{\rm d}_{\nu} (k) + I^{\rm s}_{\nu} (k) + I^{\rm f}_{\nu} (k) + I^{\rm e}_{\nu} (k) + I^{\rm n}_{\nu} (k) \ ,\label{eq:a1.3} 
\end{equation}
where the superscripts c, d, s, f, e, and n represent the CMB, dust, synchrotron, free-free, AME, and noise, respectively. Combining Eqs.~\eqref{eq:a1.2} and ~\eqref{eq:a1.3} we find
\begin{align}
&[\alphaInu]_{353}^{\rm 1T} = N \sum_{k=1}^{N_{\rm pix}} [ \hatI^{\rm c} (k) + \hatI^{\rm d}_{\nu} (k) + \hatI^{\rm s}_{\nu} (k) + \hatI^{\rm f}_{\nu} (k) + \hatI^{\rm e}_{\nu} (k) + \hatI^{\rm n}_{\nu} (k) ] \nonumber \\
& \times [ \hatI^{\rm c} (k) + \hatI^{\rm d}_{353} (k) + \hatI^{\rm s}_{353} (k) + \hatI^{\rm f}_{353} (k) + \hatI^{\rm e}_{353} (k) + \hatI^{\rm n}_{353} (k) ] \label{eq:a1.4} \ ,
\end{align}
where $N=(1/\sum_{k=1}^{N_{\rm pix}} \ \hatI_{353}(k)^2)$ is the normalization factor. At HFI frequencies, we can neglect the contribution of the synchrotron, free-free, AME, and noise within our global mask. This reduces Eq.~\eqref{eq:a1.4} to
\begin{align}
[\alphaInu]_{353}^{\rm 1T}& = N \sum_{k=1}^{N_{\rm pix}} \ \left ( \hatI^{\rm c} (k) + \hatI^{\rm d}_{\nu} (k) \right ) \ \times \ \left( \hatI^{\rm c} (k) + \hatI^{\rm d}_{353} (k) \right ) \nonumber \\
& = N \ \sum_{k=1}^{N_{\rm pix}} \left [ \ \hatI^{\rm c} (k) \left ( \hatI^{\rm c} (k) + \hatI^{\rm d}_{353} (k) \right ) + \ \hatI^{\rm d}_{\nu} (k) \left ( \hatI^{\rm c} (k) + \hatI^{\rm d}_{353} (k) \right ) \right ] \nonumber \\
& = \alpha(c_{353}) + N \sum_{k=1}^{N_{\rm pix}} \hatI^{\rm d}_{\nu} (k) \left ( \hatI^{\rm c} (k) + \hatI^{\rm d}_{353} (k) \right ) .
\end{align}
The CMB contribution, $\alpha(c_{353})$, is the CC coefficient obtained by correlating the CMB signal map with the $353$\,GHz template, which is independent of frequency. The dust emission at a given frequency is a scaled version of $353$\,GHz dust emission, $ \hatI^{\rm d}_{\nu} (k) = \alpha^{\rm d}_{\nu} \ \hatI^{\rm d}_{353} (k)$, where $\alpha^{\rm d}_{\nu}$ is the mean dust SED over the given sky patch. The CC coefficient is then 
\begin{align}
[\alphaInu]_{353}^{\rm 1T}& = \alpha(c_{353}) + \alpha^{\rm d}_{\nu} \ N \sum_{k=1}^{N_{\rm pix}} \hatI^{\rm d}_{353} (k) \left ( \hatI^{\rm c} (k) + \hatI^{\rm d}_{353} (k) \right ) \nonumber \\
& = \alpha(c_{353}) + \alphaInu (d_{353}) \ . \label{eq:a1.5}
\end{align}
where $\alphaInu (d_{353})$ is proportional to the mean dust SED. The colour ratio \RIT\ is then given by
\begin{align}
\RIT& = \dfrac{[\alpha^{\rm I}_{353}]_{353}^{\rm 1T} -[\alpha^{\rm I}_{100}]_{353}^{\rm 1T} }{[\alpha^{\rm I}_{217}]_{353}^{\rm 1T} -[\alpha^{\rm I}_{100}]_{353}^{\rm 1T} } \nonumber \\
& = \dfrac{\alpha^{\rm d}_{353} - \alpha^{\rm d}_{100} }{\alpha^{\rm d}_{217} - \alpha^{\rm d}_{100}} \ . \label{eq:a1.6}
\end{align}
In the ratio the scaling between the $\alphaInu (d_{353})$ and $\alpha^{\rm d}_{\nu}$ goes away. The colour ratio only depends on the dust spectral properties and not on the CMB signal. The extension of Eq.~\eqref{eq:a1.5} in the presence of AME, synchrotron, and free-free emission at the \wmap\ and LFI frequencies is given by Eq.~\eqref{eq:4.2}. 

The CMB contribution, $\alpha(c_{353})$, in the presence of inverse noise-weighting can be written as
\begin{equation}
\alpha(c_{353}) = \frac{\sum_{k=1}^{N_{\rm pix}} w_{\nu}(k) \ \hatI^{\rm c} (k) \left ( \hatI^{\rm c} (k) + \hatI^{\rm d}_{353} (k) \right )}{ \sum_{k=1}^{N_{\rm pix}} \ w_{\nu}(k) \ \hatI_{353}(k)^2} \ , 
\end{equation}
where $w_{\nu}$ is a weighting factor given by $w_{\nu}=(1/\sigma_{I_{\nu}}^2)$. If the weighting factor depends on the frequency, the CMB contribution is not strictly constant in \kcmb\ units. This effect can reach up to 2\,\%, as the weighting factors for the \wmap\ and \planck\ maps are quite different. This is not negligible compared to the dust emission at microwave frequencies. That is the reason why we do not use inverse noise-weighting in our \chiI\ minimization. 

In Sect.~\ref{sec:pol_synch} we compute the frequency dependence of synchrotron emission correlated with dust. The mean spectrum of this component is given by
\begin{align}
\langle \alphaInu (s_{353})\rangle& = N \left < \sum_{k=1}^{N_{\rm pix}} \hatI^{\rm s}_{\nu} (k) \left ( \hatI^{\rm c} (k) + \hatI^{\rm d}_{353} (k) \right ) \right > \nonumber \\
& = N \left < \sum_{k=1}^{N_{\rm pix}} \hatI^{\rm s}_{\nu} (k) \ \hatI^{\rm d}_{353} (k) \right > \ ,
\end{align}
assuming the CMB chance correlation term with synchrotron emission is zero over all the sky patches. We detect $\langle \alphaInu (s_{353})\rangle$ with high-significance in our analysis (Sect.~\ref{sec:pol_synch}), which cannot be just a chance correlation term. One would expect such a correlation, since synchrotron emission arises from the same ISM as dust emission. 

Similarly for polarization, we minimize \chiP\ between the data and $353$\,GHz Stokes \StokesQ\ and \StokesU\ maps, as given by Eq.~\eqref{eq:4.5}. The polarization CC coefficient is then given by
\begin{equation}
[\alphaPnu]_{353}^{\rm 1T} = N_{\rm P} \sum_{i=1}^2 \ \sum_{k=1}^{N_{\rm pix}} \left [ \hatQ_{\nu}(k) \ . \ \hatQ^i_{353}(k) + \hatU_{\nu}(k) \ . \ \hatU^i_{353}(k) \right ]\ , \label{eq:a1.7} 
\end{equation}
where $N_{\rm P}$ is the normalization factor for polarization. Following the same logic as described for intensity, the polarization CC coefficient at the HFI frequencies can be written as
\begin{align}
[\alphaPnu]_{353}^{\rm 1T}&= \alpha^{\rm P}(c_{353}) \nonumber \\
& + \ N_{\rm P} \sum_{i=1}^2 \ \sum_{k=1}^{N_{\rm pix}} \left [ \hatQ^{\rm d}_{\nu}(k) \ \left ( \hatQ^{\rm c} (k) + \hatQ^{\rm d}_{353}(k) + \hatQ_{353}^{\rm {n}_i} (k) \right ) \right ] \nonumber \\
& +\ N_{\rm P} \sum_{i=1}^2 \ \sum_{k=1}^{N_{\rm pix}} \left [ \hatU^{\rm d}_{\nu}(k) \ \left ( \hatU^{\rm c} (k) + \hatU^{\rm d}_{353}(k) + \hatU_{353}^{\rm {n}_i} (k) \right ) \right ]. \label{eq:a.1.8}
\end{align}
Assuming the dust polarization at a given frequency is a scaled version of $353$\,GHz dust polarization yields $\hatQ^{\rm d}_{\nu}(k) = \alpha_{\nu}^{\rm d} \ \hatQ^{\rm d}_{353}(k)$ and $\hatU^{\rm d}_{\nu}(k) = \alpha_{\nu}^{\rm d} \ \hatU^{\rm d}_{353}(k)$.
Putting this back to into Eq.~\eqref{eq:a.1.8} gives
\begin{equation}
[\alphaPnu]_{353}^{\rm 1T} =\alpha^{\rm P}(c_{353}) + \alphaPnu(d_{353}) \ .
\end{equation}
The polarized colour ratio does not depend on the CMB like Eq.~\eqref{eq:a1.5}. The polarized CMB contribution $\alpha^{\rm P}(c_{353})$ is strictly constant in \kcmb\ units if we do not apply any noise weighting, similar to the intensity analysis. To deal with the noise, we first smooth all the maps to 1\deg\ resolution and then perform correlation over local patches on the sky. To compute the uncertainty on the CC coefficients, we rely on Monte Carlo simulations, as discussed in Appendix~\ref{sec:simul}.

\section{Simulations}\label{sec:simul}

This appendix presents the simulations of the sky emission in intensity and polarization at HFI frequencies that we use to test the CC analysis. The intensity and polarization 
emission components are listed in Table~\ref{tab:A.1}. The simulations use a simplified model of dust emission in intensity and polarization that is good enough to provide a realistic 
framework to test the CC analysis. They are computed on \healpix\ pixels at $\Nside=128$ with a 1\deg\ Gaussian beam. The Monte Carlo simulations serve two specific purposes. 
First, we use them to check that the CC analysis does not introduce any bias on our estimations of the mean dust spectral indices in intensity and polarization. Second, they provide 
realistic uncertainties on the CC coefficients, which we use in the spectral fit to separate out the dust and the CMB emission (Sect.~\ref{sec:cmb_est}).

\subsection{Intensity}\label{sec:em_simul}

At HFI frequencies, the main diffuse emission components are the thermal dust, free-free, CMB, and CIB emission. The simulations also include instrumental noise.
We now describe how we simulate each of these components.

The \hi\ column density from the LAB survey \citep{Kalberla:2005} is taken as a proxy for thermal dust emission. We normalize the \hi\ data to a suitable amplitude to match the 
observed \planck\ data at $353$\,GHz and extrapolate to the other HFI frequencies using an MBB spectrum with a fixed spectral index $\betad=1.5$ and temperature $\Td=19.6$\,K 
over the whole sky.  The \hi\ data provide only a partial description of the thermal dust emission, as quoted in \citet{planck2013-XVII}. We include an additional dust component, 
spatially uncorrelated with the \hi\ data, to mimic the residuals present after adopting the IR-\hi\ correlation at 857\,GHz. The additional dust-like emission is assumed to have an 
$\ell^{-3}$ power spectrum, with a normalized amplitude of $4\pi\sigma_{857}^2$ for $\ell=2$, where $\sigma_{857}$ is the residual at 857\,GHz after applying the IR-\hi\ correlation 
and removing the CIB contribution \citep{planck2013-XVII}. The amplitude of the uncorrelated \hi\ emission is normalized at 857\,GHz, taken from \cite{planck2013-XVII}, and scaled 
to the HFI frequencies assuming $\beta_{\rm R}=2.0$ for a dust temperature of $T_{\rm R}=19.6$\,K. We use the DDD \halpha\ map as a proxy for free-free emission, which we 
compute at HFI frequencies for a spectral index $\beta_{\rm f} = -2.14$ (in \krj\ units) \citep{planck2013-XIV} and an electron temperature $T_{\rm e}=7000$\,K 
\citep{Dickinson:2003}. No dust extinction correction is applied to the DDD \halpha\ map.

For the CMB, we compute Gaussian realizations of the CMB sky from the theoretical power spectrum of the \planck\ best-fit model \citep{planck2013-p08}.  The CIB emission is 
generated using the best-fit model of CIB anisotropies at $353$\,GHz obtained directly from the \planck\ data \citep{planck2013-pip56}. We assume 100\,\% correlated CIB across all 
the HFI frequencies, assuming an MBB spectrum with $\beta_{\rm CIB}=1.3$ and $T_{\rm CIB} = 18.4$\,K. The Gaussian realizations of the instrumental noise are obtained at each 
frequency, using the noise variance maps \citep{planck2013-p03}. The noise realizations are simulated at the full resolution of the \planck\ data, before smoothing to 1\deg\ 
resolution and reducing the  pixelization from $\Nside = 2048$ to $128$. 
 
We compute 1000 realizations of sky maps of the additional dust component, together with the CMB and CIB anisotropies. Independent realizations of the instrumental noise are generated for each sky simulation at a given frequency. The dust component computed from the \hi\ map and the free-free emission traced by the \halpha\ map are kept fixed.

\begin{table*}[tmb]
\begingroup
\newdimen\tblskip \tblskip=5pt
\caption{\label{tab:A.1} The ancillary data sets and models used in the Monte Carlo simulations. }
\nointerlineskip
\vskip -3mm
\setbox\tablebox=\vbox{
   \newdimen\digitwidth 
   \setbox0=\hbox{\rm 0} 
   \digitwidth=\wd0 
   \catcode`*=\active 
   \def*{\kern\digitwidth}
   \newdimen\signwidth 
   \setbox0=\hbox{+} 
   \signwidth=\wd0 
   \catcode`!=\active 
   \def!{\kern\signwidth}

\halign{
\hbox to 1.2in{#\leaderfil}\tabskip 0.7em&
\hfil #\hfil&
\hfil #\hfil&
\hfil #\hfil&
\hfil #\hfil\tabskip=0pt\cr
\noalign{\doubleline}
\omit Components\hfil& Tracer& Model& Parameters& References\cr
\noalign{\vskip 4pt\hrule\vskip 6pt}
CMB& TT spectrum& &\planck\ best-fit parameters& \citet{planck2013-p08}\cr
Thermal dust& \hi\ template& MBB& $(\beta_{\rm d}, \Td)$ = ($1.5, 19.6$\,K)& \citet{planck2013-XVII}\cr 
IR-\hi\ excess& Residual at 857\,GHz& MBB& $(\beta_{\rm R},T_{\rm R})$ = ($2.0, 19.6$\,K)& \citet{planck2013-XVII}\cr 
Free-free& \halpha\ map& PL& $(\beta_{\rm f}, T_{\rm e})$ = ($-2.14, 7000$\,K)& \citet{Dickinson:2003}\cr
CIB& CIB spectrum& MBB& $(\beta_{\rm CIB},T_{\rm CIB})$ = ($1.3, 18.4$\,K)& \citet{planck2013-pip56}\cr
Statistical noise& Variance maps& & (\StokesI\StokesI, \StokesI\StokesQ, \StokesI\StokesU, \StokesQ\StokesQ,\StokesQ\StokesU, \StokesU\StokesU)& \citet{planck2013-p03}\cr
Polarized thermal dust& Model& MBB& $(\beta_{\rm d}, \Td)$ = ($1.6, 19.6$\,K)&\cr
\noalign{\vskip 5pt\hrule\vskip 3pt}}}
\endPlancktablewide
\par
\endgroup
\end{table*}

We analyse the 1000 simulated maps with the CC method applied to the \planck\ intensity data. We compute the mean and standard deviation of the \RIT\ values for each sky patch. 
Both are plotted in Fig.~\ref{fig:A.1} versus the local dispersion of the $353$\,GHz template, \sigmaI. There is no bias on the estimation of the mean dust spectral index \betadmmI.
We recover a mean value equal to the index of 1.5 we used for the main \hi-correlated dust component. The uncertainties on \RIT, and hence on \betadmmI, are associated with noise, CIB anisotropies, free-free emission, and the additional dust component. The $1\sigma$ dispersion of \betadmmI\ across sky patches for a given Monte Carlo realization is 0.02. This is smaller than the scatter of 0.07 measured for the \planck\ data. We interpret the difference as evidence for a small intrinsic dispersion in the spectral index of the dust emission.

\subsection{Polarization}\label{sec:pol_simul}

The simulations of the polarized sky at HFI frequencies include polarized CMB, thermal dust emission, and noise. We compute 1000 realizations of the CMB Stokes \StokesQ\ and 
\StokesU\ maps using the best-fit \planck\ model \citep{planck2013-p08}, smoothed to 1\deg\ resolution at \healpix\ resolution $\Nside=128$. Random realizations of Gaussian noise 
\StokesQ\ and \StokesU\ maps are generated at each pixel using the $3 \times 3$ noise covariance matrix defined at $\Nside=2048$. The noise maps are then smoothed to 1\deg\ 
resolution and projected on to a \healpix\ map at $\Nside=128$. We generate independent realizations of the instrumental noise to mimic the detector sets at $353$\,GHz 
(Sect.~\ref{sec:systematics}).

For polarized thermal dust emission, we use the following model:
\begin{equation}
[\StokesQ^{\rm d}_{\nu} \ , \StokesU^{\rm d}_{\nu}] = \ \polfrac_{\rm d} \left( \frac{\nu}{\nuref} \right)^{\betadmmP} \frac{B_{\nu}(\Td)}{B_{\nuref}(\Td)}\ I_{\nuref}^{\rm d} \ [\cos 2\polang_{\rm d} \ , \sin 2\polang_{\rm d}] \ .
\end{equation}
Here $\polfrac_{\rm d}$, $\StokesI^{\rm d}_{\rm ref}$, and $\polang_{\rm d}$ are the polarization fraction, the dust intensity at reference frequency, and the polarization angle, respectively. The reference frequency is $\nuref=353$\,GHz.
We fix $ \polfrac_{\rm d}$ to a constant value of 10\,\% over the whole sky. The $\StokesI^{\rm d}_{\rm ref}$ map is that obtained by \citet{planck2013-p06b} from the spectral fit to the 
high frequency \planck\ and \iras\ $100~\mu$m data, with an MBB model. For the dust polarization, we use an MBB spectrum with $\betadmmP=1.6$ and $\Td=19.6$\,K, constant 
over the whole sky. We derive $\polang_{\rm d}$ from the 1\deg\ smoothed \planck\ Stokes maps using the relation
\begin{equation}
\polang_{\rm d} = - 0.5 \times \text{atan2}\left( \StokesU^{\rm obs}_{\nu_2} -\StokesU^{\rm obs}_{\nu_1},\StokesQ^{\rm obs}_{\nu_2} -\StokesQ^{\rm obs}_{\nu_1}\right) \ .
\end{equation}
We choose $\nu_1$ and $\nu_2$ as the 143 and $353$\,GHz, respectively. The difference between the two frequencies removes the CMB contribution. 

\begin{figure}[h!]
\begin{tabular}{c}
\includegraphics[width=8.8cm]{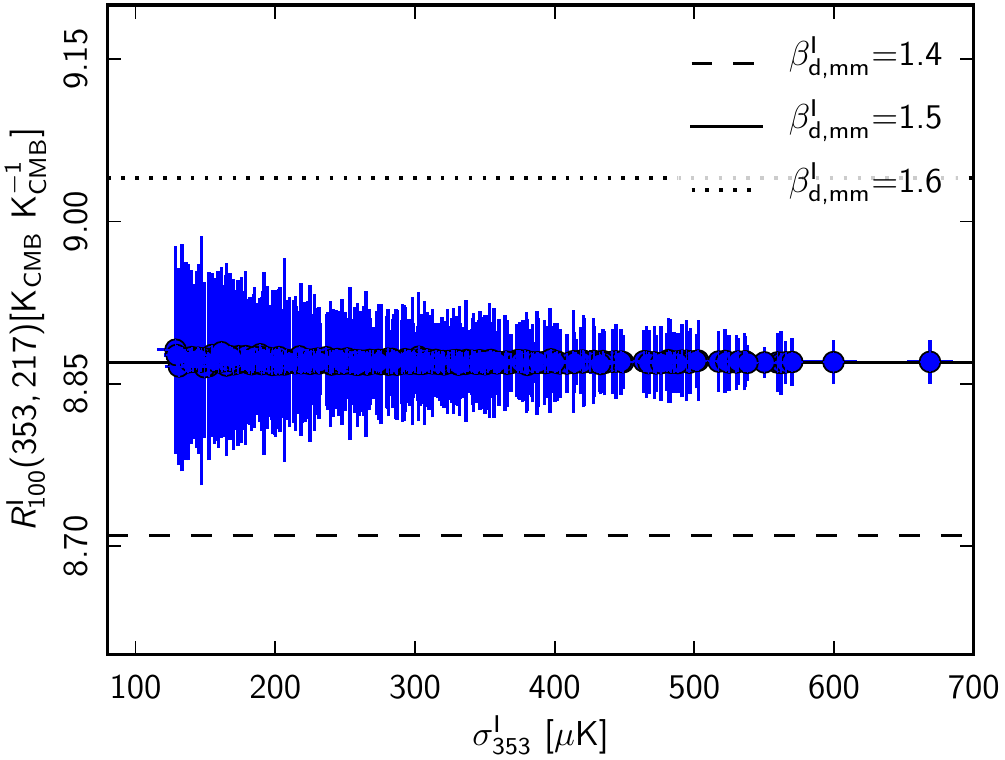} \\
\includegraphics[width=8.8cm]{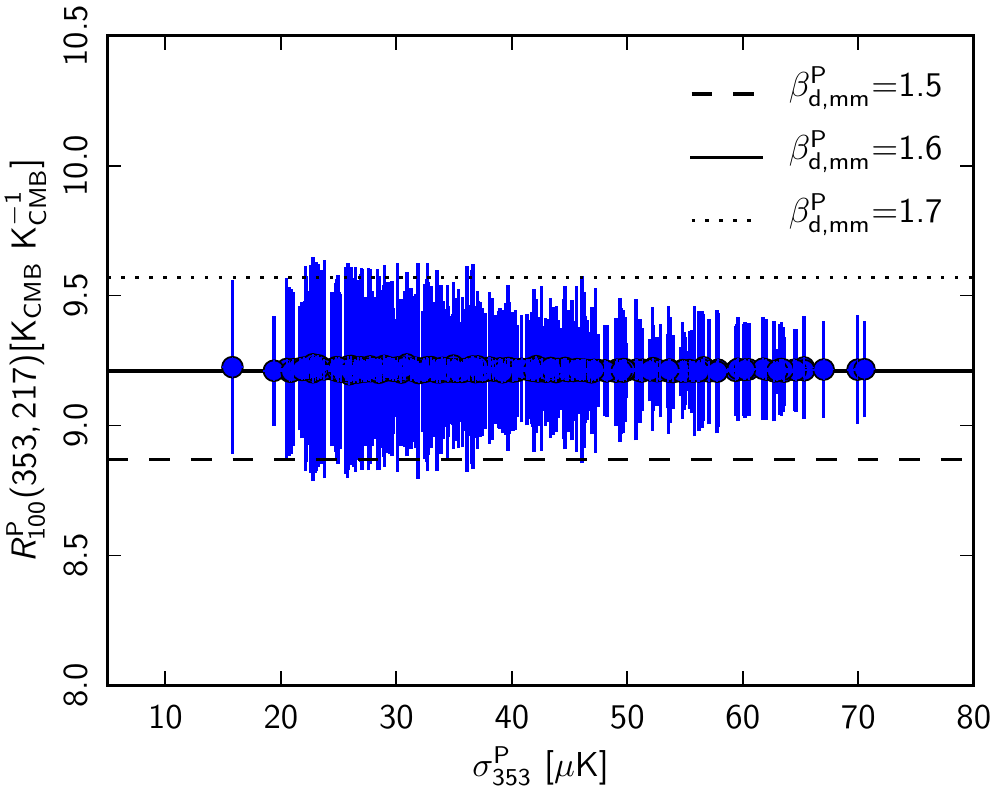}
\end{tabular}
\caption{\RI\ colour ratios from the Monte Carlo simulations for intensity (\emph{top}) and polarization (\emph{bottom}). The two plots show that the CC analysis does not introduce any bias on the estimation of \betadmm.}
\label{fig:A.1}
\end{figure}

We analyse 1000 polarized simulated maps using the CC analysis as applied to the \planck\ data. We compute the mean and the standard deviation of the \RIP\ for each sky patch.
The plot of \RIP\ versus \sigmaP\ is shown in Fig.~\ref{fig:A.1}. We find no bias in the estimation of \RIP\ and hence in the measurement of \betadmmP.  The $1\sigma$ dispersion of \betadmmP\ across sky patches for a given simulation is 0.07. 
The $1\sigma$ dispersion of \betadmmP\ from the simulations is smaller compared than that measured from the \planck\  data, because we use a simplified white noise model. However, some of the dispersion may come from the intrinsic dispersion of the polarized dust spectral index, and also additional Galactic polarized emission components, which we neglect in the simulations.

\section{Mean dust SED with dust extinction and scattering correction on the \halpha\ template}\label{sec:dust_scattering}

\begin{figure}
\includegraphics[width=8.8cm]{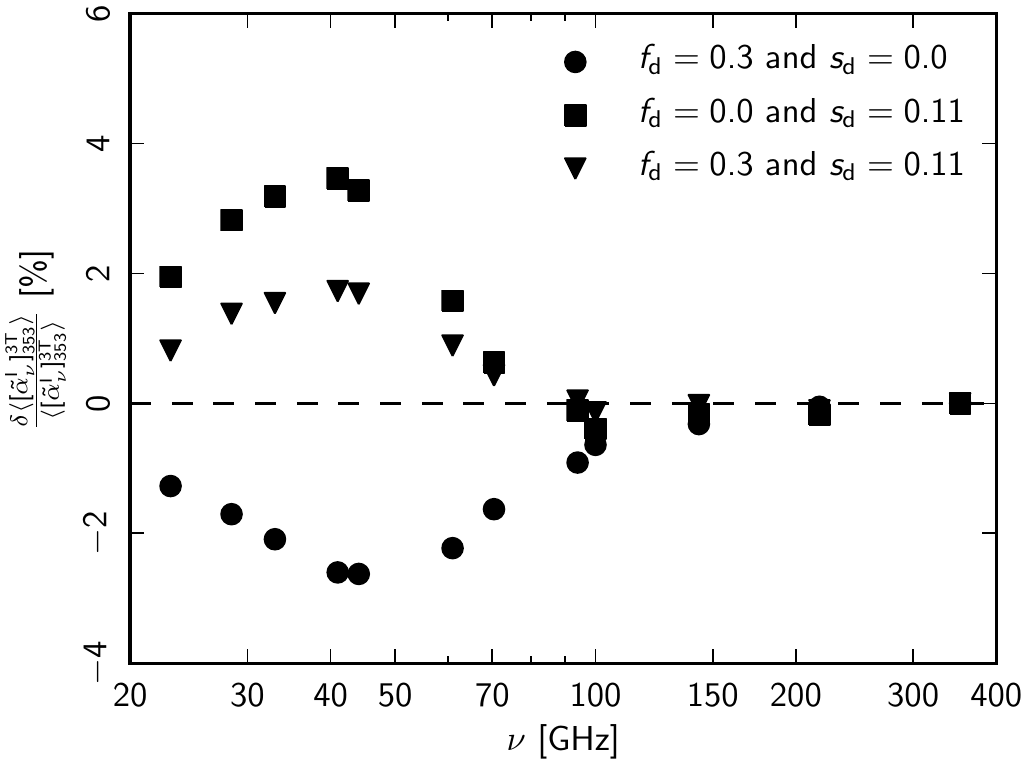} 
\caption{Fractional change in the mean dust SED with respect to the reference dust SED presented in this paper, for different combinations of \ffd\ and \sd\ corrections on the DDD \halpha\ template. }
\label{fig:B1}
\end{figure}

The mean dust SED for intensity presented in this paper is obtained using the three-template fit with no extinction and dust scattering correction from the DDD \halpha\ template. 
The effect of dust extinction (\ffd) on the \halpha\ template is described in equation~3 of \citet{Dickinson:2003}, whereas the effect of dust scattering (\sd) on the \halpha\ template is 
described in equation~26 of \citet{Bennett:2012}. The mean measured value of \sd\ is 0.11 R(\MJysr)$^{-1}$ in high Galactic latitude regions 
\citep{Lehtinen:2010,Witt:2010,Seon:2012, Brandt:2012,Bennett:2012}. To check the impact of the \ffd\ and \sd\ corrected \halpha\ template on the mean dust SED, we repeat the 
analysis with different combinations of \ffd\ and $s_{\rm d}$. The three different combinations of \ffd\ and \sd\ corrected \halpha\ templates we choose are: 
$\ffd = 0.3$ and $\sd = 0.0$ R(\MJysr)$^{-1}$; $\ffd = 0.0$ and $\sd = 0.11$ R(\MJysr)$^{-1}$; and $\ffd = 0.3$ and $\sd = 0.11$ R(\MJysr)$^{-1}$. The fractional change in the mean 
dust SED with respect to the reference dust SED ($\ffd = 0.0$ and $\sd = 0.0$ R(\MJysr)$^{-1}$) is  presented in Fig.~\ref{fig:B1}.  At higher frequencies ($\nu \ge 100$\,GHz), the 
impact of both dust extinction and scattering is negligible. However at frequencies $\nu \le 50$\,GHz, the fractional change on the mean SED can go as high as $\pm$ 4\,\%.  The 
\ffd\ and \sd\ parameters are degenerate, although their effect on the derived best-fit parameter of models DI+AI and DI+AII, listed in Table~\ref{tab:7.1}, is very small.

\section{Power spectra of the templates}\label{sec:power_spectra}

In this section, we compute the temperature power spectra of the three templates at 1\deg\ resolution: 408\,MHz; DDD \halpha; and $353$\,GHz dust template. The \planck\ 
$353$\,GHz map contains a significant component of CMB anisotropies. Taking the \smica\ map \citep{planck2013-p06} as a proxy for the CMB map, we remove its contribution from 
the $353$\,GHz total map. These spectra are combined with the SEDs from this  paper to compute the contributions of each emission component to the microwave sky emission as a 
function of angular scales in \citet{planck2013-p01} (see their Figs.~27 and 28).

\begin{figure}[h!]
\begin{tabular}{c}
\includegraphics[width=8.8cm]{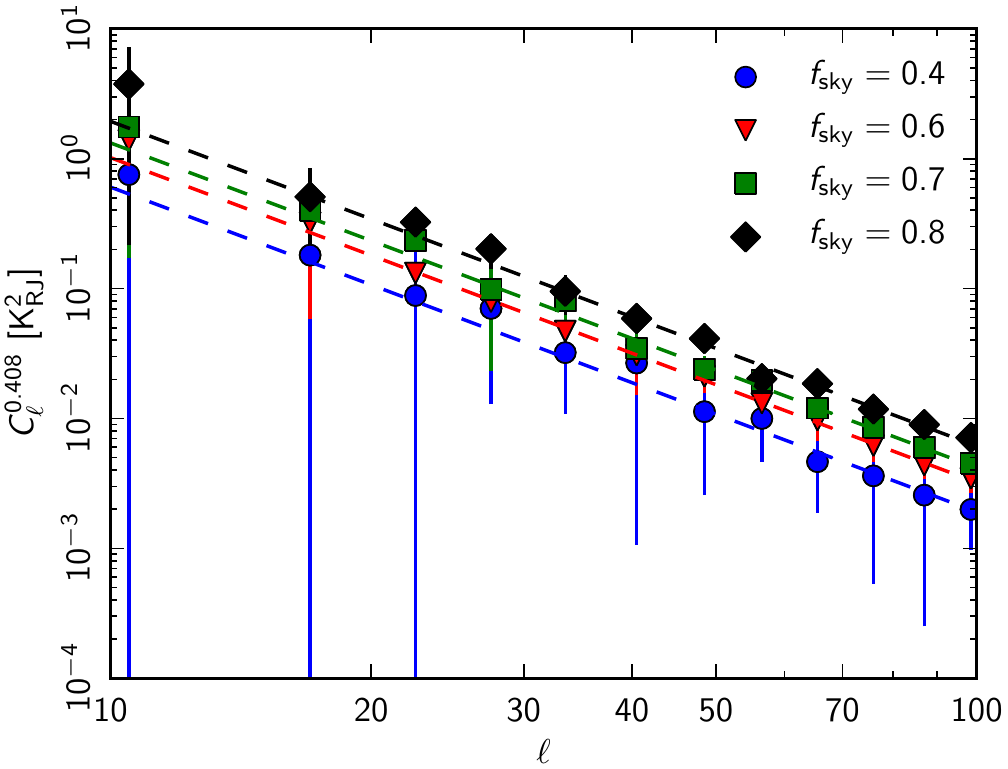} \\
\includegraphics[width=8.8cm]{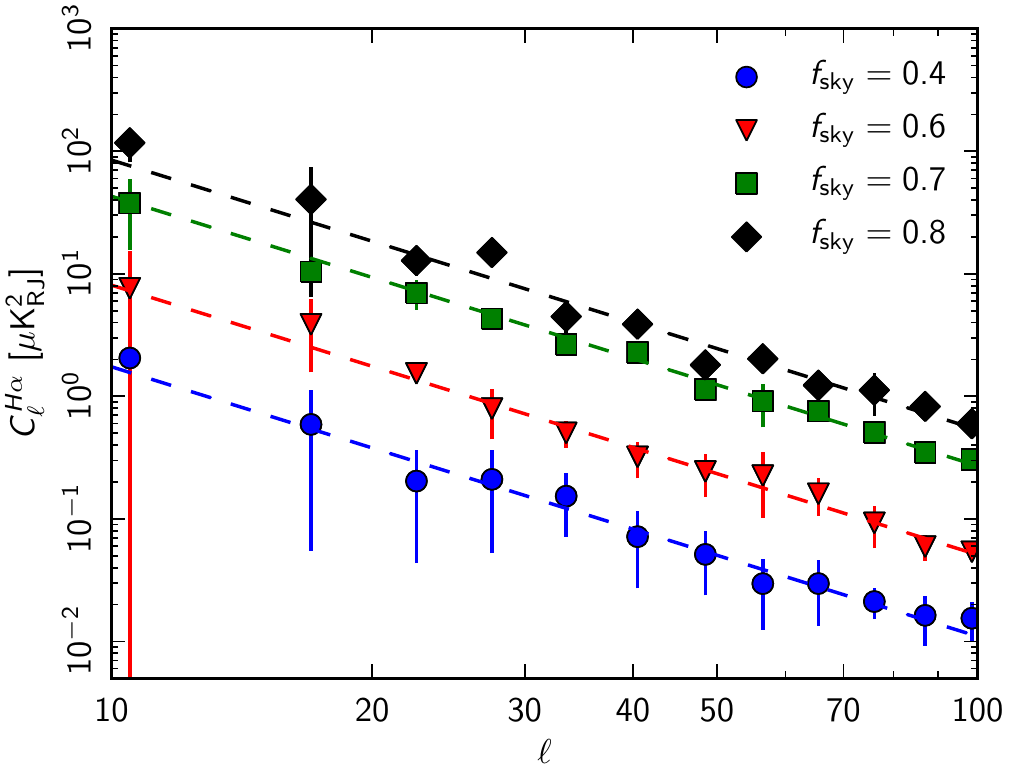} \\
\includegraphics[width=8.8cm]{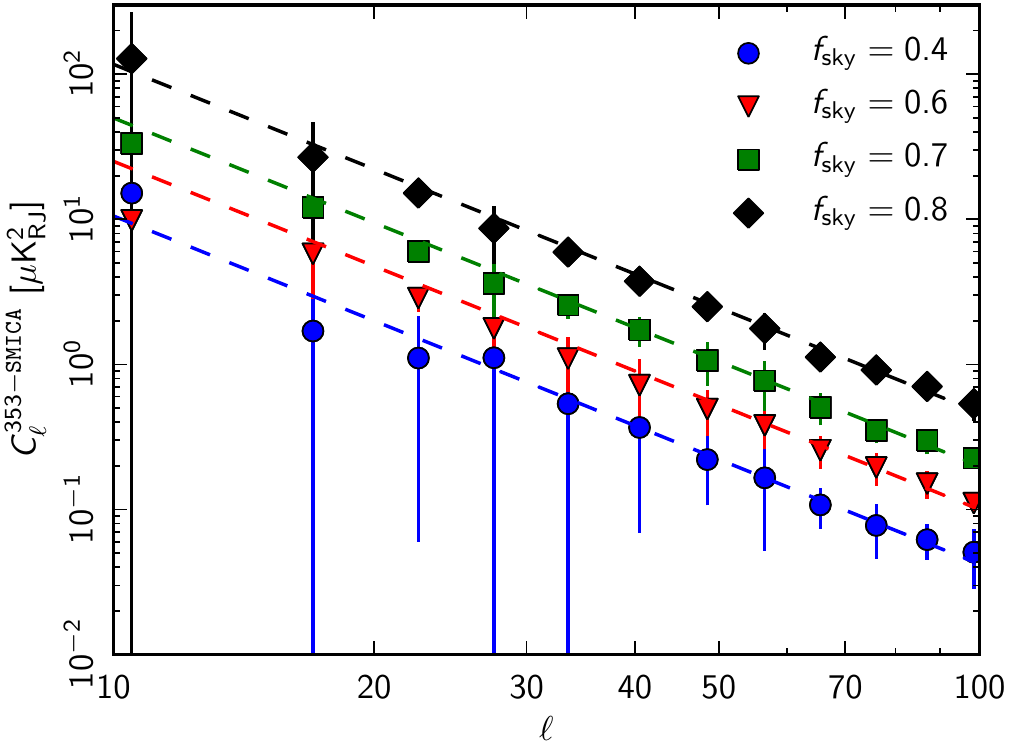}
\end{tabular}
\caption{Power spectra of the 408\,MHz, DDD \halpha, and \smica\-subtracted $353$\,GHz templates, smoothed to 1\deg\ resolution for different Galactic masks (or \fsky). }
\label{fig:D1}
\end{figure}

For the computation of the power spectra, we consider the four diffuse Galactic masks based on the percentage of the sky retained (\fsky), i.e., G40, G60, G70 and G80 
\citep{planck2013-p08}. The same set of masks have been used in the likelihood analysis of the 2013 \planck\ data release \citep{planck2013-p08}. The power spectra are 
computed only at low multipoles ($\ell <$ 100) with ${\tt PolSpice}$ v2.9.0 \citep{Chon:2004}, corrected for the masking, beam, and pixel window effect. Figure~\ref{fig:D1} presents 
binned power spectra of the three templates: 408\,MHz, DDD \halpha, and \smica-subtracted $353$\,GHz maps as a function of the Galactic masks. The uncertainties on the binned 
power spectra include only the statistical variance and not the cosmic variance.

At low multipoles, $\ell <$ 100, the three power spectra are well-fit with a power-law model. Using this assumption, the measured power spectra are written as $C_{\ell} = \ A \times (\ell/100)^{\alpha}$.  Here $A$ represents the normalized amplitude at $\ell=100$ and $\alpha$ represents the slope of the power-law for a given template. We fix $\alpha$ based 
on the measured spectra and only fit for the amplitudes as a function of the Galactic masks. We find that the slope of the 408\,MHz spectra over all the Galactic masks is consistent 
with $-2.5$. In case of DDD \halpha\ template is $-2.2$ over the masks and the same for the \smica-subtracted $353$\,GHz template is $-2.4$. The results of the power-law fit for the 
three templates and different Galactic masks are shown as a dashed lines in Fig.~\ref{fig:D1}. The amplitudes of each of the templates as a function of the Galactic masks (or \fsky) 
are listed in Table~\ref{tab:D.1}. 

The amplitudes of the given templates vary nonlinearly as a function of \fsky. They can be fitted with a second-order polynomial in a log$A$ -- log\fsky\ plane. Combining the $\nu$, $\ell$, and \fsky\ dependence, we analytically model the power spectra of the diffuse synchrotron, free-free, and dust emission components for intensity. For amplitude normalization, 
we made an assumption on the nature of the synchrotron and free-free emission. We assume a single power-law model for the synchrotron emission from 408\,MHz to microwave 
frequencies, $\nu \le 353$\,GHz. For free-free emission, we assume a single power-law model at microwave frequencies, with a mean electron temperature of $7000$\,K 
\citep{Dickinson:2003}. The power spectra of the three diffuse emission components, in $\mu {\rm K}_{\rm RJ}^2$ units, are
\footnotesize
\begin{align}
\Cl^{\rm f} \ =& \ 0.068 \times \left ( \frac{f_{\rm sky}}{0.6} \right)^{\left [6.10 \ + 3.90 \ln \left ({f_{\rm sky}/0.6} \right) \right ]} \times \left( \frac{\ell}{100}\right)^{-2.2} \times \left( \frac{\nu}{\nub}\right)^{-4.28}, \\
\Cl^{\rm s} \ =& \ 2.96 \times 10^{9} \times \left ( \frac{f_{\rm sky}}{0.6} \right)^{\left [2.12 \ + 2.67 \ln \left ({f_{\rm sky}/0.6} \right) \right ]} \times \left( \frac{\ell}{100}\right)^{-2.5} \times \left( \frac{\nu}{\nuc}\right)^{-6.0}, \\
\Cl^{\rm d} \ =& \ 0.086 \times \left ( \frac{f_{\rm sky}}{0.6} \right)^{\left [4.60 \ + 7.11 \ln \left ({f_{\rm sky}/0.6} \right) \right ]} \times \left( \frac{\ell}{100}\right)^{-2.4} \times \mathcal{D_{\nu}},
\end{align}
\normalsize
where $\nub=23$\,GHz, $\nuc=0.408$\,GHz, $\mathcal{D_{\nu}}$ is a spectral model of the dust emission given by one of the two models presented in Eqs.~\eqref{eq:7.2} and~\eqref{eq:7.3}. The derived analytical model of these power spectra are valid in the frequency range 20 to $353$\,GHz, and for \fsky\ between 0.4 and 0.8.

\begin{table}[tmb]
\begingroup
\newdimen\tblskip \tblskip=5pt
\caption{\label{tab:D.1} Amplitudes of the power spectra, normalized at $\ell$\,=\,100, as a function of \fsky. }
\nointerlineskip
\vskip -3mm
\tiny
\setbox\tablebox=\vbox{
   \newdimen\digitwidth 
   \setbox0=\hbox{\rm 0} 
   \digitwidth=\wd0 
   \catcode`*=\active 
   \def*{\kern\digitwidth}
   \newdimen\signwidth 
   \setbox0=\hbox{+} 
   \signwidth=\wd0 
   \catcode`!=\active 
   \def!{\kern\signwidth}
\halign{
#\hfil\tabskip 1.2em&
\hfil #\hfil\tabskip 3.2em&
\hfil #\hfil&
\hfil #\hfil&
\hfil #\hfil\tabskip=0pt\cr
\noalign{\doubleline}
\omit& & \multispan{3}\hfil Amplitudes \hfil\cr
\noalign{\vskip -3pt}
\omit& & \multispan{3}\hrulefill\cr
\noalign{\vskip 2pt}
Gal.& \fsky& $A_{0.408}$ & $A_{{\rm H}\alpha}$ & $A_{353-\smica}$ \cr
masks & &  [$10^{9}$ $\mu {\rm K}_{\rm RJ}^2$]&  [$\mu {\rm K}_{\rm RJ}^2$]&  [$\mu {\rm K}_{\rm RJ}^2$]\cr
\omit & & ($\alpha=-2.5$)& ($\alpha=-2.2$)& ($\alpha=-2.4$)\cr
\noalign{\vskip 4pt\hrule\vskip 6pt}
G40& 0.40& $1.913\pm0.492$& $0.011\pm0.002$& $0.042\pm0.007$\cr
G60& 0.60& $3.223\pm0.314$& $0.051\pm0.003$& $0.101\pm0.009$\cr 
G70& 0.70& $4.187\pm0.379$& $0.271\pm0.016$& $0.199\pm0.016$\cr 
G80& 0.80& $6.139\pm0.489$& $0.536\pm0.026$& $0.466\pm0.029$\cr
\noalign{\vskip 5pt\hrule\vskip 3pt}}}
\endPlancktablewide
\par
\endgroup
\end{table}

\raggedright
\end{document}